\RequirePackage{fix-cm}
\RequirePackage{rotating}
\documentclass[smallcondensed]{svjour3}     % onecolumn (ditto)
\smartqed  % flush right qed marks, e.g. at end of proof

\AtBeginDocument{%
	\setlength{\oddsidemargin}{\dimexpr(\paperwidth-\textwidth)/2-1in}%
	\setlength{\evensidemargin}{\oddsidemargin}%
	\setlength{\topmargin}{%
		\dimexpr(\paperheight-\textheight)/2-\headheight-\headsep-1in}%
}
\usepackage{graphicx}
\usepackage[numbers]{natbib}
\usepackage{url}
\usepackage{array,tabularx}
\usepackage{graphicx}
\usepackage{subfigure}
\usepackage{amsmath}
\usepackage[ruled,vlined,linesnumbered]{algorithm2e}
\usepackage{siunitx}
\usepackage{mathtools}
\usepackage{amsfonts}
\usepackage{placeins}
\usepackage{booktabs}
\usepackage{enumitem}
\usepackage{booktabs}
\usepackage[]{algorithm2e}
\usepackage{adjustbox}
\usepackage{multirow}
\usepackage{array, makecell}
\usepackage{stfloats}
\usepackage{mathrsfs}
\usepackage{fixmath}
\usepackage{tabu}
\usepackage{verbatim}
\usepackage{stackengine}
\usepackage{tikz}
\usepackage[flushleft]{threeparttable}
\usetikzlibrary{arrows,positioning}
\usepackage[bordercolor=white,backgroundcolor=gray!30,linecolor=black,colorinlistoftodos]{todonotes}

\usepackage{xcolor}
\usepackage{rotating}
\usepackage{tabularx}
\usepackage{diagbox}
\usepackage{footnote}
\usepackage{threeparttable}
\DeclareMathOperator*{\argmin}{arg\,min}
\DeclareMathOperator*{\argmax}{arg\,max}

\journalname{Under review as a paper at Journal of Autonomous Agents and Multi-Agent Systems}
\begin{document}

\title{Automatic Calibration of Dynamic and Heterogeneous Parameters in Agent-based Model%\thanks{Grants or other notes
%about the article that should go on the front page should be
%placed here. General acknowledgments should be placed at the end of the article.}
}
%\subtitle{Do you have a subtitle?\\ If so, write it here}

%\titlerunning{Short form of title}        % if too long for running head

\author{Dongjun Kim         \and
        Tae-Sub Yun \and
        Il-Chul Moon
}

%\authorrunning{Short form of author list} % if too long for running head

\institute{Dongjun Kim \at
              \email{dongjoun57@kaist.ac.kr}           %  \\
%             \emph{Present address:} of F. Author  %  if needed
           \and
           Tae-Sub Yun \at
              \email{andy2402@kaist.ac.kr}
           \and
           Il-Chul Moon \at
              \email{icmoon@kaist.ac.kr}
}

\date{ }

\maketitle

\begin{abstract}
While simulations have been utilized in diverse domains, such as urban growth modeling, market dynamics modeling, etc; some of these applications may require validations based upon some real-world observations modeled in the simulation, as well. This validation has been categorized into either qualitative face-validation or quantitative empirical validation, but as the importance and the accumulation of data grows, the importance of the quantitative validation has been highlighted in the recent studies, i.e. digital twin. The key component of quantitative validation is finding a calibrated set of parameters to regenerate the real-world observations with simulation models. While this parameter calibration has been fixed throughout a simulation execution, this paper expands the static parameter calibration in two dimensions: dynamic calibration and heterogeneous calibration. First, dynamic calibration changes the parameter values over the simulation period by reflecting the simulation output trend. Second, heterogeneous calibration changes the parameter values per simulated entity clusters by considering the similarities of entity states. We experimented the suggested calibrations on one hypothetical case and another real-world case. As a hypothetical scenario, we use the Wealth Distribution Model to illustrate how our calibration works. As a real-world scenario, we selected Real Estate Market Model because of three reasons. First, the models have heterogeneous entities as being agent-based models; second, they are economic models with real-world trends over time; and third, they are applicable to the real-world scenarios where we can gather validation data.
\keywords{Simulation Validation \and Parameter Calibration \and Machine Learning \and Digital Twin \and Agent Based Model}
\end{abstract}

\section{Introduction}
\label{sec:Introduction}

Simulation has been useful in market modeling \cite{bonabeau2002agent}, traffic management \cite{naiem2010agent}, and urban planning \cite{hosseinali2013agent}, and these simulations are real-world based, validated simulations on well-defined scenarios. Because of the detail and the fitness to the real-world, the simulation becomes a meaningful tool for design, management, and analyses of modeled real-world. While modelers accept the fundamental requirement of the validation \cite{beisbart2018computer}, many simulation models are validated through parameter calibrations that can be easily invalidated as time progresses or heterogeneity arises in the simulations. Some modelers manually calibrate the parameters to improve the fitness, but this manual effort is often limited to the inital state of the simulation.

A simulation world is naturally diverging from the real-world as the simulation progresses, and the parameter calibration should accommodate this natural divergence over the simulation period. If we utilize handpicked or heuristically calibrated parameters, we cannot mitigate the natural divergence as often as needed. Hence, a common practice in the field is calibrating simulation parameters at the up-front or over the predetermined period \cite{thiele2014facilitating} , and the calibrated parameters are used by ignoring the simulation divergence from the real-world in the simulation run-time. Therefore, if we intend to accommodate the parameter calibration as often as needed, we need to design an automatic calibration method that can resolve the natural divergence of simulation from the real-world.

This automatic calibration requires two functionalities: \textit{when to calibrate} and \textit{how to calibrate}. As the simulation diverges from the real-world, the automatic calibration needs to determine when is the best time to calibrate by itself, i.e. calibrations for every simulation timestep in extreme cases. Moreover, if we consider the heterogeneity of simulated entities in the model as the source of divergence, the automatic calibration needs to determine which groups of entities to be calibrated. After these decisions on \textit{when to calibrate}, the automatic calibration requires a component of \textit{how to calibrate}. For instance, some parameters are temporal and simulation-widely applicable, and other parameters are static and selected entity-widely applicable.

This paper categorizes the calibrated parameters into two groups: dynamic parameters and heterogeneous parameters. First, the dynamic parameters are calibrated temporally, and the parameters are shared across simulated entities at a certain timestep. To remedy the divergence over simulation time, we extract unobservable regimes \cite{hamilton2016regime} that explain the temporal dynamics. In dynamic calibration, a hidden structure (regime) is extracted from the deviation of observations and simulation results by a variant of Hidden Markov Model (HMM) \cite{fox2008hdp, bishop2006pattern}. Then, the regimes separately calibrate the dynamic parameters by taking the validation level into account. These dynamic calibrations occur iteratively, so the iterative calibration cycle would result in improving the fitness to the validation set.

Second, the heterogeneous parameters are calibrated for selected entities with the parameters. To mitigate the simulation divergence from agent heterogeneity \cite{rahmandad2008heterogeneity}, we customize a static parameter by assigning different values for different agent sub-populations to impose heterogeneity through the parameter. A hidden structure, or agent sub-population, in this case is extracted from agent-level simulation results. Particularly, a cluster model groups agents with similar agent-level state variables. To extract agent sub-populations, due to the curse of dimension, we first reduce the dimension of agent-level state variables by applying a latent representation learning algorithm, Variational Autoencoder \cite{kingma2013auto}. Afterwards, a probabilistic mixture model \cite{bishop2006pattern, teh2010dirichlet} is used to obtain hidden sub-populations of simulation entities. After the clustering, we adopt the surrogate model calibration methodology, Gaussian Process based Bayesian Optimization \cite{snoek2012practical}, applied to reduce the simulation divergence.

The suggested calibration methodologies, dynamic calibration and heterogeneous calibration, could be considered in a single framework that the methods utilize hidden structures extracted from simulation state variables. According to our framework, hidden structures fundamentally determines the target simulation validity. Hence, for instance, dynamic calibration fits the dynamic trend with capturing dynamically varying hidden regimes, and heterogeneous calibration fits the agent heterogeneity by uncovering agent sub-populations. Before we discuss the framework, we limit ourselves to the in-sample validation \cite{windrum2007empirical}. In-sample validation focuses on constructing a highly descriptive model with calibrated parameters, and we expect the in-sample validated models to repeat the real-world scenario in the simulation.

\section{Previous Research}
\label{sec:PreviousResearch}

\tabulinesep=0.8mm
\renewcommand{\arraystretch}{1.3}

Parameter calibration can be viewed as an inverse problem of estimating optimal simulation parameters that make the most consistent result with real-world observations:
\begin{equation}\label{eq:StaticCalibration}
\mathcal{\tilde{P}}=\argmin_{\mathcal{P}}d(E_{\omega \in \Omega}[\mathcal{S}(\mathcal{M}(\mathcal{P};\omega))],\mathcal{D})\footnote{$d(E_{\omega\in\Omega}[\mathcal{S}(\mathcal{M}(\mathcal{P};\omega))],\mathcal{D})$ represents the distance between the validation data $\mathcal{D}$ with the expectation over $\Omega$ of summarizing (with a function of $\mathcal{S}$) the output results ($\mathcal{O}$) from executing an agent based model (as a function of $\mathcal{M}$) with a sample random path $\omega$ and a parameter $\mathcal{P}$.},
\end{equation}
where $\mathcal{P}$ is the set of parameters to calibrate and $\mathcal{\tilde{P}}$ is the optimal parameters.

\begin{table}[!htbp]
	\renewcommand{\arraystretch}{1.3}
	\caption{Description of notations are presented in the following table.}
	\label{DescriptionofNotations}
	\centering
	%\resizebox{\textwidth}{!}{
	\begin{tabularx}{\columnwidth}{l|l} % <-- Alignments: 1st column left, 2nd middle and 3rd right, with vertical lines in between
		\toprule
		\textbf{Notation} & \textbf{Description} \\
		\hline
		%		$f$ & Simulation error function \\
		%		$MS$ & Simulation model structure \\
		%		$IS$ & Simulation input scenario \\
		$\mathcal{M}$ & Agent-based model \\
		$\delta$ & Optimization algorithm\\
		$d$ & Simulation error distance measure \\
		$\mathcal{P}$ & Simulation input parameters to optimize \\
		$\mathcal{P}_{dyn}$ & Simulation dynamic parametrs to optimize \\
		$\mathcal{P}_{het}$ & Simulation heterogeneous parameters to optimize \\
		$\mathcal{O}$ & Simulation output state variables \\
		$\Omega$ & Sample path space of stochastic simulation \\
		\hline
		$\mathcal{D}$ & Empirical validation summary statistics\\
		$\mathcal{S}$ & Simulation summary statistics function\\
		$\mathcal{D}_{dyn}$ & Empirical validation summary statistics data used in dynamic calibration\\
		$\mathcal{D}_{het}$ & Empirical validation summary statistics data used in heterogeneous calibration\\
		$\mathcal{S}_{dyn}$ & Dynamic calibration summary statistics function \\
		$\mathcal{S}_{het}$ & Heterogeneous calibration summary statistics function\\
		$\mathcal{S}_{agent}$ & Agent-level simulation result\\
		$\mathcal{S}_{latent}$ & The representational compressed data of $\mathcal{S}_{agent}$ \\
		$\mathcal{G}$ & Aggregated data of simulation error in heterogeneous calibration \\
		\hline
		$N_{dyn}$ & The number of dynamic calibration target parameters \\
		$N_{het}$ & The number of heterogeneous calibration target parameters \\
		$C_{cal}$ & The number of calibration framework iterations \\
		$C_{dyn}$ & The number of dynamic calibration iterations \\	
		$C_{het}$ & The number of heterogeneous calibration iterations \\
		$S_{dyn}$ & The dimension of summary statistics in dynamic calibration\\
		$S_{het}$ & The dimension of summary statistics in heterogeneous calibration\\
		$K_{dyn}$ & The number of hidden regimes in HMM \\
		$K_{het}$ & The number of agent sub-populations\\
		$A$ & The number of agents \\
		$T$ & The number of total simulation time \\
		$R$ & The number of simulation replications for each candidate hypothesis \\
		$I$ & The number of candidate hypotheses \\
		$U$ & The number of merged regimes \\
		$Att$ & The number of selected agent attributes for agent clustering \\
		$H$ & The dimension of hidden representation of agent attributes \\
		\bottomrule
	\end{tabularx}
	%}
\end{table}

In detail, there are three steps in the parameter calibration \cite{hartig2011statistical}. The first step is choosing summary statistics $\mathcal{D}$ from observations that are selected as emprirical validation data. Total $S$ number of summary statistics are considered, with each of summary statistic is the characteristic quantity of the observational distribution. Simulation summary statistics are generated from the simulation output $\mathcal{O}$ with summary statistics function $\mathcal{S}$, given by $\mathcal{S}(\mathcal{M}(\mathcal{P};\omega)=\mathcal{O})$, by packing up the output distribution into few principal features, where the agent-based model is $\mathcal{M}:\mathcal{P}\times\Omega\rightarrow\mathcal{O}$, as described in Tab. \ref{DescriptionofNotations}. For the next step, the simulation distance formula, $d$, will be defined to measure the closeness between the simulation result $\mathcal{S}$ and the summary statistics $\mathcal{D}$ \cite{rand2012does, marks2013validation}. Simulation stochasticity is represented by $\omega$, which is a sampled path from the sample path space $\Omega$. As the final step, the modeler applies an iterative stochastic function optimization algorithm to reduce the simulation average error, $d(E_{\omega\in\Omega}[\mathcal{S}(\mathcal{M}(\mathcal{P};\omega))],\mathcal{D})$.

Since the optimization lies at the core of the calibration process \cite{thiele2014facilitating, brenner2006practical}, modelers have categorized optimization algorithms for simulations into the four taxonomies \cite{fu2002optimization} listed below:
\begin{enumerate}
	\item \textit{Category 1} finds an optimal solution of stochastic function (under certain condition).
	\item \textit{Category 2} finds an optimal solution of deterministic function (under certain condition).
	\item \textit{Category 3} finds an optimal solution with probability $p$ less than 1.
	\item \textit{Category 4} finds a practically good solution stably.
\end{enumerate}
These categories are generalized criteria, so each optimization algorithm is assessed by the criteria in the following survey.

\begin{table}[!htbp]
	\renewcommand{\arraystretch}{1.3}
	\caption{Previous researches on simulation static parameter calibration are listed. The suggested calibration methodologies in this paper are listed in the last two columns.}
	\label{PreviousResearch}
	\centering
	%	\resizebox{0.5\textwidth}{!}{
	\begin{tabular}{|l|l|l|l|l|l|}
		%\toprule
		
		\hline
		& \multicolumn{1}{p{2cm}|}{The name of algorithms} & Search Method & \multirow{1}{*}[-0.25em]{\makecell[l]{Global\\optimality}} & Advantage & Disadvantage \\
		\hline
		\multirow{3}{*}[-2.2em]{\parbox{1.5cm}{Design of\\Experiments}} & \multicolumn{1}{p{2cm}|}{Full factorial \cite{thiele2014facilitating}} & \multicolumn{1}{p{3cm}|}{Grid Search} & \multicolumn{1}{p{1cm}|}{Yes} & \multicolumn{1}{p{3cm}|}{Always find the global optimum within grid distance} & \multicolumn{1}{p{3cm}|}{Computationally expensive} \\\cline{2-6}
		& \multicolumn{1}{p{2cm}|}{Latin hypercube \cite{thiele2014facilitating}} & \multicolumn{1}{p{3cm}|}{Space filling search} & \multicolumn{1}{p{1cm}|}{No} & \multicolumn{1}{p{3cm}|}{Sampling from equally probable axis-aligned hyperplanes} & \multicolumn{1}{p{3cm}|}{Computationally expensive} \\\cline{2-6}
		& \multicolumn{1}{p{2cm}|}{Random \cite{thiele2014facilitating}} & \multicolumn{1}{p{3cm}|}{Random search} & \multicolumn{1}{p{1cm}|}{No} & \multicolumn{1}{p{3cm}|}{Exploration} & \multicolumn{1}{p{3cm}|}{No exploitation} \\
		\hline
		\multirow{2}{*}[-2.1em]{\parbox{1.5cm}{Derivative\\-based\\Algorithms}} & \multicolumn{1}{p{2cm}|}{Gradient Descent with finite difference} & \multicolumn{1}{p{3cm}|}{Gradient Descent} & \multicolumn{1}{p{1cm}|}{No} & \multicolumn{1}{p{3cm}|}{Model-free algorithm} & \multicolumn{1}{p{3cm}|}{Multiple evaluation required for an update} \\\cline{2-6}
		& \multicolumn{1}{p{2cm}|}{Gradient Descent with IPA \cite{suri1987infinitesimal}} & \multicolumn{1}{p{3cm}|}{Gradient Descent} & \multicolumn{1}{p{1cm}|}{No} & \multicolumn{1}{p{3cm}|}{Single evaluation required for an update} & \multicolumn{1}{p{3cm}|}{Limited applicability, high variance} \\
		\hline
		\multirow{2}{*}[-4em]{\parbox{1.5cm}{Heuristic\\Algorithms}} & \multicolumn{1}{p{2cm}|}{Simulation Annealing \cite{hara2013configuring}} & \multicolumn{1}{p{3cm}|}{Local search with cooling mechanism} & \multicolumn{1}{p{1cm}|}{No} & \multicolumn{1}{p{3cm}|}{Able to escape from local minimum} & \multicolumn{1}{p{3cm}|}{Need to control the cooling rate} \\\cline{2-6}
		& \multicolumn{1}{p{2cm}|}{Particle Swarm Optimization \cite{noel2004simulation}} & \multicolumn{1}{p{3cm}|}{Update particles' positions and velocities towards current best solution} & \multicolumn{1}{p{1cm}|}{No} & \multicolumn{1}{p{3cm}|}{Model-free optimization applicable to high dimensional parameters} & \multicolumn{1}{p{3cm}|}{No guarantee on the convergence to the global optimum} \\\cline{2-6}
		& \multicolumn{1}{p{2cm}|}{Genetic Algorithm \cite{stonedahl2010evolutionary}} & \multicolumn{1}{p{3cm}|}{Selection, crossover, and mutation} & \multicolumn{1}{p{1cm}|}{No} & \multicolumn{1}{p{3cm}|}{Robust in optimization performance, model-free} & \multicolumn{1}{p{3cm}|}{No guarantee on the convergence to the global optimum} \\
		\hline
		\multirow{3}{*}[-7em]{\parbox{1.5cm}{Sampling\\-based\\Algorithms}} & \multicolumn{1}{p{2cm}|}{Rejection Sampling \cite{tavare1997inferring}} & \multicolumn{1}{p{3cm}|}{Rejection Sampling} & \multicolumn{1}{p{1cm}|}{Yes} & \multicolumn{1}{p{3cm}|}{Eventually converge to the target distribution} & \multicolumn{1}{p{3cm}|}{Highly inefficient if the prior is dissimilar with the target distribution, only feasible for the calibration task of few parameters} \\\cline{2-6}
		& \multicolumn{1}{p{2cm}|}{Monte-Carlo Markov Chain \cite{marjoram2003markov}} & \multicolumn{1}{p{3cm}|}{Random walk from the previous evaluated particles} & \multicolumn{1}{p{1cm}|}{Yes} & \multicolumn{1}{p{3cm}|}{Eventually converge to the target distribution, much faster than rejection sampling} & \multicolumn{1}{p{3cm}|}{Sampled points are highly correlated, slow convergence rate} \\\cline{2-6}
		& \multicolumn{1}{p{2cm}|}{Sequential Monte Carlo \cite{sisson2007sequential}} & \multicolumn{1}{p{3cm}|}{Sampling from a learned proposal distribution} & \multicolumn{1}{p{1cm}|}{Yes} & \multicolumn{1}{p{3cm}|}{Avoid sampling inefficiency through learning of intermediate distribution, less prone to get stuck in regions of low probability} & \multicolumn{1}{p{3cm}|}{Choice of the number of particles and the forward-backword kernels affects the performance} \\\cline{2-6}
		\hline
		\multirow{2}{*}[-5em]{\parbox{1.5cm}{Model\\-based\\Algorithms}} & \multicolumn{1}{p{2cm}|}{Gaussian Process Regression-based Bayesian Optimization \cite{tresidder2012acceleration}} & \multicolumn{1}{p{3cm}|}{GPR as surrogate, Bayesian Optimization as parameter search} & \multicolumn{1}{p{1cm}|}{Yes} & \multicolumn{1}{p{3cm}|}{Theoretic analysis, fast convergence} & \multicolumn{1}{p{3cm}|}{No standard criteria on the selection of kernel and acquisition function} \\\cline{2-6}
		& \multicolumn{1}{p{2cm}|}{XGBoost-based Nelder Mead Optimization \cite{lamperti2018agent}} & \multicolumn{1}{p{3cm}|}{XGBoost as surrogate, Nelder Mead as parameter search} & \multicolumn{1}{p{1cm}|}{No} & \multicolumn{1}{p{3cm}|}{Applicable to high dimension} & \multicolumn{1}{p{3cm}|}{Parametric surrogate model} \\\cline{2-6}
		\hline
		\multirow{2}{*}[-3em]{\parbox{1.5cm}{Proposed Framework \tiny{(Including interactions between dynamic and heterogeneous calibrations)}}} & \multicolumn{1}{p{2cm}|}{Proposed Dynamic Calibration} & \multicolumn{1}{p{3cm}|}{Sampling from a posterior distribution for each of merged regime} & \multicolumn{1}{p{1cm}|}{No} & \multicolumn{1}{p{3cm}|}{Able to calibrate the dynamic parameter} & \multicolumn{1}{p{3cm}|}{No theoretical convergence} \\
		\cline{2-6}
		& \multicolumn{1}{p{2cm}|}{Proposed Heterogeneous Calibration} & \multicolumn{1}{p{3cm}|}{GPR-based Bayesian Optimization} & \multicolumn{1}{p{1cm}|}{Yes} & \multicolumn{1}{p{3cm}|}{Able to calibrate the heterogeneous parameter} & \multicolumn{1}{p{3cm}|}{Injecting heterogeneity induces an optimization with higher dimension than original static calibration with the same parameters} \\
		\hline
		
		%\bottomrule
	\end{tabular}
	%	}
\end{table}

\subsection{Calibration with Design of Experiment}
\label{sec:DOE}
Besides of the algorithm-based adaptive calibration, the simplest systematic calibration is using the standard design of experiments, or DOE \cite{lee2015complexities}. However, the classical DOE algorithms have pre-determined input parameters, so those do not adaptively generate the parameters. In addition, the DOEs are not specifically designed to handle the interference between the input parameters, which affects the simulation outputs, collectively.

The full factorial search, in Tab. \ref{PreviousResearch}, is powerful if there are few parameters to calibrate \cite{thiele2014facilitating}. However, the Latin hypercube, a near-uniformly spce filling search method by sampling from the equally probable axis-aligned hyperplanes, could be an alternative of full factorial search if there are too many grids to evaluate. The random search mainly acts as exploring the search space.

\subsection{Calibration with Derivative-based Algorithms}
\label{sec:Derivative-based}
The optimization algorithm could be a derivative-based calibration with gradient descents to minimize expected simulation errors. This derivative-based optimization belongs to the category of \textit{Category 4} in the above taxonomy. Since the derivative of simulation error is not attainable, the derivative-based calibration requires to use a derivative approximation techniques such as finite difference or Infinitesimal Perturbation Analysis (IPA) \cite{suri1987infinitesimal, ho1990infinitesimal}. While the finite difference is attainable, this method is not appropriate for the calibration task with the expensive evaluation cost since it requires a number of evaluations, proportional to the optimization dimension, to get a derivative at a point. IPA approximates the derivative with a single function evaluation, but this single evaluation makes IPA too volatile to be used as a calibration method. In addition, IPA is less applicable than the variants of finite difference methods.

\subsection{Calibration with Heuristic Algorithms}
\label{sec:Heuristic}
Among derivative-free optimization algorithms, Heuristic algorithm is frequently used for the calibration task. Heuristic algorithms search a next design point $\mathcal{P}_{C+1}$ with a current parameter $\mathcal{P}_{C}$ by $\mathcal{P}_{C+1}=\delta(\mathcal{P}_{C})$, where $\delta$ is the optimization algorithm. Most of Heuristic algorithms, due to its memory-less property and simple selection process, are not proven to find global optimum. However, some Heuristic algorithms with simple idea work better than sophisticated designed optimization algorithms in the practice \cite{stonedahl2010evolutionary, stonedahl2010finding}, which makes the Heuristic algorithms to be \textit{Category 4}.

While some research applies either Simulated Aneealing \cite{hara2013configuring, kirkpatrick1983optimization} or Particle Swarm Optimization \cite{noel2004simulation, kennedy2010particle}; most of calibration articles, which construct calibration as an optimization problem, use Genetic Algorithm \cite{calvez2005automatic, yang2009agent, nannen2006method, heppenstall2007genetic, stonedahl2010finding, whitley1994genetic} as an optimization solver. There are four main reasons for using Genetic Algorithm \cite{nguyen2014review}. First, Genetic Algorithm is robust on the task of global optimization for highly nonlinear multi-modal discontinuous functions. Second, Genetic Algorithm is designed to align well with parallel computations in multi-processor computers. Parallelizability is especially advantageous in a computationally expensive simulation optimization. Furthermore, unlike standard DOEs, Genetic Algorithm is suitable in treating functions with input parameters intricately entangled in its effect on outputs. Lastly, both continuous and discrete variables could be calibrated using Genetic Algorithm, which makes both simulation parameter $\mathcal{P}$ and simulation structure to be optimized at one hand.

\subsection{Calibration with Sampling-Based Algorithms}
\label{sec:Sampling-based}
In statistical inference algorithms, the Approximate Bayesian Computation(ABC) \cite{beaumont2010approximate, tavare1997inferring, marjoram2003markov, sisson2007sequential} is a class of sampling-based algorithms used when the likelihood, $p(\mathcal{S}(\mathcal{M}(\mathcal{P};\omega)=\mathcal{O})=\mathcal{D}|\mathcal{P})$, is intractible \cite{jabot2013easy}. In ABC, the likelihood is approximately defined as $p(\mathcal{S}(\mathcal{M}(\mathcal{P};\omega))=\mathcal{D}|\mathcal{P})\simeq c\cdot p(d(\mathcal{S}(\mathcal{M}(\mathcal{P};\omega)),\mathcal{D})<\tau|\mathcal{P})$, where $\mathcal{S}$ is a summary statistics function, $c$ is a normalization constant, and $\tau$ is a pre-difined threshold. The likelihood might be analytically calculated in a simple stochastic simulation, but it is intractible in the case of agent-based simulation \cite{hartig2011statistical, thiele2014facilitating}, where the simple agent-level behavior merged up to the emergent system-level behavior. Also, this calculation is infeasible if the agent behaviors are the collection of discrete logics.  Some sampling-based ABC algorithms use only the previous parameters $\mathcal{P}_{C}$ to generate the next set of design points $\mathcal{P}_{C+1}$, by $\mathcal{P}_{C+1}=\delta(\mathcal{P}_{C})$, and this generated parameters may form a stationary distribution which could be an inferred distribution that satisfies the calibration task. Hence, under the condition of converging to the stationary distribution, these sampling based ABC is categorized into either \textit{Category 1} or \textit{Category 2}. 

In spite of this optimization property, ABC inference algorithms suffers from the slow convergence rate since the model requires the tens of thousands of point evaluations \cite{hartig2011statistical, thiele2014facilitating} in the case of large scale agent-based simulation. The proposed dynamic calibration algorithm in this article borrows an essential idea of Sequantial Monte Carlo ABC algorithm \cite{sisson2007sequential}, in that the next set of evaluation points, a.k.a particles, are sampled from the proposal distribution generated from the current evaluation points with likelihoods.

\subsection{Calibration with Model-Based Algorithms}
\label{sec:Model-based}
Surrogate model-based optimization algorithms \cite{tresidder2012acceleration, lamperti2018agent} are derivative-free optimization algorithms to calibrate with two iterative steps: predicting response surface, $p(d(\mathcal{S}(\mathcal{M}(\mathcal{P};\omega)),\mathcal{D})|\mathcal{P})$, and setting new evaluation design parameter, $\mathcal{P}_{C+1}=\delta(\mathcal{P}_{1},...,\mathcal{P}_{C})$. The variants of regression models, including parametric and nonparametric models, could be used in the construction of a response surface. In large scale simulation, nonparametric kernel-based Gaussian process regression, or a.k.a. \textit{kriging}, \cite{rasmussen2006gaussianprocessesformachinelearning} would be flexible in estimating the response surface when only a small number of evaluations are obtained. Gaussian process regression is a Bayesian model that predicts a response surface, $p(d(\mathcal{S}(\mathcal{M}(\mathcal{P};\omega)),\mathcal{D})|\mathcal{P})$, as a normal distribution, $\mathcal{N}(\mu(\mathcal{P}),\sigma^{2}(\mathcal{P}))$. The predictive variance, $\sigma(\mathcal{P})$, measures the uncertainty of the predictive mean $\mu(\mathcal{P})$. As a parameter generation algorithm, Bayesian optimization \cite{schonlau1997computer}, using an acquisition function, with a Gaussian process regression is one of the most efficient tools to find the global optimum in terms of function evaluations \cite{lizotte2008practical}. The proposed heterogeneous calibration could be categorized as one of surrogate model-based algorithm.

Theoretical analysis, in particular, on Bayesian optimization with Gaussian pocess regression has been studied \cite{mockus2012bayesian, vazquez2010convergence, grunewalder2010regret, bull2011convergence, ryzhov2016convergence}, and these analysis provides the convergence of the stochastic function with conditions\cite{bull2011convergence}. Thus, the calibration with Gaussian process regression could be categorized in either \textit{Category 1}, \textit{Category 2}, or \textit{Category 3} by the specifics of the conditions. Having said that, the practical conditions would be different from the theoretical analyses, so we need to study the feasibility with an actual system with an implemented calibration model.

\section{Preliminaries}
\label{sec:Preliminaries}
This section introduces machine learning algorithms that we used for our calibration framework to extract hidden structures. These machine learning algorithms are used as components in our calibration framework. We present a set of clustering and latent representation models because the calibration needs to capture the divergence trend between the validation and the simulation results. Afterwards, we introduce a nonparametric response surface model and its corresponding optimization model because the calibration needs to suggest the next experimental point.

\subsection{Clustering Models for Divergence Trend Detection}
\label{sec:Clustering}
This subsection introduces multple clustering algorithms, Hidden Markov Model (HMM) \cite{bishop2006pattern}, Gaussian Mixture Model (GMM) \cite{bishop2006pattern} and Dirichlet Process Mixture Model (DPMM) \cite{teh2010dirichlet} in Section~\ref{sec:TemporalClustering} and~\ref{sec:MixtureClustering}. Also, we present a latent structure modeling method, Variational Autoencoder (VAE) \cite{kingma2013auto}, in Section~\ref{sec:VAE}. 

\subsubsection{Temporal Clustering Models (HMM)}
\label{sec:TemporalClustering}

HMM \cite{bishop2006pattern} is a statistical graphical model that has an input as time-dependent data, and an output as time-dependent regimes. The graphical architecture of HMM consists of two random variables: 1) the latent variable $z_{t}$ for the hidden regime, and 2) the observable variable $O_{t}$ where the graphical relations $z_{t-1}\rightarrow z_{t}$ and $z_{t}\rightarrow O_{t}$ are assumed. The below is the generative process of HMM.
\begin{equation}\label{eq:HMM}
\begin{aligned}
& z_{1} \sim Categorical(\mathbold{\pi}_{init}) \\
& z_{t}|z_{t-1} \sim Categorical(\mathbold{\pi}_{z_{t-1}}) \\
& O_{t}|z_{t} \sim F(\theta_{z_{t}})
% & O_{t}|z_{t} \sim \mathcal{N}(\mu_{z_{t}},\sigma^{2}_{z_{t}}), \\
\end{aligned}
\end{equation}
$\mathbold{\pi}_{init}$ is the initial probability of clustering; $\mathbold{\pi}_{z_{t-1}}$ is the transition probability; and $\mathbold{\pi}_{z_{t-1}=i}=p(z_{t}|z_{t-1}=i)$. $Categorical$ denotes a categorical distribution; $F$ is the emission probability distribution; and $\theta_{z_{t}}$ is the emission parameter, which specify the type and the shape of the distribution on observation variable $O_{t}$. If an observation is assumed to follow the Gaussian distribution, then the emission distribution $F$ becomes the Gaussian distribution, with the emission parameter $\theta_{z_{t}}$ of the mean $\mu_{z_{t}}$ and the standard deviation $\sigma_{z_{t}}$. The Baum-Welch algorithm, which is the Expectation-Maximization algorithm on HMM, estimates the parameters and the hidden regime assignments.

\subsubsection{Clustering Mixture Models (GMM, DPMM)}
\label{sec:MixtureClustering}
In this subsection, we introduce an entity-wise static hidden structure extraction algorithm required in heterogeneous calibration. We define that a group has agents who share similarities in their state variables. Mixture models follows a probabilistic approach for representing the presence of sub-populations within the overall population by assuming the observation data is distributed multi-modally. The mixture models have a categorical cluster assignment variable, $z_{i}$, as a latent variable for inference, which follows the categorical distribution of $\mathbold{\pi}_{k}$. Each mixture component has the emission probability parameter, $\phi_{k}$, and the emission probability distribution, $F$. The parametric model fixes the number of hidden groups, and the nonparametric model automatically finds the optimal number of sub-populations.

Gaussian Mixture Model (GMM) \cite{bishop2006pattern} is a parametric clustering algorithm that the observation distribution is assumed to follow the mixture of the Gaussian distributions. The emission parameter, $\phi_{k}$, is consisted of the mean $\mu_{k}$ and the standard deviation $\sigma_{k}$. We have the following generative process of GMM:
\begin{equation}\label{eq:GMM}
\begin{aligned}
%& \pi \sim Dirichlet(\frac{\alpha}{N_{c}} \textbf{1}_{N_{c}}) \\
& z_{a}\sim Categorical(\mathbold{\pi}) \\
%& \phi_{k} \sim G_{0} \\
%& O_{i}|z_{i}\sim F(\phi_{z_{i}})
& O_{a}|z_{a}\sim \mathcal{N}(\mu_{z_{a}},\sigma^{2}_{z_{a}})
\end{aligned}
\end{equation}
Again, the Expectation-Maximization algorithm is used to maximize the likelihoods. In the Expectation step, $z_{a}$ is assigned by the expectation of the likelihoods. In Maximization step, we maximize the likelihoods by optimizing parameters, such as $\mathbold{\pi}_{k}$, $\mu_{k}$, and $\sigma_{k}$ in GMM.

A measure theoretic view of the parametric bayesian mixture distribution is $G^{N_{c}}=\sum_{k=1}^{N_{c}}\pi_{k}\delta_{\theta_{k}}$, where $\{\pi_{k}\}_{k=1}^{N_{c}}|\gamma\sim Dir(\gamma/N_{c},...,\gamma/N_{c})$ and $\{\theta_{k}\}_{k=1}^{N_{c}}\sim G_{0}$. Dirichlet process \cite{teh2010dirichlet} is a prior for the nonparametric mixture clustering model, in which the data-adaptive optimal number of clusters are automatically extracted. As the mixture weight, $\{\pi_{k}\}_{k=1}^{\infty}$, is no more finite dimension if the model is nonparametric, the infinite weight prior is not a distribution of a vector, Dirichlet distribution, rather it is a distribution of measures, Dirichlet process. Dirichlet Process \cite{teh2010dirichlet} $DP(\gamma)$ is a prior distribution of the infinitized mixture weight $\{\pi_{k}\}_{k=1}^{\infty}$ that makes the mixture distribution into the infinite weighted sum $G=\sum_{k=1}^{\infty}\pi_{k}\delta_{\theta_{k}}$. The infinite weight, $\mathbold{\pi}$ from the Dirichlet process, is constructed by the stick-breaking process, $\pi_{k}=\pi_{k}'\prod_{l=1}^{k-1}(1-\pi_{l}')$ with $\pi_{k}'\sim Beta(1,\gamma)$. In general, base distribution $G_{0}$ and emission distribution $F$ are conjugate priors to make posterior computation tractable. The below specifies the generative process of DPMM.
\begin{equation}\label{eq:DPMM}
\begin{aligned}
& \mathbold{\pi} \sim GEM(\alpha) \\
& z_{a} \sim Categorical(\mathbold{\pi}) \\
& \phi_{k} \sim G_{0} \\
& O_{a} | z_{a} \sim F(\phi_{z_{a}})
\end{aligned}
\end{equation}
The parameters of DPMM are inferenced by maximizing the posterior distribution. A Gibbs sampling algorithm is used in maximizing a posterior in our implementation. At each iteration, the Gibbs sampling algorithm updates $z_{a}$ with fixed hidden structures of other assignments. After iterations, the Gibbs sampling method is known to find the equilibrium distribution.

\subsubsection{Variational Autoencoder}
\label{sec:VAE}
Agent-level state variables often have a large dimension that causes failure in extracting meaningful clusters due to the curse of dimensionality. Assuming the data is distributed normally, as the dimension, $n$, gets larger, most of the mass lies in the ellipsoidal shell of radius $\sqrt{\text{trace}(\Sigma)}$ measured by the Mahalanobis distance, where $\Sigma$ is the covariance of the Gaussian distribution. This requires about $2^{\text{O}(n)}$ sampled data points in order to get a single point with distance less than $\sqrt{\text{trace}(\Sigma)}/2$ from the center of the distribution \cite{dasgupta1999learning}. Therefore, we introduce a latent representation learning algorithm to compress the high dimensional data into low dimensional data.

Variational autoencoder \cite{kingma2013auto} is a probabilistic latent variable model that extracts the latent representation of data. The structure of Variational Autoencoder is consisted of two neural networks: an encoder network and a decoder network. An encoder network extracts the hidden latent representation from the input of the original data. The extracted representation propagated to the decoder network to regenerate the original data.
The vanilla version of Variational Autoencoder assumes the encoder distribution to have a standard Gaussian distribution $\mathcal{N}(0,I)$ as a prior.

VAE parametrizes the posterior distribution with the neural network parameters. To make the reconstructed output similar to the original input, the log-likelihood of the Variational Autoencoder should be maximized. However, since the log likelihood is intractable to calculate, an alternative loss function, Evidence Lower BOund (ELBO), is suggested as a supporting lower bound function. In detail, the below formula shows ELBO.
\begin{equation}\label{eq:VAE}
\begin{aligned}
\log p(x)&= \log\int_{z}p(x|z)p(z)dz\\
&=\log\int_{z}q(x|z)\frac{p(x|z)p(z)}{q(z|x)}dz\\
&\ge \int_{z}q(x|z)\log\frac{p(x|z)p(z)}{q(z|x)}dz\\
&=E_{q}[\log p(x|z)]-KL(q||p)
\end{aligned}
\end{equation}
The reconstruction term, $E_{q}[\log p(x|z)]$, calculates the expectation of the log-likelihood when the posterior latent representation follows a variational distribution of $q$. The regularization term, Kullback-Leibler divergence $KL(q||p)$, measures how much information is lost by using the encoder distribution, $q$, to represent the prior distribution, $p$. With a stochastic gradient descent method, the neural network parameters are learned.

\subsection{Gaussian Process Regression-based Bayesian Optimization}
\label{sec:GPR-based_BO}
After a sub-population of agents is obtained, heterogeneous calibration becomes an optimization problem with a heterogeneity-embedded loss function. This subsection introduces the surrogate based algorithm to optimize heterogeneous calibration problem, Gaussian process regression-based Bayesian optimization. Gaussian process regression (GPR) \cite{rasmussen2006gaussianprocessesformachinelearning} is a Bayesian nonparametric regression algorithm that estimates a response variable, or simulation error in this paper, as a black-box function of dependant variables, or simulation parameters in our case. If $\mathcal{P}$ is an arbitrary input parameter, GPR estimates the predictive output distribution as a Gaussian distribution of $\mathcal{N}(\mu_{err}(\mathcal{P}),\sigma_{err}^{2}(\mathcal{P}))$.

Suppose the data $\mathcal{G}=\big\{\big(\mathcal{P}_{c'},d(E_{R}[\mathcal{S}(\mathcal{M}(\mathcal{P}_{c'};\omega_{r}))],\mathcal{D})\big)\big\}_{c'=1}^{c}$ is given, where $d(E_{R}[\mathcal{S}(\mathcal{M}(\mathcal{P}_{c'};\omega_{r}))],\mathcal{D})$ is the average of summarizing (with a function of $\mathcal{S}$) output results ($\mathcal{O}$) from executing an agent based model (as a function of $\mathcal{M}$) with a sample random path $\omega_{r}$ and a parameter $\mathcal{P}_{c'}$ with $R$ replications. Then, the predictive posterior distribution is specified in the below.
\begin{equation}\label{eq:GaussianProcessPosterior}
p(E_{\omega\in\Omega}[\mathcal{S}(\mathcal{M}(\mathcal{P};\omega))]|\mathcal{G}) \sim \mathcal{N}(\kappa^{T}cov_{c}^{-1}F,K_{\mathcal{P},\mathcal{P}}-\kappa^{T}cov_{c}^{-1}\kappa)
\end{equation}
$\kappa=[K_{\mathcal{P},\mathcal{P}_{1}},...,K_{\mathcal{P},\mathcal{P}_{c}}]^{T}$; $F=\big[d(E_{R}[\mathcal{S}(\mathcal{M}(\mathcal{P}_{1};\omega_{r}))],\mathcal{D}),...,d(E_{R}[\mathcal{S}(\mathcal{M}(\mathcal{P}_{c};\omega))],\mathcal{D})\big]^{T}$; and $(cov_{c})_{i,j}=K_{\mathcal{P}_{i},\mathcal{P}_{j}}+\frac{1}{\beta}\delta_{\{i=j\}}$. Here, $K_{\mathcal{P},\mathcal{P}'}$ is the kernel distance between $\mathcal{P}$ and $\mathcal{P}'$. The kernel distance is selected by considering the shape of the true function and the meaning of each input variable.

The kernel function, $K$, plays the key role in GPR. Stationary kernel functions, such as Squared Exponential kernel and Matern kernel, are well-studied in theory and experiments. A kernel function has hyperparameters to determine the exact shape of distances where the hyperparameters could be learned by maximizing the log likelihood from data. Recently, a fully Bayesian treatment of kernel hyperparameters showed its usefulness \cite{snoek2012practical, wu2017bayesian}.

The Bayesian optimization \cite{snoek2012practical, brochu2010tutorial, frazier2018tutorial, lizotte2008practical, sasena2002exploration, preuss2018global} is an adaptive optimization tool that selects a new design point to optimize the response. It finds the next evaluation point by optimizing an acquisition function, which is generated using $\mu_{err}(\cdot)$ and $\sigma_{err}(\cdot)$. This acquisition function optimization problem is represented in the below.
\begin{equation}
\mathcal{P}_{c+1}=\argmax_{\mathcal{P}\in \mathbb{H}}AF(\mathcal{P}|\mathcal{G})
\end{equation}
This is called an inner optimization problem, where $AF$ is an acquisition function, and $\mathbb{H}=\big\{\mathcal{P}\big|\lVert\mathcal{P}-\mathcal{P}_{c'}\rVert_{2}\ge r_{0} \text{ for all }\mathcal{P}_{c'}\in\mathcal{G}\big\}$ with given radius $r_{0}\ge0$. $\mathbb{H}$ prevents a calibration optimization being stuck in a local optimum by forcing an exploration. A single calibration iteration contains an optimization of an easy-to-optimize inner problem.

The exploration-exploitation ratio affects the convergence speed of the calibration problem. The purely exploitation search algorithm is using the predictive mean acquisition function, $\mu_{err}(\cdot)$. Minimizing the predictive mean will find a next evaluation point with a pure exploitation. For the pure exploration search algorithm, the predictive variance acquisition function, $\sigma_{err}(\cdot)$, is used. In this case, maximizing this function will find a design parameter that has never been visited before.

%Expected Improvement is an Acquisition Function that uses both predictive mean and variance, which makes parameter generation step as the mixture of exploration and exploitation. Expected Improvement Function at $x$ measures how much minimum value will be improved by testing parameter $x$. By maximizing Expected Improvement, next parameter is generated. Expected Improvement gain its popularity by having both theoretical analysis \cite{mockus2012bayesian, vazquez2010convergence, grunewalder2010regret, bull2011convergence} and experimental success \cite{snoek2012practical, lizotte2008practical, schonlau1997computer}. 

The Expected Improvement (EI) is a mixture of the exploration and the exploitation. The below formula is an acquisition function of EI.
\begin{equation} \label{eq:ExpectedImprovement}
\begin{aligned}
\text{EI}(\mathcal{P}|\mathcal{G})=&\int_{0}^{\infty}P(y(x)\le y_{min}-m)dm\\
=&(\min_{c'}{d(E_{R}[\mathcal{S}(\mathcal{M}(\mathcal{P}_{c'};\omega))],\mathcal{D})}-\mu_{err}(\mathcal{P}))
\Phi\bigg(\frac{\min_{c'}{d(E_{R}[\mathcal{S}(\mathcal{M}(\mathcal{P}_{c'};\omega_{r}))],\mathcal{D})}-\mu_{err}(\mathcal{P})}{\sigma_{err}(\mathcal{P})}\bigg)\\
+&\sigma_{err}(\mathcal{P})\phi\bigg(\frac{\min_{c'}{d(E_{R}[\mathcal{S}(\mathcal{M}(\mathcal{P}_{c'};\omega_{r}))],\mathcal{D})}-\mu_{err}(\mathcal{P})}{\sigma_{err}(\mathcal{P})}\bigg)
\end{aligned}
\end{equation}
$\phi$ is a probability density function, and $\Phi$ is a cumulative density function of the standard Gaussian distribution, $\mathcal{N}(0,1)$. The integration in Eq. \ref{eq:ExpectedImprovement} is marginalized out as the addition of two terms, where the first term (exploitation) is the predicted difference between the minimum of simulation errors, $\min_{c'}{d(E_{R}[\mathcal{S}(\mathcal{M}(\mathcal{P}_{c'};\omega_{r}))],\mathcal{D})}$, and the predicted mean, $\mu_{err}(\mathcal{P})$, at the current point, $\mathcal{P}$, penalized by the probability of improvement, $\Phi(\{\min_{c'}{d(E_{R}[\mathcal{S}(\mathcal{M}(\mathcal{P}_{c'};\omega_{r}))],\mathcal{D})}-\mu_{err}(\mathcal{P})\}/\sigma_{err}(\mathcal{P}))$. The second term (exploration) represents the uncertainty of the surrogate model predictions. Therefore, EI is a balanced exploration and exploitation method in its nature. By maximizing the EI acquisition function, the next design parameter, $\mathcal{P}_{c+1}$, is generated. EI gained its popularity with both theoretical analyses \cite{mockus2012bayesian, vazquez2010convergence, grunewalder2010regret, bull2011convergence, ryzhov2016convergence} and experimental successes \cite{snoek2012practical, lizotte2008practical, schonlau1997computer}.

EI is one-step Bayes optimal \cite{wu2016parallel}, but EI becomes heuristic when we consider multi-steps. In calibration tasks, EI often fails to find a global optimum parameter in the following reasons. First, the effect of hyperparameters is not well-studied. If we estimate the hyperparameters via MLE \cite{snoek2012practical}, EI finds the basin of attraction only for certain circumstances because the hyperparameters are changing at every iterations \cite{bull2011convergence}. Meanwhile, if we adopt a fully Bayesian treatment of hyperparameter \cite{snoek2012practical, wu2017bayesian}, then the theoretic analysis on asymptotic consistency of EI in continuous parameter space is not equipped \cite{wu2017bayesian}. In practice, EI, with either Bayesian treatment or pointwise treatment on hyperparameters, fails to detect the basin of attraction and stops exploring, which makes no improvement on the response surface after iterations.

More importantly, looking at the shape of the true error function, a well-modeled simulation is likely to have a sub-optimal plateau near the global minimum. As the simulation model becomes more insensitive on its input parameters, the sub-optimal plateau area becomes bigger. On the sub-optimal plateau, if we assume the predicted difference $\min_{c'}{d(E_{R}[\mathcal{S}(\mathcal{M}(\mathcal{P}_{c'};\omega_{r}))],\mathcal{D})}-\mu_{err}(\mathcal{P})$ to be constant, denoted by $\mu_{0}$, then the derivative of EI w.r.t. $\sigma_{err}$ is specified in the below.
\begin{equation}\label{eq:ExpectedImprovementDerivative}
\begin{aligned}
&\frac{\partial}{\partial\sigma_{err}}\bigg[\mu_{0}\Phi\Big(\frac{\mu_{0}}{\sigma_{err}}\Big)+\sigma_{err}\phi\Big(\frac{\mu_{0}}{\sigma_{err}}\Big)\bigg]\\
&=-\Big(\frac{\mu_{0}}{\sigma_{err}}\Big)^{2}\phi\Big(\frac{\mu_{0}}{\sigma_{err}}\Big)+\phi\Big(\frac{\mu_{0}}{\sigma_{err}}\Big)\bigg[1+\Big(\frac{\mu_{0}}{\sigma_{err}}\Big)^{2}\bigg]\\
&=\phi\Big(\frac{\mu_{0}}{\sigma_{err}}\Big)
\end{aligned}
\end{equation}
The above formula is turned out to be positive in any cases. This suggests that EI is an increasing function of $\sigma_{err}$, so EI will explore on the sub-optimal plateau that it finds $\mathcal{P}_{c+1}$ as a next design point that makes $\sigma_{err}(\mathcal{P}_{c+1})$ the biggest. In these reasons, some articles investigate the portfolio of the mixture of acquisition functions \cite{hoffman2011portfolio} to accelerate optimization, which we follow in this calibration task.

Besides of using EI in its fundamental form, the ratio of the exploitation term and the exploration term could be controlled by adding weight $w_{c}$ as the below.
\begin{equation} \label{eq:WeightedExpectedImprovement}
\begin{aligned}
&\text{w-EI}(\mathcal{P}|\mathcal{G})\\
&=(1-w_{c})\times\big(\min_{c'}{d(E_{R}[\mathcal{S}(\mathcal{M}(\mathcal{P}_{c'};\omega))],\mathcal{D})}-\mu_{err}(\mathcal{P})\big)
\Phi\bigg(\frac{\min_{c'}{d(E_{R}[\mathcal{S}(\mathcal{M}(\mathcal{P}_{c'};\omega))],\mathcal{D})}-\mu_{err}(\mathcal{P})}{\sigma_{err}(\mathcal{P})}\bigg)\\
&+w_{c}\times\sigma_{err}(\mathcal{P})\phi\bigg(\frac{\min_{c'}{d(E_{R}[\mathcal{S}(\mathcal{M}(\mathcal{P}_{c'};\omega))],\mathcal{D})}-\mu_{err}(\mathcal{P})}{\sigma_{err}(\mathcal{P})}\bigg)
\end{aligned}
\end{equation}
We call this acquisition function as weighted Expected Improvement (w-EI) \cite{sobester2005design}.

Parallel Bayesian Optimization algorithms \cite{wu2016parallel, wu2017bayesian, wang2016parallel} could enhance the calibration speed dramatically, that they could find a set of multiple heterogeneous parameters $\mathcal{P}_{C+1},\mathcal{P}_{C+2},...,\mathcal{P}_{C+L}$ to test. The distributed parallel simulation then tests $L$ different set of heterogeneous parameters and yield $L$ different simulation errors. Parallel Bayesian Optimization also supports asynchronized situation where different simulation setup takes different error evaluation time.

\section{Proposed Calibration Methodology}
\label{sec:CalibrationMethodology}
This section describes our calibration framework for agent-baed models with a validation dataset. The description starts from the overall illustration of the calibration framework. Then, we describe two detailed calibration components, which are \textit{dynamic calibration} and \textit{heterogeneous calibration}, in turn. 

\subsection{Overall Calibration Framework}
Our calibration framework intends to optimize the parameters of a generic agent based model in order to reduce an error function between a validation dataset and a set of stochastic simulation summary statistics. Some of these components are already defined in the previous literature in the previous section. For instance, we represent the agent based model as $\mathcal{M}$; parameter set as $\mathcal{P}$; an error function as $d$; a validation dataset as $\mathcal{D}$; and a set of stochsatic simulation output summary statistics as $\mathcal{S}(\mathcal{M}(\mathcal{P};\omega)=\mathcal{O})$.

\begin{figure*}
	\centering
	\includegraphics[width=17cm]{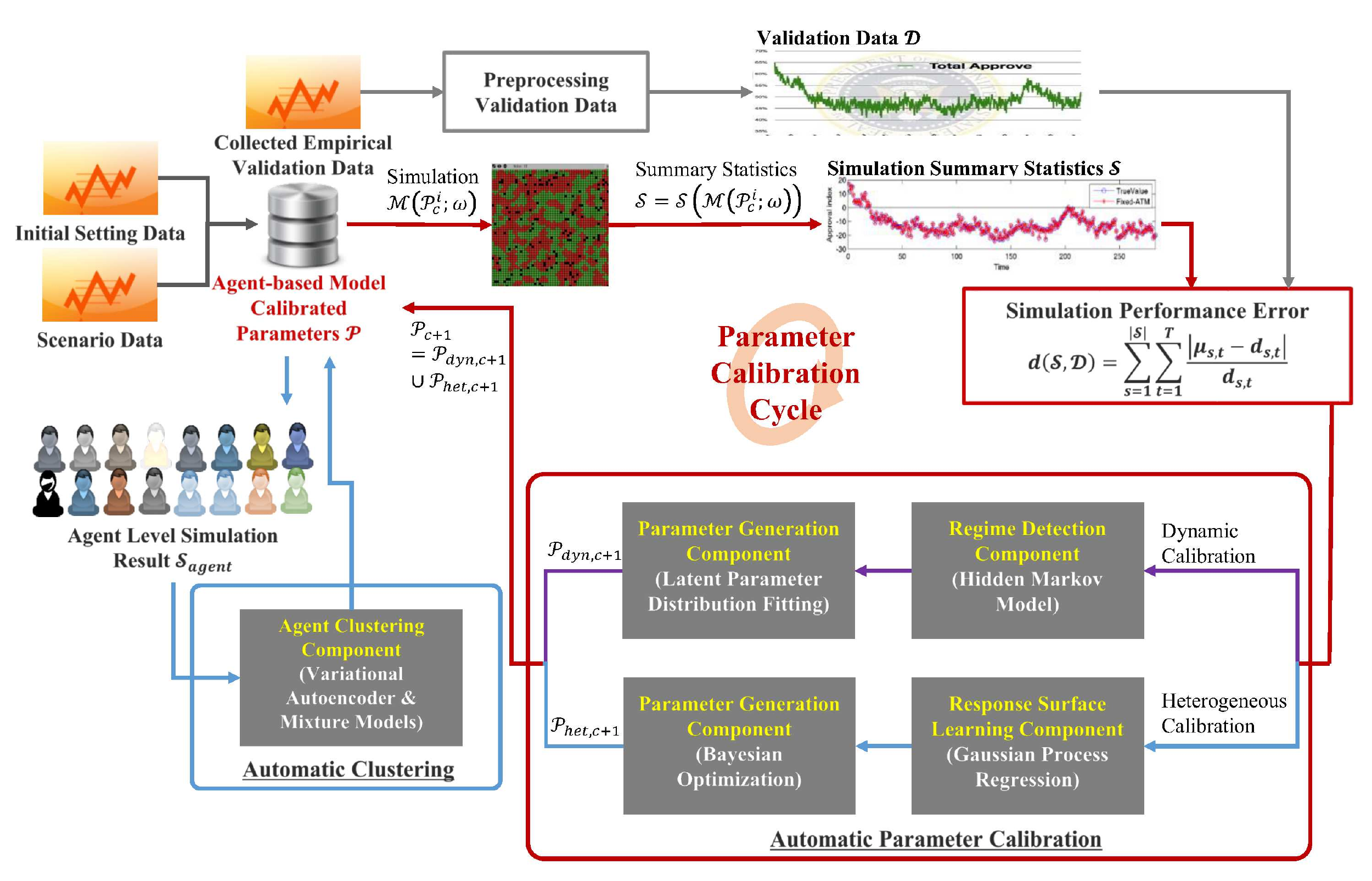}
	\caption{The suggested calibration framework overview is presented. \textit{Agent Clustering Component} is executed at the beginning of the calibration, and next the calibration iteration starts. For each iteration, either dynamic calibration or heterogeneous calibration are activated. Dynamic calibration contains two components: \textit{Regime Detection Component} and \textit{Parameter Generation Component}. Heterogeneous calibration in the iteration also contains two components: \textit{Response Surface Learning Component} and \textit{Parameter Generation Component}}
	\label{fig:overview}
\end{figure*}

As all of the surveyed calibration methods can be executed in an iterative setting, our model assumes an iterative optimization of $\mathcal{P}$, which is illustrated in Fig. \ref{fig:overview} and Alg. \ref{alg:CalibrationFramework}. Before we start our iterations, we create a set of agent clusters because our heterogeneous calibration will be applied to each group of agents, rather than an individual agent due to the time complexity, see line 3 in Alg. \ref{alg:CalibrationFramework}. Also, we assume that we obtained a pre-processed validation dataset $\mathcal{D}$ that is a set of temporal sequence values.  Afterwards, we start the calibration iterations from a randomly initialized parameter. Having said that, our calibration would better work if the initial parameter is close to the optimal parameter $\mathcal{P}^{*}$, and we provide this trade-off in Sec. \ref{sec:DynamicCalibrationResultsTestCase1}.

Our iterations consist of one minor step and two major components with machine learning models. 
\begin{enumerate}
\item Our iteration starts from a minor step of calculating $d(E_{R}[\mathcal{S}(\mathcal{M}(\mathcal{P}_{c};\omega_{r}))],\mathcal{D})$ from the most recent simulation outputs $\mathcal{M}(\mathcal{P}_{c};\omega_{r})$ with the most recent parameter $\mathcal{P}_{C}$, see line 15 in Alg. \ref{alg:DynamicCalibration} and line 26 in Alg. \ref{alg:HeterogeneousCalibration}. The function $d$ is instantiated for each application case, and $d$ can be either simple Euclidean distance, KL divergence, negative log-likelihoods, etc. We specify $d$ for each of our examples in Section \ref{sec:DynamicCalibration} and \ref{sec:HeterogeneousCalibration}. 
\item After calculating $d$, we perform either \textit{dynamic calibration} or \textit{heterogeneous calibration}, which are the two major components of the calibration framework. The proposed calibration framework alternates each calibration with pre-determined iterations, to stabilize the estimation of each parameter: $C_{dyn}$ iterations for the dynamic component, and $C_{het}$ iterations for the heterogeneous component.
\begin{enumerate}
\item Dynamic calibration takes the result of $d$, $d(E_{R}[\mathcal{S}_{dyn}(\mathcal{M}(\mathcal{P}_{c};\omega_{r}))],\mathcal{D}_{dyn})$, simulation mean, $E_{R}[\mathcal{S}_{dyn}(\mathcal{M}(\mathcal{P}_{c};\omega_{r}))]$, and the current dynamic parameter candidate hypotheses, $\mathcal{P}_{dyn,c}^{i}$, to perform the regime detection, see line 17 in Alg. \ref{alg:DynamicCalibration}, and to generate the next dynamic parameter hypotheses $\mathcal{P}_{dyn,c+1}^{i}$, see line 18 in Alg. \ref{alg:DynamicCalibration}.
\item Heterogeneous calibration accepts the result of $d$, $d(E_{R}[\mathcal{S}_{het}(\mathcal{M}(\mathcal{P}_{c};\omega_{r}))],\mathcal{D}_{het})$, and $\mathcal{P}_{het,c}$ to learn the response surface, see line 28 in Alg. \ref{alg:HeterogeneousCalibration}, and to generate the heterogeneous parameter $\mathcal{P}_{het,c+1}$, see line 29 in Alg. \ref{alg:HeterogeneousCalibration}. 
\end{enumerate}
\item Since the union of $\mathcal{P}_{dyn,c+1}^{i}$ and $\mathcal{P}_{het,c+1}$ becomes a new parameter input $\mathcal{P}_{c+1}^{i}$ of $i^{\text{th}}$ candidate hypothesis for the next iteration, our iterative calibration becomes a loop that measures $d$ and optimizes $\mathcal{P}$, alternatively. 
\end{enumerate}
The following sections show the details of two major components.

%\footnote{$E_{\omega_{1} ... \omega_{R} \in \Omega}[S_{agent}(O(M(\omega_{r} ,P^{t})))]$ represents the expectation, or simply average, of summarizing (with a function of $S_{agent}$) output results ($O$) from executing an agent based model (as a function of $M$) with a sample random path $\omega_{r}$ and a parameter $P^{t}$ with $R$ replications. }

\begin{algorithm*}
	\DontPrintSemicolon
	\LinesNumbered
	\SetKwFunction{F}{\texttt{CalibrationFramework}}
	\SetKwInOut{Input}{input}\SetKwInOut{Output}{output}
	\Input{Input parameter combination $\mathcal{P}^{in}=\mathcal{P}_{dyn}^{in}\cup \mathcal{P}_{het}^{in}$}
	\Output{Calibrated parameter combination $\mathcal{P}^{out}=\mathcal{P}_{dyn}^{out}\cup\mathcal{P}_{het}^{out}$}
	\BlankLine
	\SetKwProg{Fn}{Function}{:}{}
	\Fn{\F{$\mathcal{P}_{dyn}^{in}\cup\mathcal{P}_{het}^{in}$}}{
		$\mathcal{P}_{dyn,0}=\mathcal{P}_{dyn}^{in}$\\
		$\mathcal{P}_{het,0}$=\textsc{AgentClustering}($\mathcal{P}_{dyn}^{in}\cup\mathcal{P}_{het}^{in}$) (see Alg. \ref{alg:HeterogeneousCalibration})\\
		\BlankLine
		\For{$c$ in range($C_{cal}$)}{
			\BlankLine
			\uIf{$0\le c-\Big[\frac{c}{C_{dyn}+C_{het}}\Big](C_{dyn}+C_{het})<C_{dyn}$}{
				\BlankLine
			$\mathcal{P}_{dyn,c+1}$=\textsc{DynamicCalibration}($\mathcal{P}_{dyn,c}\cup\mathcal{P}_{het,c}$) (see Alg. \ref{alg:DynamicCalibration})\\
			$\mathcal{P}_{het,c+1}=\mathcal{P}_{het,c}$\\
		}
	\BlankLine
		\uElseIf{$C_{dyn}\le c-\Big[\frac{c}{C_{dyn}+C_{het}}\Big](C_{dyn}+C_{het})<C_{dyn}+C_{het}$}{
			\BlankLine
			$\mathcal{P}_{het,c+1}$=\textsc{HeterogeneousCalibration}($\mathcal{P}_{dyn,c}\cup\mathcal{P}_{het,c}$) (see Alg. \ref{alg:HeterogeneousCalibration})\\
			$\mathcal{P}_{dyn,c+1}=\mathcal{P}_{dyn,c}$\\
	}
	}
\BlankLine
		Set $\mathcal{P}^{out}=\mathcal{P}_{dyn}^{opt}\cup\mathcal{P}_{het}^{opt}$ to have the lowest simulation error\\
		\BlankLine
		\KwRet $\mathcal{P}_{dyn}^{out}\cup\mathcal{P}_{het}^{out}$
	}
	\caption{Calibration Framework Algorithm}\label{alg:CalibrationFramework}
\end{algorithm*}

\subsection{Dynamic Parameter Calibration}
\label{sec:DynamicCalibration}

Dynamic parameter calibration \cite{moon2018data} is an inverse problem of finding optimal parameters that yields the most consistent simulation summary statistics, $S_{dyn}$, with given validation empirical data, $D_{dyn}$. To further illustrate the calibration algorithm, we state two assumptions. First, we assume that we have a fixed parameter set of $\mathcal{P}_{het}$, which is $\mathcal{P}\backslash\mathcal{P}_{dyn}$. This assumption limit this algorithm to finding only $\mathcal{P}_{dyn}$, which is a subset of $\mathcal{P}$, without the joint consideration with $\mathcal{P}_{het}$. Second, we assume that the dynamic parameter is assigned to each timestep as a switching regime. Formally, if we assume to calibrate $N_{dyn}$ number of dynamically switching parameter, $\mathcal{P}_{dyn}=\{\mathcal{P}_{dyn}^{n,t}|n=1\dots,N_{dyn},t=1,\dots,T\}$. We describe our optimization task as the below.
\begin{equation}\label{eq:DynamicCalibration}
\mathcal{P}_{dyn}^{*} = \argmin_{\mathcal{P}_{dyn}}d(E_{\omega\in\Omega}[\mathcal{S}_{dyn}(\mathcal{M}(\mathcal{P}_{dyn}\cup\mathcal{P}_{het};\omega))],\mathcal{D}_{dyn})
\end{equation}
%This dynamic parameter calibration is a natural generalization of the static parameter calibration, since the restriction of supporting a parameter space into a static parameter set with Eq. \ref{eq:DynamicCalibration} is the static calibration problem. 

We treat the dynamic parameters for each timestep, $\mathcal{P}_{dyn}^{n,t}$, as the static input determined prior to the simultion, so the dynamic parameters become the composition of such parameters multiplied by the execution timesteps, $T$. This can be viewed as turning the dynamic parameters into the static parameters by increasing the number of parameters, which could be viewed as an identical problem of Eq. \ref{eq:DynamicCalibration}. However, this increment of parameter numbers is significant if the simulation becomes a long execution. Therefore, we limit the increment by assuming that the dynamic parameters are identical for identified temporal regimes over the simulation executions. This assumption enables limiting the multiplication of dynamic parameter numbers to the given regime number, $K_{dyn}$.

Because of the above reasoning, our calibration algorithm divides simulation timesteps into a number of regimes to generate the dynamic parameters for each regime, separately. Also, we separate well-fitted regimes from poorly-fitted regimes. This separation enables that the poorly-fitted regimes can explore farther parameter spaces while the well-fitted regimes can exploit the bounded parameter spaces.

\subsubsection{Dynamic Calibration Performance}
\label{sec:DynamicCalibrationPerformance}

This subsection explains line 12-15 in Alg. \ref{alg:DynamicCalibration}, which measures the performance of dynamic calibration. Dynamic calibration uses a posterior distribution inferred from $I$ number of candidate hypotheses, $\mathcal{P}_{dyn,c}^{i}$, to generate a new set of parameters, $\mathcal{P}_{dyn,c+1}^{i}$ for $(c+1)^{\text{th}}$ iteration. To obtain statistically stable simulation summary statistics, we run simulation $R$ times for each $i^{\text{th}}$ candidate hypothesis, see line 13 in Alg. \ref{alg:DynamicCalibration}. Then, the simulation generates state variables $\mathcal{O}^{i}_{r}=\mathcal{M}(\mathcal{P}_{dyn,c}^{i}\cup\mathcal{P}_{het};\omega_{r})$ of $i^{\text{th}}$ candidate hypothesis are post-processed to generate the simulation summary statistics means as follows.
\begin{equation}
\label{eq:SimulationSummaryStatisticsMean}
E_{R}[\mathcal{S}_{dyn}(\mathcal{O}^{i}_{r})]=\{\mu_{s,t}^{i}|s=1,\dots,|S_{dyn}|,t=1,\dots,T,i=1,\dots,I\}.
\end{equation}

The distance function $d$ in dynamic calibration is the negative likelihood function, which requires the simulation variance. The simulation variance is set to be $\sigma_{s,t}^{i}=0.1\times\mu_{s,t}^{i}$ in this article. Then, the likelihood of true observation $\mathcal{D}_{dyn}=\big((d_{s,t})_{s=1}^{S}\big)_{t=1}^{T}$ being observed from the normal distribution with the output mean $\mu^{i}_{s,t}$ and the variance $\sigma^{i}_{s,t}$ is calculated as below.
\begin{equation} \label{eq:Likelihood}
L_{s,t}^{i}=\mathcal{N}(d_{s,t}|\mu_{s,t}^{i},\sigma_{s,t}^{i})
\end{equation}
The joint likelihood over all the summary statistics is $L_{t}^{i}=\prod_{s=1}^{S_{dyn}}{L_{s,t}^{i}}$. The implementation version utilizes the log-likelihood of Eq. \ref{eq:Likelihood}.

\begin{algorithm*}
	\DontPrintSemicolon
	\LinesNumbered
	\SetKwFunction{F}{\textsc{RegimeDetection}}
	\SetKwFunction{G}{\textsc{ParameterGeneration}}
	\SetKwFunction{H}{\textsc{DynamicCalibration}}
	\SetKwFunction{FMain}{}
	\SetKwInOut{Input}{Input}\SetKwInOut{Output}{Output}
	\Input{$n^{\text{th}}$ Dynamic parameter $\mathcal{P}_{dyn,c}^{n,i}$ in $i^{\text{th}}$ hypothesis at $c^{\text{th}}$ iteration}
	\Input{Simulation summary statistics mean $\mu_{s,t}^{i}$}
	\Output{Merged regimes $MR_{u}$}
	\BlankLine
	\SetKwProg{Fn}{Function}{:}{}
	\Fn{\F{$\mathcal{P}_{dyn,c}^{n,i},\mu_{s,t}^{i}$}}{
		\For{$i$ in range($I$)}{
			Apply Hidden Markov Model to obtain the regime detection results for the $i^{\text{th}}$ candidates}
		Merge regime detection results over the candidate parameters to split the simulation time into merged regimes $MR_{u}$
		
	\BlankLine
		\KwRet $MR_{u}$,
	}
\BlankLine
	\SetKwInOut{Input}{Input}\SetKwInOut{Output}{Output}
	\Input{$n^{\text{th}}$ dynamic parameter hypotheses $\mathcal{P}_{dyn,c}^{i}$ at $c^{\text{th}}$ iteration}
	\Input{Likelihoods $L_{s,t}^{i}$}
	\Input{Merged regimes $MR_{u}$}
	\Output{$n^{\text{th}}$ calibrated parameter hypotheses $\mathcal{P}_{dyn,c+1}^{n,i}$}
	\BlankLine
	\SetKwProg{Fn}{Function}{:}{}
	\Fn{\G{$\mathcal{P}_{dyn,c}^{n,i},L_{s,t}^{i},MR_{u}$}}{
		\For{$u$ in range($U$)}{
			Estimate the beta distribution $\text{Beta}(\alpha_{u},\beta_{u})$, using normalized parameters and normalized joint likelihoods\\
		}
		Generate the next parameter hypotheses $\mathcal{P}_{dyn,c+1}^{n,i}$ from estimated distribution $\text{Beta}(\alpha_{u},\beta_{u})$ \\
		\BlankLine
		\KwRet $\mathcal{P}_{dyn,c+1}^{n,i}$ \\
	}
\BlankLine
	\SetKwInOut{Input}{Input}\SetKwInOut{Output}{Output}
	\Input{Parameter combination, $\mathcal{P}_{dyn,c}\cup\mathcal{P}_{het,c}$, of the $c^{\text{th}}$ calibration}
	\Output{Calibrated dynamic parameter $\mathcal{P}_{dyn,c+1}$}
	\BlankLine
	\SetKwProg{Fn}{Function}{:}{}
	\Fn{\H{$\mathcal{P}_{dyn,c}\cup\mathcal{P}_{het,c}$}}{
	\For{$i$ in range($I$)}{
		Run simulation $R$ times with the $i^{\text{th}}$ candidate hypothesis, $\mathcal{P}_{c}^{i}=\mathcal{P}_{dyn,c}^{i}\cup\mathcal{P}_{het,c}$
	}
	Obtain simulation mean $\mu_{s,t}^{i}$ and variance $\sigma_{s,t}^{i}$ \\
	Obtain the likelihoods $L_{s,t}^{i}=\mathcal{N}(d_{s,t}|\mu_{s,t}^{i},\sigma_{s,t}^{i})$ \\
	\For{$n$ in range($N_{dyn}$)}{
		$MR_{u}=$\texttt{RegimeDetection}($\mathcal{P}_{dyn,c}^{n,i},\mu_{s,t}^{i}$) (see Section \ref{sec:DynamicAlgorithmOverview})\\
		$\mathcal{P}_{dyn,c+1}^{n,i}=$\texttt{ParameterGeneration}($\mathcal{P}_{dyn,c}^{n,i},L_{s,t}^{i},MR_{u}$) (see Section \ref{sec:DynamicParameterGeneration})
	}
	\KwRet $\mathcal{P}_{dyn,c+1}$
	}
	\caption{Dynamic Calibration}\label{alg:DynamicCalibration}
\end{algorithm*}

\subsubsection{Regime Detection Component}
\label{sec:DynamicAlgorithmOverview}

After calculating the likelihoods, we apply the regime detection algorithm of HMM to obtain temporal clustering results by regarding the deviations of the simulation result, $\mu_{s,t}^{i}$, and the validation data, $d_{s,t}$, as observed variables in Eq. \ref{eq:HMM}, introduced in Section ~\ref{sec:TemporalClustering}, as below.
\begin{equation} \label{eq:Observation}
O_{t}=\Big(\mu_{1,t}^{i}-d_{1,t},\dots,\mu_{|S_{dyn}|,t}^{i}-d_{|S_{dyn}|,t}\Big)
\end{equation}
The temporal clustering results are $R_{t}^{i}\in\{1,\dots,K_{dyn}\}$ for each of $i^{\text{th}}$ candidate hypothesis. After obtaining regimes of all the candidate hypotheses, we define merged regimes $\{\text{MR}_{u}\}_{u=1}^{U}$ as the partition of simulation timesteps that satisfy the following conditions: if $t_{1},t_{2}\in \text{MR}_{u}$,
\begin{equation}\label{eq:MergedRegimeClassifier}
(R_{t_{1}}^{1},...,R_{t_{1}}^{I})=(R_{t_{2}}^{1},...,R_{t_{2}}^{I})
\end{equation}

We utilize $\text{MR}$ as the final regime detection result, which answers to the question of \textit{when to calibrate}. Also, $\text{MR}$ becomes the input to the algorithm of answering \textit{how to calibrate}.

\subsubsection{Parameter Generation Component}
\label{sec:DynamicParameterGeneration}

We generate the next set of paramter hypotheses from each merged regime, $\text{MR}$. In detail, from the merged regime $\text{MR}_{u}$, we infer the posterior distribution of the parameter space, $\text{Beta}(\alpha_{u},\beta_{u})$, to sample the next parameter candidate hypotheses $\mathcal{P}_{dyn,c+1}^{i}$. To estimate the $u^{\text{th}}$ beta distribution parameters, we use the current parameter values $\{P_{dyn,c}^{n,t,i}|t\in \text{MR}_{u},i=1,\dots,I\}$ and the likelihoods
$\{L_{t}^{i}|t\in \text{MR}_{u},i=1,\dots,I\}$.
In particular, the parameter and the likelihood values are normalized to estimate the beta distribution parameters $\alpha_{u}$ and $\beta_{u}$ via maximum likelihood estimation. The parameter is normalized by
\begin{equation}
\hat{P}_{dyn,c}^{n,t,i}=\frac{P_{dyn,c}^{n,t,i}-\underline{P}_{dyn}^{n}}{\overline{P}_{dyn}^{n}-\underline{P}_{dyn}^{n}},
\end{equation}
where $\overline{P}_{dyn}^{n}$ and $\underline{P}_{dyn}^{n}$ are the maximum and the minimum of the $n^{\text{th}}$ parameter range, respectively. The likelihood is normalized by
\begin{equation}
\hat{L}_{t}^{i} = \frac{L_{t}^{i}}{\sum_{i=1}^{I}\sum_{t\in \text{MR}_{u}}L_{t}^{i}}.
\end{equation}
Then, the estimated beta distribution, $\text{Beta}(\alpha_{u},\beta_{u})$, generates a next set of parameters by either sampling or \textit{Mode Selection}, which is explained, later.

If a simulation is far from the validation for all the candidate hypotheses, additional parameter exploration is executed in order to avoid searching prohibitive region. If the likelihoods are less than given threshold for all the candidate hypotheses and for all the simulation timestep $t\in \text{MR}_{u}$,
\begin{equation}\label{eq:PoorlyFitted}
L_{t}^{i}<\frac{\min_{t}L_{t}^{i}+\text{ratio}\times\max_{t}L_{t}^{i}}{1+\text{ratio}},
\end{equation}
then the normalized likelihoods are set to be equal, $\hat{L}_{t}^{i}=1/(I\times|MR_{u}|)$. This equally distributed likelihoods make the estimated beta distribution equally distributed, and the next parameter will be randomly picked at the $u^{\text{th}}$ merged regime. The ratio is decreased in this article, with $\text{ratio}=0.9^{c}$.

Given a posterior distribution, $\text{Beta}(\alpha_{u},\beta_{u})$, we have three options to generate new parameters:
\begin{enumerate}
	\item \textit{Sampling by Time}: the $n^{\text{th}}$ dynamic parameters $P_{dyn,c+1}^{n,t,i}$ are sampled from $\text{Beta}(\alpha_{u},\beta_{u})$ for each $t\in\text{MR}_{u}$ and $i=1,\dots,I$.
	\item \textit{sampling by Regime}: the $n^{\text{th}}$ dynamic parameters $P_{dyn,c+1}^{n,t,i}$ are set to be constant for all $t\in\text{MR}_{u}$, where the constant is sampled from $\text{Beta}(\alpha_{u},\beta_{u})$ for each $i=1,\dots,I$.
	\item \textit{Mode Selection}: the $n^{\text{th}}$ dynamic parameters $P_{dyn,c+1}^{n,t,i}$ are sampled from the modes of $\text{Beta}(\alpha_{u},\beta_{u})$, where the modes are meaningful moments of the distribution. In this article, we use $\big\{\mu_{u}^{\beta}+\big(i-\frac{I+1}{2}\big)\sigma_{u}^{\beta}\big|i=1,...,I\big\}$ to be the set of modes, where $\mu_{u}^{\beta}$ and $\sigma_{u}^{\beta}$ are the mean and the standard deviation of the Beta distribution, respectively.
\end{enumerate}

\textit{Sampling by Time} searches the most rapidly switching parameters among three generation processes. However, the small parameter gap at the front timesteps affects to the simulation dynamics afterwards, and this dependency hinders dynamic calibration to estimate accurate time-varying parameters afterwards. Furthermore, \textit{Sampling by Time} is expected to estimate the overfitted, unrealistic dynamic parameters. On the other hand, \textit{Sampling by Regime} searches parameters with an identical value at the identical regimes, which makes the calibrated parameter to be less fluctuating. Besides of the sampling-based methods, \textit{Mode Selection} focuses on exploitation in parameter generations. Exploration is implemented when the likelihoods are below the threshold, and \textit{Mode Selection} do not explore the parameter space, additionally. We experiment all parameter generating rules in Section \ref{sec:TestCase1}.

\subsection{Heterogeneous Parameter Calibration}
\label{sec:HeterogeneousCalibration}

Heterogeneous calibration resolves the simulation divergence arisen from the agent heterogeneity by transforming the simulation result, $\mathcal{S}_{het}$, to the validation data, $\mathcal{D}_{het}$, through calibrating heterogeneous parameters. Before introducing the algorithm, we present two assumptions. First, we assume that we have a fixed set of dynamic parameters, $\mathcal{P}_{dyn}$, which are the complement of heterogeneous parameters from the set of model parameters, $\mathcal{P}$. Heterogeneous calibration focuses on calibrating the heterogeneous parameters, $\mathcal{P}_{het}$. Second, the heterogeneous parameters, $\mathcal{P}_{het}$, are static parameters differentiated by agent sub-populations. The heterogeneous calibration solves the following inverse problem.
\begin{equation}\label{eq:HeterogeneousCalibration}
\mathcal{P}_{het}^{*} = \argmin_{\mathcal{P}_{het}}d(E_{\omega\in\Omega}[\mathcal{S}_{het}(\mathcal{M}(\mathcal{P}_{het}\cup\mathcal{P}_{het};\omega))],\mathcal{D}_{het})
\end{equation}
The heterogeneous parameters are not differentiated by agents, since the above optimization problem becomes infeasible. This infeasibility comes from the huge dimensionality of the problem, where the optimization dimension, $A\times N_{het}$, is the multiplication of the number of agents, $A$, by the number of heterogeneous parameters, $N_{het}$. The second assumption enables the problem to feasible by reducing the dimension, $K_{het}\times N_{het}$, where $K_{het}$ is the number of sub-populations.

\subsubsection{Heterogeneous Calibration Performance}
\label{sec:HeterogeneousCalibrationPerformance}

In heterogeneous calibrations of our experiments, the simulation error function, $d$, is the Mean Absolute Percentage Error (MAPE), but this selection can change per application case. The experimental error, $d(E_{R}[\mathcal{S}_{het}(\mathcal{M}(\mathcal{P};\omega_{r}))],\mathcal{D}_{het})$, corresponds to the noisy observation of the true error, $d(E_{\omega\in\Omega}[\mathcal{S}_{het}(\mathcal{M}(\mathcal{P};\omega))],\mathcal{D}_{het})$. We generate the next design point $\mathcal{P}_{het,c+1}$ by Bayesian optimization, with a predicted surrogate response surface obtained from the gathered data $\mathcal{G}=\Big\{\big(\mathcal{P}_{het,c'},d(E_{R}[\mathcal{S}_{het}(\mathcal{M}(\mathcal{P}_{dyn}\cup\mathcal{P}_{het,c'};\omega_{r}))],\mathcal{D}_{het})\big)\Big\}_{c'=1}^{c}$.

\subsubsection{Agent Clustering Component}
\label{sec:HeterogeneousAgentClustering}
A clustering algorithm divides agents into several sub-populations, using the simulation agent-level results, $S_{agent}$. Since the classical clustering algorithms, such as GMM and DPMM, suffers from the curse of large dimensionality, we compress the data through a latent representation learning algorithm, VAE (see Section \ref{sec:VAE}). The vanilla VAE has an encoder-decoder neural network, $\text{Enc}:\mathbf{R}^{Att\times T}\rightarrow \mathbf{R}^{H}$ and $\text{Dec}:\mathbf{R}^{H}\rightarrow\mathbf{R}^{Att\times T}$, where the two functions, embedded in a neural network, are learned by maximizing the ELBO, given in Eq. \ref{eq:VAE}. The latent representation of an agent, $LR^{a}=\{LR_{h}^{a}|h=1,\dots,H\}$ when $a \in \{1...A\}$, is the output of the encoder-part network of the agent-level simulation results of the agent.

The next step executes the mixture models, introduced in Eq. \ref{eq:GMM} or~\ref{eq:DPMM}, with the latent representation of agents, $\{LR^{1},\dots,LR^{A}\}$.
\begin{equation}
O_{a}=(LR_{1}^{a},\dots,LR_{H}^{a}),
\end{equation}
We test both parametric mixture model, Eq. \ref{eq:GMM}, and nonparametric mixture model, Eq. \ref{eq:DPMM}, in our calibration.

\begin{algorithm*}
	\LinesNumbered
	\DontPrintSemicolon
	\BlankLine
	\SetKwProg{Fn}{Function}{:}{}
	\SetKwFunction{F}{\textsc{AgentClustering}}
	\SetKwFunction{G}{\textsc{ResponseSurface}}
	\SetKwFunction{H}{\textsc{ParameterGeneration}}
	\SetKwFunction{I}{\textsc{HeterogeneousCalibration}}
	\SetKwFunction{FMain}{}
	\SetKwInOut{Input}{Input}\SetKwInOut{Output}{Output}
	\Input{Union of calibrated dynamic parameters and initial heterogeneous parmaeters $\mathcal{P}_{dyn}\cup\mathcal{P}_{het}$}
	\Output{Heterogeneity embedded initial heterogeneous parameters $\mathcal{P}_{0}$}
	\BlankLine
	\Fn{\F{$\mathcal{P}_{dyn}\cup\mathcal{P}_{het}$}}{
		Run simulation $R$ times with input parameters, $\mathcal{P}_{dyn}\cup\mathcal{P}_{het}$, to obtain agent-level simulation result $S_{agent}$\\
		Apply latent representation extraction algorithm (VAE) on agent-level simulation result to obtain compressed representations \\
		Apply clustering algorithm (GMM or DPMM) on the latent representations to obtain agent sub-populations assignments\\
		Generate the heterogeneous parameters, $\mathcal{P}_{het}$, by assigning different parameter values in different agent sub-populations \\
		\BlankLine
		\KwRet $\mathcal{P}_{het}$
	}
	\BlankLine
	\SetKwInOut{Input}{Input}\SetKwInOut{Output}{Output}
	\Input{parameter-error aggregated data $\mathcal{G}$}
	\BlankLine
	\SetKwProg{Fn}{Function}{:}{}
	\Fn{\G{$\mathcal{G}$}}{
		Learn Kernel hyperparameters in Gaussian process regression to fit the given data $\mathcal{G}$\\
	}
	\BlankLine
	\SetKwInOut{Input}{Input}\SetKwInOut{Output}{Output}
	\Output{Next heterogeneous parameter $\mathcal{P}_{het,C+1}$}
	\BlankLine
	\SetKwProg{Fn}{Function}{:}{}
	\Fn{\H{}}{
		\uIf{$C<C_{0}$}{Randomly select $\mathcal{P}_{het,C+1}$ as the next parameter inputs}
		\uElseIf{$C\ge C_{0}$}{
			Sample $\xi$ from uniform distribution with support $[0,1]$\\
			\uIf{$\xi<\xi_{rand}$}{Randomly select $\mathcal{P}_{het,C+1}$ as the next parameter inputs}
			\uElseIf{$\xi_{rand}\le\xi<\xi_{rand}+\xi_{PV}$}{Maximize the Predictive Variance to obatin a new parameter set $\mathcal{P}_{het,C+1}$}
			\uElseIf{$\xi_{rand}+\xi_{PV}\le\xi<\xi_{rand}+\xi_{PV}+\xi_{PM}$}{Minimize the Predictive Mean to obtain a new parameter set $\mathcal{P}_{het,C+1}$}
			\uElseIf{$\xi\ge\xi_{rand}+\xi_{PV}+\xi_{PM}$}{
				Maximize the weighted Expected Improvement with cooling weight $\omega_{C}=0.99^{C-C_{dyn}}/2$ to obtain a new parameter set $\mathcal{P}_{het,C+1}$}
		}
		\BlankLine
		\KwRet $\mathcal{P}_{het,C+1}$
	}
\BlankLine
\SetKwInOut{Input}{Input}\SetKwInOut{Output}{Output}
\Input{Parameter combination, $\mathcal{P}_{dyn,c}\cup\mathcal{P}_{het,c}$, of the $c^{\text{th}}$ iteration}
\Output{Calibrated heterogeneous parameter, $\mathcal{P}_{het,c+1}$, for the next iteration}
\BlankLine
	\SetKwProg{Fn}{Function}{:}{}
	\Fn{\I{$\mathcal{P}_{dyn,c}\cup\mathcal{P}_{het,c}$}}{
	Select the best candidate of the dynamic parameter, $\mathcal{P}_{dyn,c}^{best}$\\
	Run simulation $R$ times with $\mathcal{P}_{c}=\mathcal{P}_{dyn,c}^{best}\cup\mathcal{P}_{het,c}$\\
	Calculate the error metric $d$ from the heterogeneous summary statistics and validation data\\
	Add simulation result $\mathcal{G}=\mathcal{G}\cup\{\mathcal{P}_{het,c},d(E_{R}[\mathcal{S}_{het}(\mathcal{M}(\mathcal{P}_{c};\omega))],\mathcal{D}_{het})\}$\\
	\texttt{ResponseSurface}($\mathcal{G}$) (see Section \ref{sec:HeterogeneousParameterInference})\\
	$\mathcal{P}_{het,c+1}=$\texttt{ParameterGeneration}() (see Section \ref{sec:ParameterGenerationProcess})\\
	\BlankLine
	\KwRet $\mathcal{P}_{het,c+1}$
	}
	\caption{Heterogeneous Calibration}\label{alg:HeterogeneousCalibration}
\end{algorithm*}

\subsubsection{Response Surface Learning Component}
\label{sec:HeterogeneousParameterInference}
Gaussian process regression estimates the predictive posterior distribution of the true error, $E_{\omega\in\Omega}[d(\mathcal{S}_{het}(\mathcal{M}(\mathcal{P}_{dyn}\cup\mathcal{P}_{het};\omega)),\mathcal{D}_{het})]$. GPR, introduced in Section \ref{sec:GPR-based_BO}, predicts the mean and the variance by
\begin{equation}\label{eq:GaussianProcessDistribution}
\begin{aligned}
\mu_{err}(\mathcal{P})&=\kappa^{T}cov_{c}^{-1}F \\
\sigma_{err}^{2}(\mathcal{P})&=K_{\mathcal{P},\mathcal{P}}-\kappa^{T}cov_{c}^{-1}\kappa,
\end{aligned}
\end{equation}
where $\kappa=[K_{\mathcal{P},\mathcal{P}_{1}},...,K_{\mathcal{P},\mathcal{P}_{c}}]^{T}$, $F=\big[E_{R}\big[d(\mathcal{S}_{het}(\mathcal{M}(\mathcal{P}_{dyn}\cup\mathcal{P}_{het,1};\omega_{r})),\mathcal{D}_{het})\big],...,E_{R}\big[d(\mathcal{S}_{het}(\mathcal{D}(\mathcal{P}\cup\mathcal{P}_{het,c};\omega_{r})),\mathcal{D}_{het})\big]\big]^{T}$, and $(cov_{c})_{i,j}=K_{\mathcal{P}_{i},\mathcal{P}_{j}}+\frac{1}{\beta}\delta_{\{i=j\}}$. Here, $K_{\mathcal{P},\mathcal{P}'}$ is the kernel distance between points $\mathcal{P}$ and $\mathcal{P}'$. We use the Matern-$5/2$ kernel as the prior covariance. The hyperparameters of kernel function are point-estimated by maximizing the likelihoods of collected data being sampled from the predictive distribution.

\subsubsection{Parameter Generation Component}
\label{sec:ParameterGenerationProcess}
As a selective search algorithm, Bayesian optimization optimizes the acquisition function to find a next evaluation design point as follows.
\begin{equation}
\label{eq:AcquisitionFunction}
\mathcal{P}_{het,c+1}=\argmax_{\mathcal{P}_{het}\in \mathbb{H}}AF(\mathcal{P}_{het}|\mathcal{G})
\end{equation}
Here, $AF$ stands for the acquisition function, and $\mathbb{H}=\big\{\mathcal{P}_{het}\big|\lVert\mathcal{P}_{het}-\mathcal{P}_{het,c}\rVert_{2}\ge r_{0} \text{ for all }\mathcal{P}_{het,c}\text{ in the domain of }\mathcal{G}\big\}$ with a given radius $r_{0}\ge0$. This adaptive parameter space, $\mathbb{H}$, forces the search algorithm to avoid the $r_{0}$-neighborhood of the previous evaluated parameters. We use L-BFGS-B algorithm to optimize the acquisition function, with $r_{0}=0$.

We apply three types of Acquisition Functions: \textit{weighted Expected Improvement}, \textit{Predictive Mean} and \textit{Predictive Variance}. We use search strategy as follows:
\begin{itemize}%[label=$\ast$]
	\item Choose random parameters for the first $C_{0}$ iterations
	\item For the next iterations,
	\begin{itemize}[label=$\ast$]
		\item With probability $\xi_{rand}$, choose random parameters
		\item With probability $\xi_{PV}$, choose parameters that maximize \textit{Predictive Variance}
		\item With probability $\xi_{PM}$, choose parameters that minimize \textit{Predictive Mean}
		\item With probability $\xi_{w-EI}$, choose parameters that maximize \textit{weighted Expected Improvement} with cooling weight $w_{c}=0.99^{c}/2$
	\end{itemize}
\end{itemize}
We use multiple acquisition functions to search parameter space without being trapped by local minima. In order to escape from local minima, the search strategy contains the exploration algorithms, such as \textit{Random Search} and \textit{Predictive Variance}, where \textit{Predictive Variance} finds the least searched parameter region. The role of exploration in heterogeneous calibration is estimating the surrogate function closer to the true error function, globally. In order to estimate the optimal parameters, the search strategy executes the exploitation by \textit{Predictive Mean} and \textit{weighted Expected Improvement}. The cooling rate, $w_{C}$, in \textit{weighted Expected Improvement} controls the weight between exploration and exploitation in Eq. \ref{eq:WeightedExpectedImprovement}, to find a narrower region in large iterations. \textit{Predictive Mean} is a pure exploitation algorithm to find the global minimum of the surrogate function.

\section{Experimental Result}
\label{sec:ExperimentalResult}

Section~\ref{sec:ExperimentalResult} introduces the experimental results of the suggested calibration framework in two test cases. First, we test the feasibility of our algorithm on Wealth Distribution Model \cite{wilensky1998netlogo}. We synthesized a validation dataset with an arbitrailiy chosen parameter set, and this synthesis limits the source of simulation divergence to random effects and the pre-determined parameters set. Second, we experiment the real-world applicability with a Real Estate Market Model which uses a real validation dataset with elaborately designed model structures and real input scenarios. In such a realistic case, the additional sources of divergence exist, including the imperfect modeling and the highly noisy socio-economics data.

Both cases experiments the calibration framework suggested in Alg. \ref{alg:CalibrationFramework}. To investigate the effects of each component in the framework, we also experiment dynamic calibration and heterogeneous calibration, seperately. We evaluate dynamic calibration without heterogeneous calibration by setting the heterogeneous parameters to be either known optimal parameters, in test case 1, or human calibrated parameters, in test case 2. Similarily, we assess heterogeneous calibration without dynamic calibration by setting the dynamic parameters to be either optimal parameters or human calibrated parameters.

\subsection{Test Case 1: Wealth Distribution ABM}
\label{sec:TestCase1}

\subsubsection{Model Description}
\label{sec:ModelDescription}

Wealth Distribution Model \cite{wilensky1998netlogo}, adapted from the sugarscape model \cite{epstein2006generative}, is an agent-based model to investigate the macroscopic wealth distribution via microscopic agent behaviors. In the model, a grid provides its wealth to agents located at the grid. At the end of each simulation timestep, agents consume their wealth to survive and move toward the wealthest neighboring grid to maximize their wealth income. Also, grids recover their wealth at the end of each timestep for future provision. The net-wealth of all grids is proportional to the dynamic parameter, \textit{Wealth Income}, and agents consume their wealth, proprortional to the heterogeneous parameter, \textit{Wealth Consumption}.

\subsubsection{Virtual Experimental Design}
\label{sec:VirtualExperimentTestCase1}

\paragraph{Synthetic Parameter Setting}
\label{sec:SyntheticParameterSetting}
Tab. \ref{tab:ListofParametersinTestCase1} presents the parameters used in Wealth Distribution ABM. The dynamic parameter, \textit{Wealth Income}, represents the seasonal effects, or the up and down of economics business cycles. The synthetic dynamic parameter is alternating between 1.5 and 0.5 by a period of ten simulation timesteps. The heterogeneous parameter, \textit{Wealth Consumption}, represents the agent level heterogeneous characteristics. Agent clusters are divided according to the initial wealth; agents with top $50\%$ in their initial wealth are clustered, and the other agents are separately clustered. The synthetic heterogeneous parameter is set to be 0.9 for the first cluster, and 0.1 for the second cluster.

\begin{table*}
	\begin{center}
		\centering
		\caption{Calibration parameters in the first test case are listed. The dynamic parameter, \textit{Wealth Income}, and the heterogeneous parameter, \textit{Wealth Consumption}, are the essential variables that control the model dynamics}
		\label{tab:ListofParametersinTestCase1}
		\begin{tabu}{||l|l|l|l|l||}
			\hline
			
			\multirow{2}{*}{\parbox{2cm}{Parameters}} & \multirow{2}{*}{\parbox{2cm}{Parameter Type}} & \multirow{2}{*}{\parbox{1.5cm}{Parameter Range}} & \multicolumn{2}{p{5cm}||}{Synthetic Parameter Setting}\\\cline{4-5}
			&&& \multicolumn{1}{p{1cm}}{Value} & \multicolumn{1}{|p{4cm}||}{Time or Cluster}\\
			%						&&& \multicolumn{1}{p{1cm}|}{Value} & \multicolumn{1}{p{3cm}||}{Time (Dynamic), Cluster (Heterogeneous)} \\
			\hline\hline
			\multirow{1}{*}{\parbox{2cm}{\textit{Wealth Income}}} & \multicolumn{1}{p{2cm}|}{Dynamic} & \multicolumn{1}{p{1.5cm}|}{0-2} & \multicolumn{1}{p{1cm}|}{1.5} & \multicolumn{1}{p{4cm}||}{1-10,21-30,41-50}\\\cline{4-5}
			&&& \multicolumn{1}{p{1cm}|}{0.5} & \multicolumn{1}{p{4cm}||}{11-20,31-40}\\
			
			\hline
			\multirow{1}{*}[-0.5em]{\parbox{2cm}{\textit{Wealth Consumption}}} & \multicolumn{1}{p{2cm}|}{Heterogeneous} & \multicolumn{1}{p{1.5cm}|}{0-1} & \multicolumn{1}{p{1cm}|}{0.9} & \multicolumn{1}{p{4cm}||}{Top $50\%$ in Initial Wealth}\\\cline{4-5}
			&&& \multicolumn{1}{p{1cm}|}{0.1} & \multicolumn{1}{p{4cm}||}{Bottom $50\%$ in Initial Wealth}\\
			\hline
			
		\end{tabu}
	\end{center}
\end{table*}

\paragraph{Summary Statistics}
\label{sec:SummaryStatistics}

The wealth inequality is the main interest in this model, which leads the validation data consisted of \textsc{High Class Wealth Average}, \textsc{Middle Class Wealth Average}, \textsc{Low Class Wealth Average}, and \textsc{Gini Index} as in Tab. \ref{tab:ListofSummaryStatisticsinTestCase1}. Here, \textsc{High Class Wealth Average} is the average wealth of the top $33\%$ agents, and \textsc{Gini Index} \cite{gini1936measure} is a index of measuring the wealth inequality. The synthetic validation data is generated from the synthetic parameter setting. The agent-level simulation result $S_{agent}$ is not used, since the agent cluster is given a-priori to the calibration task in Wealth Distribution ABM to test the performance of the suggested Bayesian optimization.% The purpose of heterogeneous calibration in Wealth Distribution ABM is to investigate the performance of the suggested Bayesian optimization.

\begin{table*}
	\begin{center}
		\centering
		\caption{The validation summary statistics in the first test case are listed. The validation data is the average of the 300 simulation replications with synthetic parameters as input parameters}
		\label{tab:ListofSummaryStatisticsinTestCase1}
		\begin{tabu}{||l|l|l||}
			\hline
			%\multicolumn{2}{||c||}{Test Case 1: Wealth Distribution ABM}\\
			%\hline\hline
			\multicolumn{1}{||p{2cm}|}{Type of Summary Statistics} & \multicolumn{1}{p{4cm}|}{Name of Summary Statistics} & \multicolumn{1}{p{7cm}||}{Variable Description}\\
			\hline\hline
			\multirow{4}{*}[-1em]{\parbox{2cm}{Validation Summary Statistics}}
			&\multicolumn{1}{p{4cm}|}{\textsc{High Class Wealth Average}} & \multicolumn{1}{p{7cm}||}{Average wealth of top 1/3 agents}\\
			\cline{2-3}
			&\multicolumn{1}{p{4cm}|}{\textsc{Middle Class Wealth Average}} & \multicolumn{1}{p{7cm}||}{Average wealth of middle 1/3 agents}\\
			\cline{2-3}
			&\multicolumn{1}{p{4cm}|}{\textsc{Low Class Wealth Average}} & \multicolumn{1}{p{7cm}||}{Average wealth of bottom 1/3 agents}\\
			\cline{2-3}
			&\multicolumn{1}{p{4cm}|}{\textsc{Gini Index}} & \multicolumn{1}{p{7cm}||}{The area ratio of the Lorenz curve to measure the wealth inequality}\\
			\hline
			
		\end{tabu}
	\end{center}
\end{table*}

\begin{table*}
	\begin{center}
		\centering
		\caption{Experimental variables for each experiment case are listed. Dynamic calibration calibrates the dynamic parameter for 100 iterations, with the given synthetic heterogeneous parameter. Heterogeneous calibration calibrates the heterogeneous parameter for 100 iterations, with the given synthetic dynamic parameter. The calibration framework calibrates all the dynamic and the heterogeneous parameters, with two subcases}
		\label{tab:ListofExperimentalVariablesinTestCase1}
		\begin{tabu}{||l|c|c|c|c|c|c|c|c|c||}
			\hline
			\diagbox{Experiments}{Variable} & $C_{cal}$ & $C_{dyn}$ & $C_{het}$ & $K_{dyn}$ & $K_{het}$ & $A$ & $T$ & $R$ & $I$\\\hline\hline
			Dynamic Calibration & 100 & 1 & 0 & 3 & 2 & 100 & 50 & 10 & 3\\\hline
			Heterogeneous Calibration & 100 & 0 & 1 & 3 & 2 & 100 & 50 & 10 & 1\\\hline
			\multirow{2}{*}{Calibration Framework} & 200 & 2 & 3 & 3 & 2 & 100 & 50 & 10 & 3\\\cline{2-10}
			& 200 & 20 & 30 & 3 & 2 & 100 & 50 & 10 & 3\\
			\hline
		\end{tabu}
	\end{center}
\end{table*}

\paragraph{Experimental Cases}
\label{sec:ExperimentalCasesTestCase1}
There are three experimental cases: dynamic calibration, heterogeneous calibration, and calibration framework of joining two calibrations into a single framework. Dynamic calibration, proposed in Section \ref{sec:DynamicCalibration}, is evaluated with suggested parameter updating schemes: \textit{Sampling by Time}, \textit{Sampling by Regime}, \textit{Mode Selection}, and \textit{Random Search}. Dynamic calibration utilizes the synthetic heterogeneous parameter, in the calibration process. The heterogeneous parameter in dynamic calibration is fixed to the synthetic parameter in Tab. \ref{tab:ListofParametersinTestCase1}. The heterogeneous calibration, proposed in Sectioin \ref{sec:HeterogeneousCalibration}, applies the surrogate-based calibration with known clustering assignments. In heterogeneous calibration, the dynamic parameter is fixed to be the synthetic parameter, presented in Tab. \ref{tab:ListofParametersinTestCase1}. The last experiment calibrates both dynamic and heterogeneous parameters to investigate the effects of the combination of two components. The two subcases are tested in Wealth Distribution Model. The first subcase calibrates the dynamic and the heterogeneous parameters alternatvely with the pre-determined iterations for dynamic calibration, $C_{dyn}=2$, and for heterogeneous calibration, $C_{het}=3$. The second subcase evaluates with $C_{dyn}=20$, and $C_{het}=30$, to investigate the effects of the pre-determined hyperparameters of the calibration framework. The model parameters are randomly initialized at the initial iteration. Tab. \ref{tab:ListofExperimentalVariablesinTestCase1} presents the list of the experimental variables in each experiment. % Also, we switch the order of calibration in the additional experiment. The experiment on the framework selects the random dynamic and the random heterogeneous parameters. Tab. \ref{tab:ListofExperimentalVariablesinTestCase1} presents the list of experimental variables in each experiment.

\paragraph{Performance Measure}
\label{sec:PerformanceMeasure}

Since we know the synthetic parameters in Wealth Distribution Model, we calculate the Mean Absolute Error (MAE) to measure the distance from the estimated dynamic parameter to the synthetic dynamic parameter. We also calculate the Euclidean error to measure the distance from the estimated heterogeneous parameter to the synthetic heterogeneous parameter. To compare the simulation output, we use Mean Absolute Percentage Error (MAPE) to measure the distance from the simulation result to the validation data in all the experiments.

\subsubsection{Dynamic Calibration Results}
\label{sec:DynamicCalibrationResultsTestCase1}

\begin{table}%[h!]
		\begin{threeparttable}
		\caption{Performance evaluations in test case 1 are listed, with the following parameter updating rules: RS for \textit{Random Search}, ST for \textit{Sampling by Time}, SR for \textit{Sampling by Regime}, MS for \textit{Mode selection}, BO for Bayesian optimization. Numbers are the mean and the standard deviation of experiments by replicating 30 times. One tailed Welch's t-tests on the suggested parameter updating rules, including ST, SR, MS and BO, are implemented with the baseline updating rule, RS.}
		\label{tab:statisticsinTestCase1}
		\begin{tabular}{||l|c|c|c|c|c|c|c|c|c||}
			\hline
			\multicolumn{10}{||c||}{Test Case 1: Wealth Distribution ABM}\\
			\hline
			\multirow{4}{*}{\parbox{2cm}{Experiments}} & \multicolumn{2}{p{2cm}|}{\multirow{3}{*}{\makecell[l]{Parameter Up-\\dating Rule}}} & \multirow{4}{*}{\parbox{1cm}{\textsc{High Class Wealth Average} MAPE}} & \multirow{4}{*}{\parbox{1cm}{\textsc{Middle Class Wealth Average} MAPE}} & \multirow{4}{*}{\parbox{1cm}{\textsc{Low Class Wealth Average} MAPE}} & \multirow{4}{*}{\parbox{1cm}{\textsc{Gini Index} MAPE}} & \multirow{4}{*}{\parbox{1cm}{Total MAPE}} & \multirow{4}{*}{\parbox{1.2cm}{Parameter Mean Absolute Error}} & \multirow{4}{*}{\parbox{1.3cm}{Parameter Euclidean Error}}\\
			&\multicolumn{1}{c}{}&&&&&&&&\\
			&\multicolumn{1}{c}{}&&&&&&&&\\\cline{2-3}
			& \multicolumn{1}{l|}{\makecell[l]{Dyn-\\amic}} & \multicolumn{1}{l|}{\makecell[l]{Hetero-\\geneous}} &&&&&&&\\
			
			\hline\hline
			
			\multirow{2}{*}{\parbox{2cm}{\scriptsize{Synthetic Parameter Experiment}}} & \multirow{2}{*}{True} & \multirow{2}{*}{True} & 0.012 & 0.016 & 0.020 & 0.015 & 0.016 & 0 & 0 \\
			&&& ($\pm$0.009) & ($\pm$0.011) & ($\pm$0.013) & ($\pm$0.009) & ($\pm$0.007) & ($\pm$0) & ($\pm$0) \\\cline{1-10}
			\multirow{6}{*}{\parbox{2cm}{Random Search}} & \multirow{2}{*}{RS} & \multirow{2}{*}{True} & 0.096 & 0.060 & 0.046 & 0.024 & 0.057 & 0.549 & 0 \\
			&&& ($\pm$0.015) & ($\pm$0.008) & ($\pm$0.008) & ($\pm$0.007) & ($\pm$0.007) & ($\pm$0.037) & ($\pm$0) \\\cline{2-10}
			& \multirow{2}{*}{True} & \multirow{2}{*}{RS} & 0.052 & 0.031 & 0.035 & 0.028 & 0.036 & 0 & 0.034 \\
			&&& ($\pm$0.041) & ($\pm$0.022) & ($\pm$0.027) & ($\pm$0.027) & ($\pm$0.018) & ($\pm$0) & ($\pm$0.020) \\\cline{2-10}
			& \multirow{2}{*}{RS} & \multirow{2}{*}{RS} & 0.140 & 0.089 & 0.077 & 0.041 & 0.087 & 0.623 & 0.339 \\
			&&& ($\pm$0.031) & ($\pm$0.018) & ($\pm$0.024) & ($\pm$0.015) & ($\pm$0.013) & ($\pm$0.060) & ($\pm$0.238)\\
			
			\hline\hline
			
			\multirow{6}{*}{\parbox{2cm}{Dynamic Calibration}} & \multirow{2}{*}{ST} & \multirow{2}{*}{True} & 0.072 & 0.045 & 0.034 & 0.018 & 0.042\tnote{*} & 0.601 & 0 \\
			&&& ($\pm$0.008) & ($\pm$0.006) & ($\pm$0.004) & ($\pm$0.004) & ($\pm$0.004) & ($\pm$0.076) & ($\pm$0) \\
			\cline{2-10}
			& \multirow{2}{*}{SR} & \multirow{2}{*}{True} & 0.088 & 0.055 & 0.043 & 0.021 & 0.052\tnote{*} & 0.537 & 0 \\
			&&& ($\pm$0.016) & ($\pm$0.013) & ($\pm$0.010) & ($\pm$0.005) & ($\pm$0.010) & ($\pm$0.104) & ($\pm$0) \\
			\cline{2-10}
			& \multirow{2}{*}{MS} & \multirow{2}{*}{True} & 0.080 & 0.048 & 0.039 & 0.021 & 0.047\tnote{*} & 0.500 & 0 \\
			&&& ($\pm$0.011) & ($\pm$0.007) & ($\pm$0.007) & ($\pm$0.005) & ($\pm$0.006) & ($\pm$0.109) & ($\pm$0) \\
			
			\hline\hline
			
			\multirow{2}{*}{\parbox{2cm}{Heterogeneous Calibration}} & \multirow{2}{*}{True} & \multirow{2}{*}{BO} & 0.025 & 0.015 & 0.018 & 0.013 & 0.018\tnote{*} & 0 & 0.021 \\
			&&& ($\pm$0.018) & ($\pm$0.010) & ($\pm$0.012) & ($\pm$0.007) & ($\pm$0.008) & ($\pm$0) & ($\pm$0.011) \\
			
			\hline\hline
			
			\multirow{2}{*}{\parbox{2cm}{Calibration Framework\tnote{a}}} & \multirow{2}{*}{MS} & \multirow{2}{*}{BO} & 0.124 & 0.085 & 0.074 & 0.044 & 0.082 & 0.462 & 0.110 \\
			&&& ($\pm$0.017) & ($\pm$0.020) & ($\pm$0.031) & ($\pm$0.021) & ($\pm$0.021) & ($\pm$0.044) & ($\pm$0.020)\\\cline{1-10}
			
			\multirow{2}{*}{\parbox{2cm}{Calibration Framework\tnote{b}}} & \multirow{2}{*}{MS} & \multirow{2}{*}{BO} & 0.113 & 0.070 & 0.073 & 0.048 & 0.076\tnote{*} & 0.501 & 0.092 \\
			&&& ($\pm$0.011) & ($\pm$0.007) & ($\pm$0.021) & ($\pm$0.021) & ($\pm$0.010) & ($\pm$0.089) & ($\pm$0.015)\\
			
			\hline

			%\multirow{}{||p{cm}|}{Dynamic Calibration} & 
			
		\end{tabular}
	\begin{tablenotes}
		\item[*] $P<0.05$
		\item[a] Calibration framework with $C_{dyn}=2$, and $C_{het}=3$
		\item[b] Calibration framework with $C_{dyn}=20$, and $C_{het}=30$
	\end{tablenotes}
	\end{threeparttable}
\end{table}

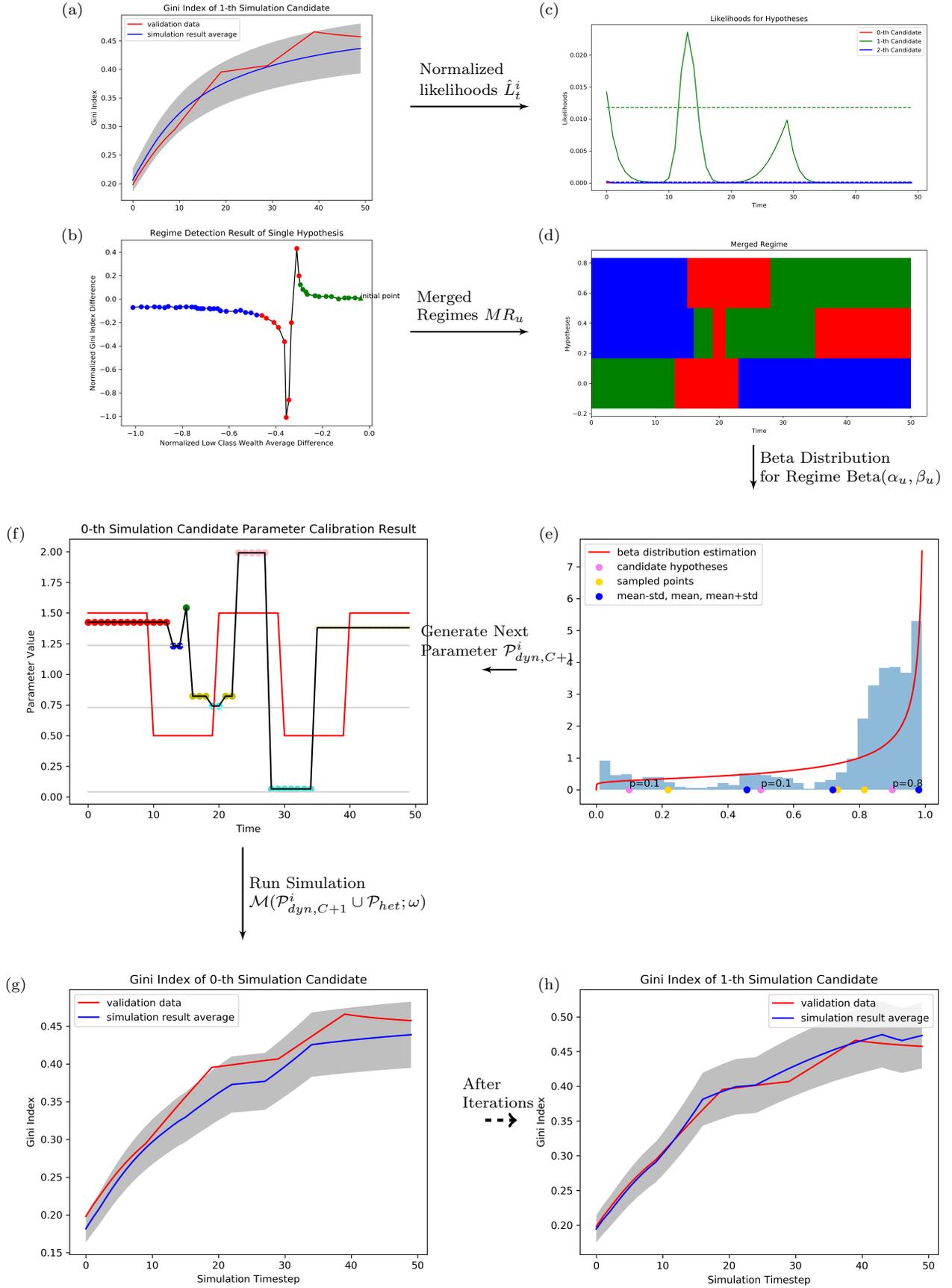
\begin{figure*}
	\begin{tikzpicture}[node distance=-3cm, auto]  
	\tikzset{
		mynode/.style={rectangle,rounded corners,draw=black, top color=white, bottom color=yellow!50,very thick, inner sep=1em, minimum size=3em, text centered},
		myarrow/.style={->, >=latex', shorten >=1pt, thick},
		mylabel/.style={text width=7em, text centered}
	}  
	\node[] (node1) at (0,2) {\includegraphics[scale=0.35]{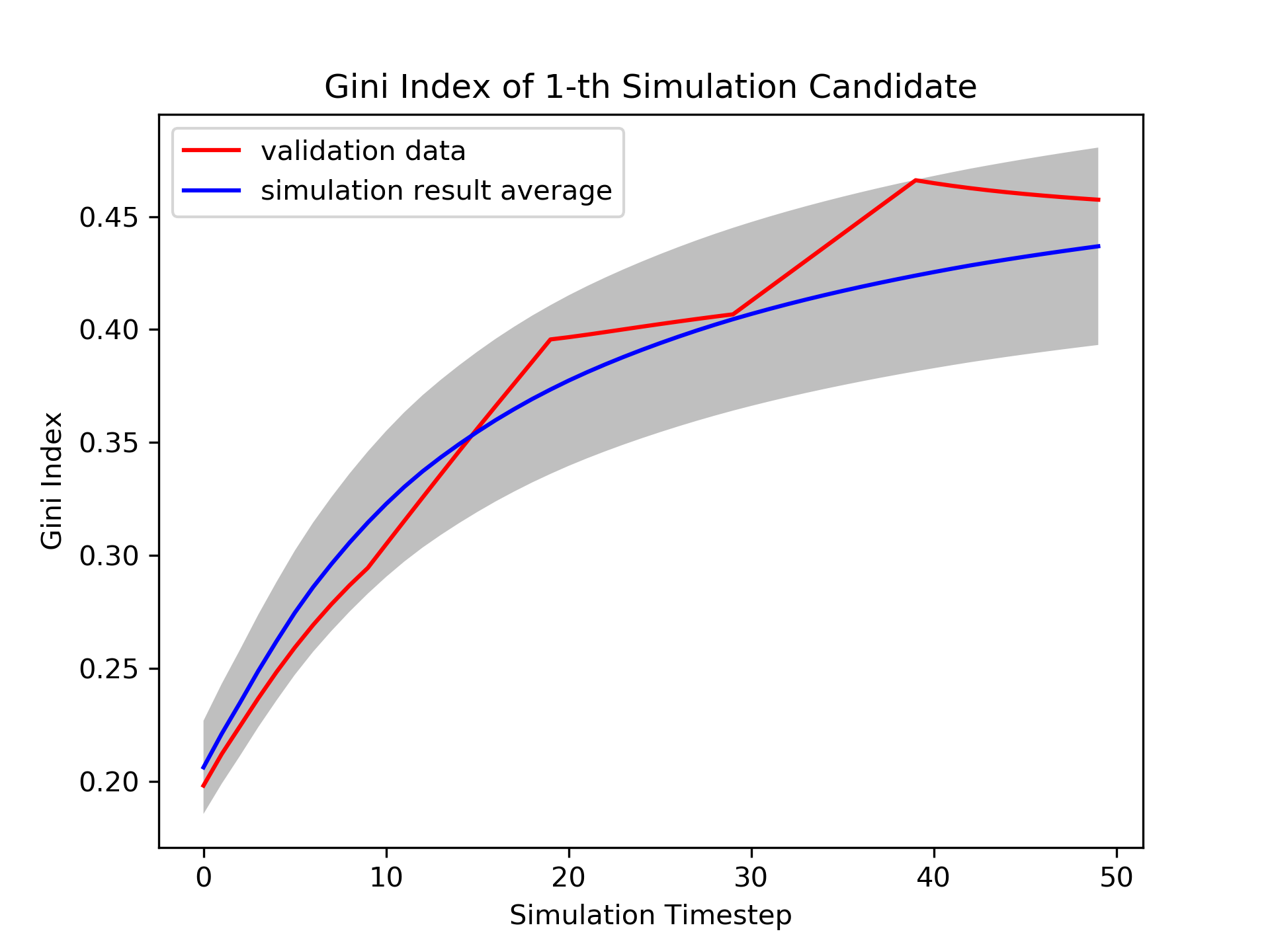}};
	\node[] (node2) at (0,-2) {\includegraphics[scale=0.35]{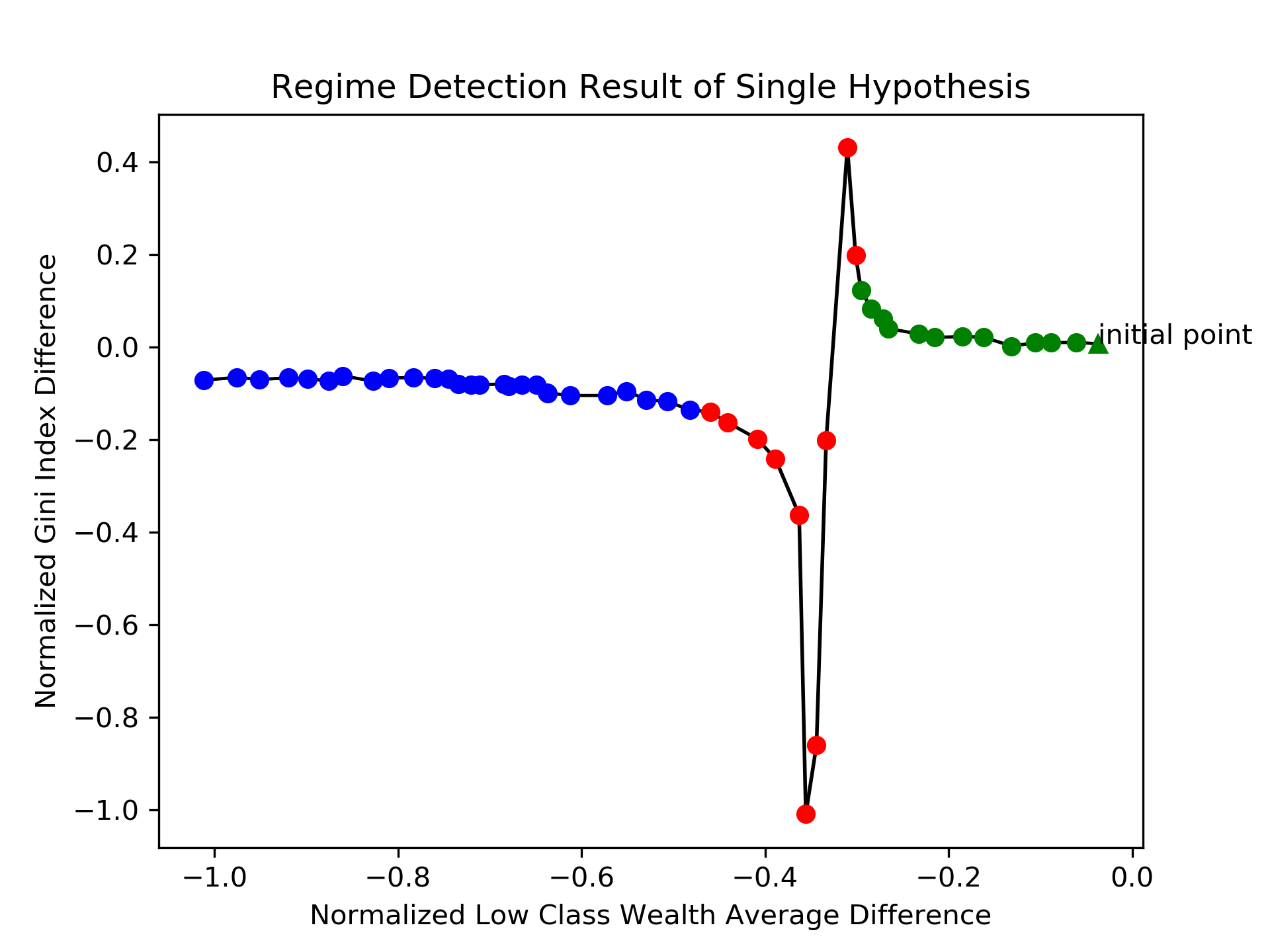}};  
	\node[] (node3) at (9,2) {\includegraphics[scale=0.3]{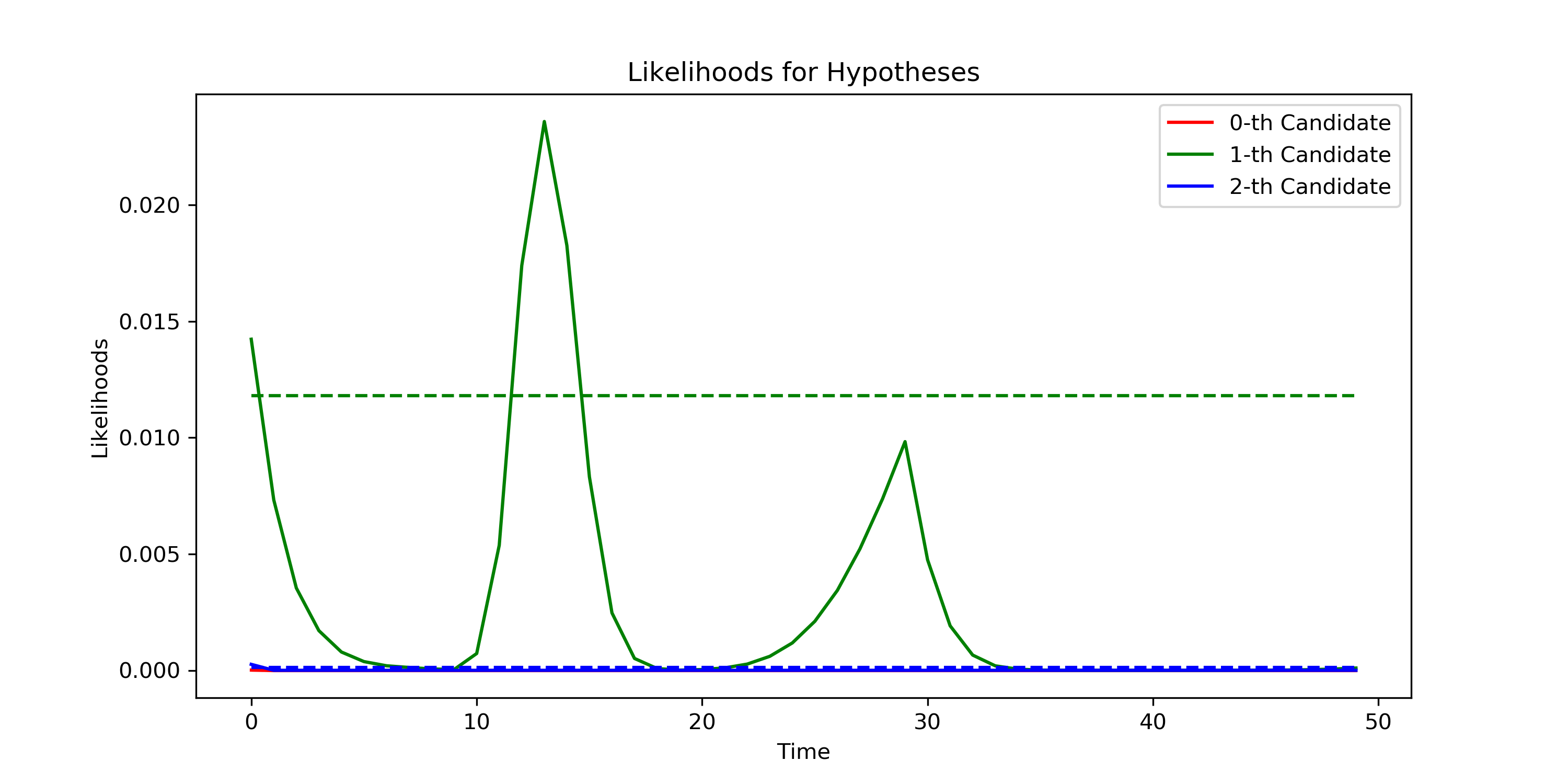}};  
	\node[] (node4) at (9,-2) {\includegraphics[scale=0.3]{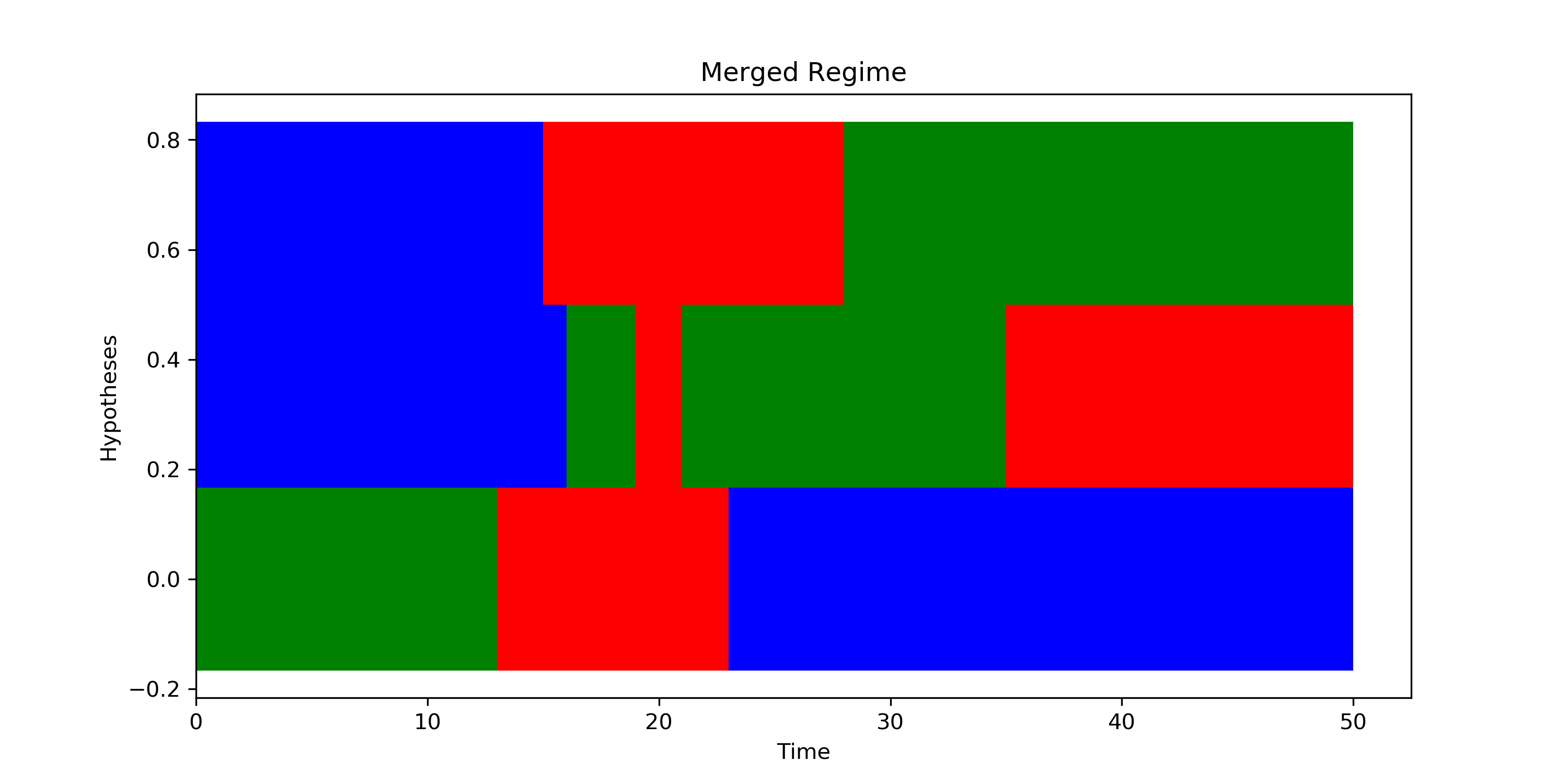}};  
	\node[] (node5) at (0,-8) {\includegraphics[scale=0.5]{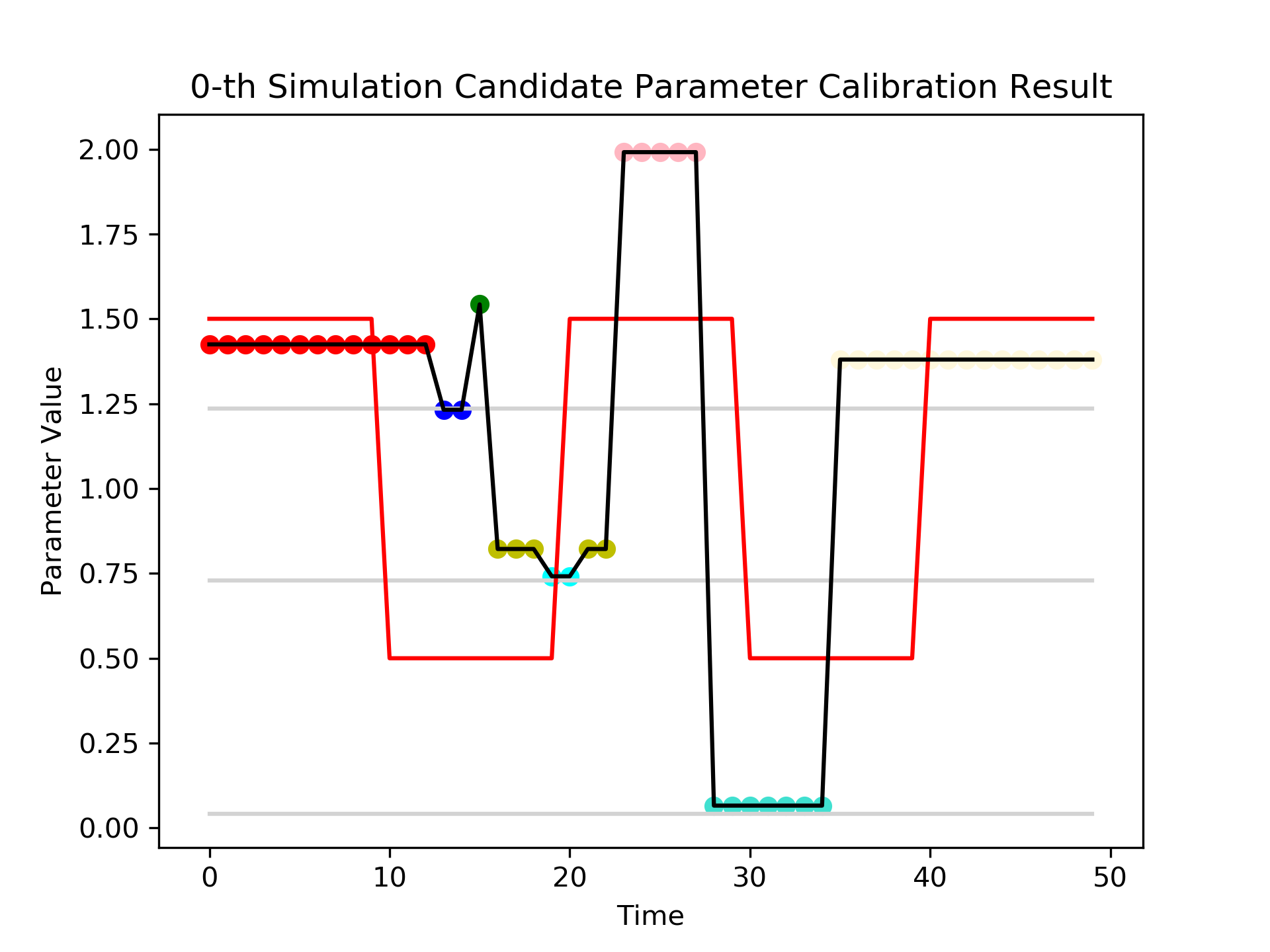}};
	\node[] (node6) at (9,-8) {\includegraphics[scale=0.5]{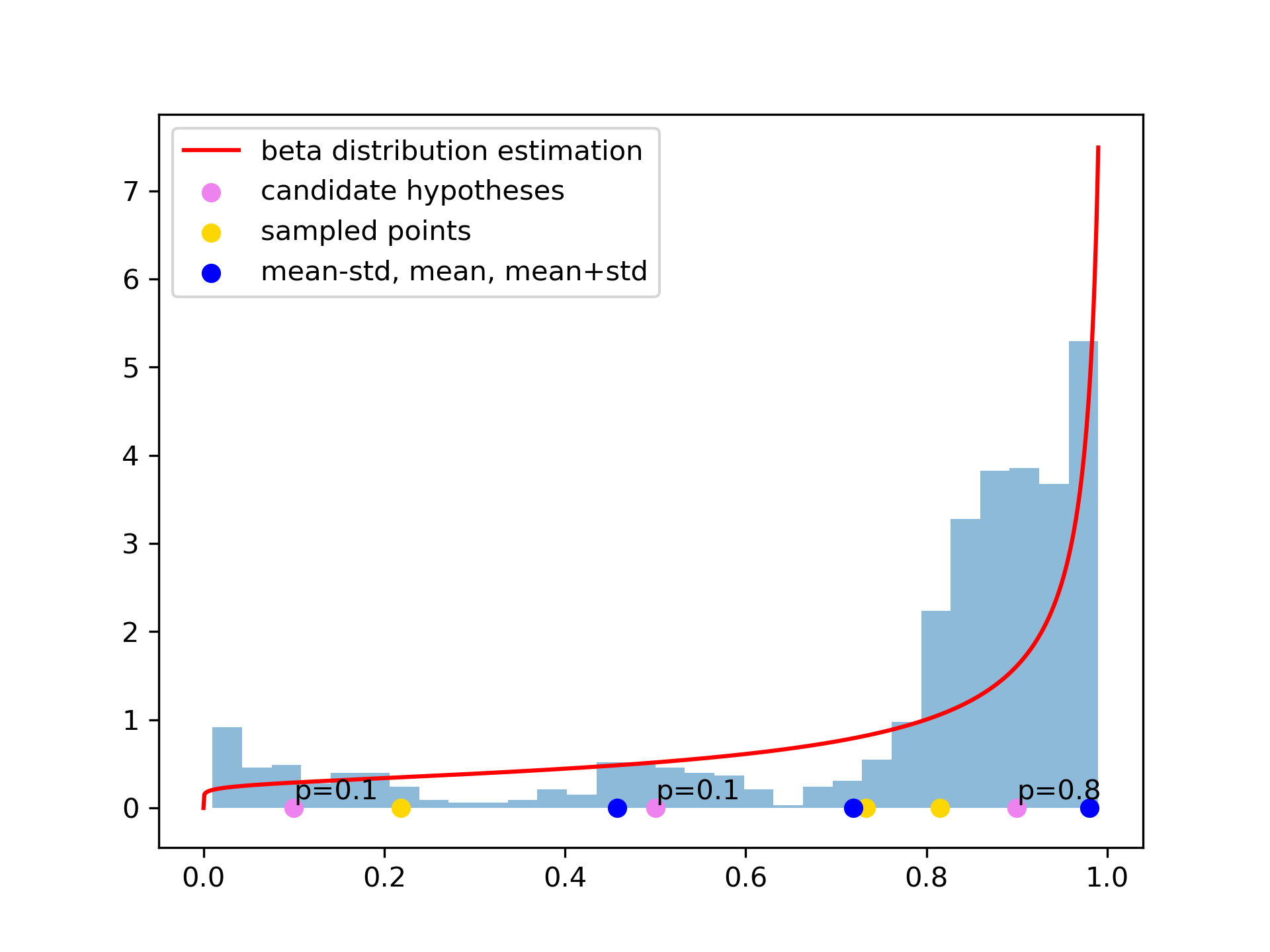}};
	\node[] (node7) at (0,-16) {\includegraphics[scale=0.5]{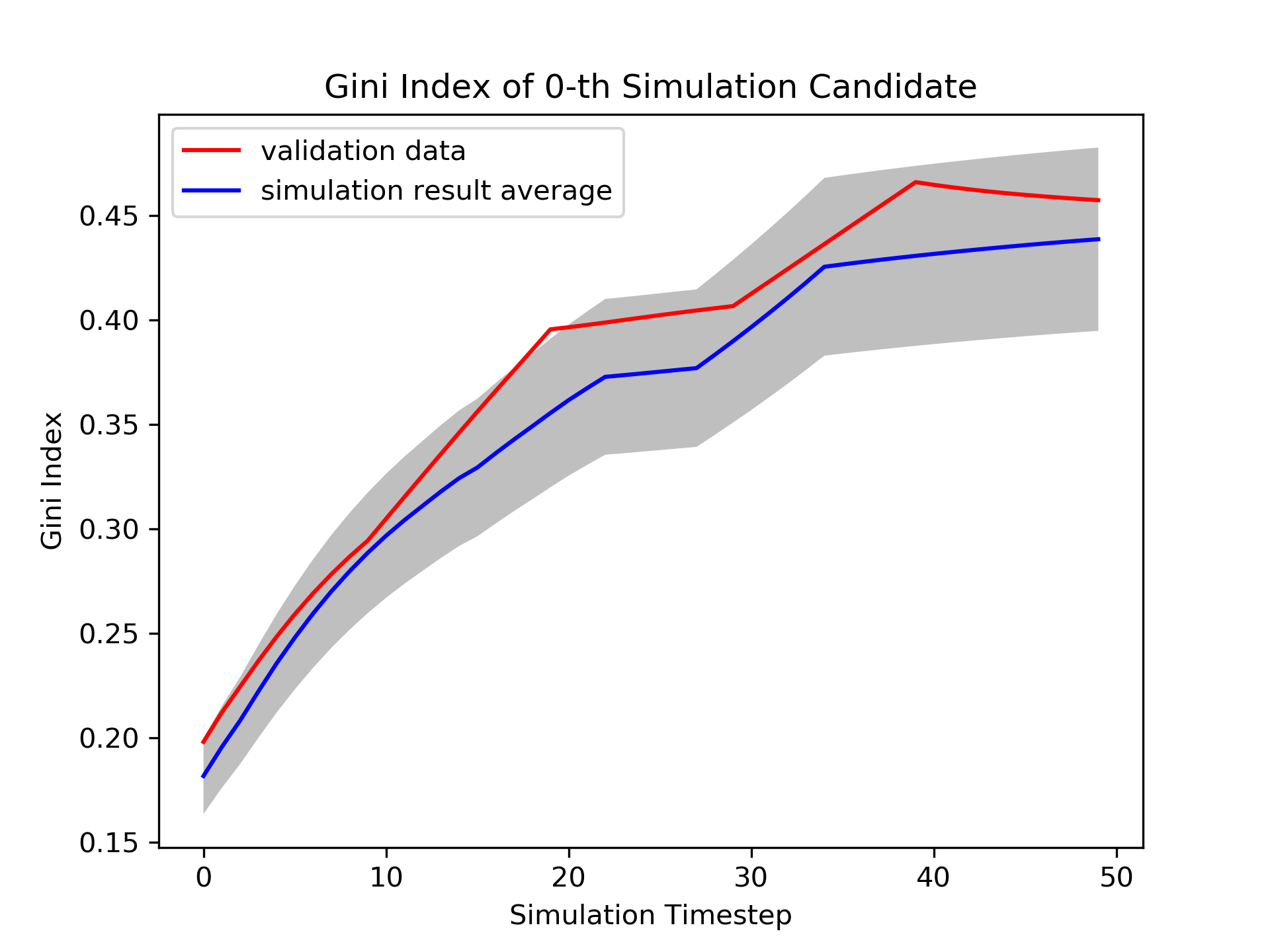}};
	\node[] (node8) at (9,-16) {\includegraphics[scale=0.5]{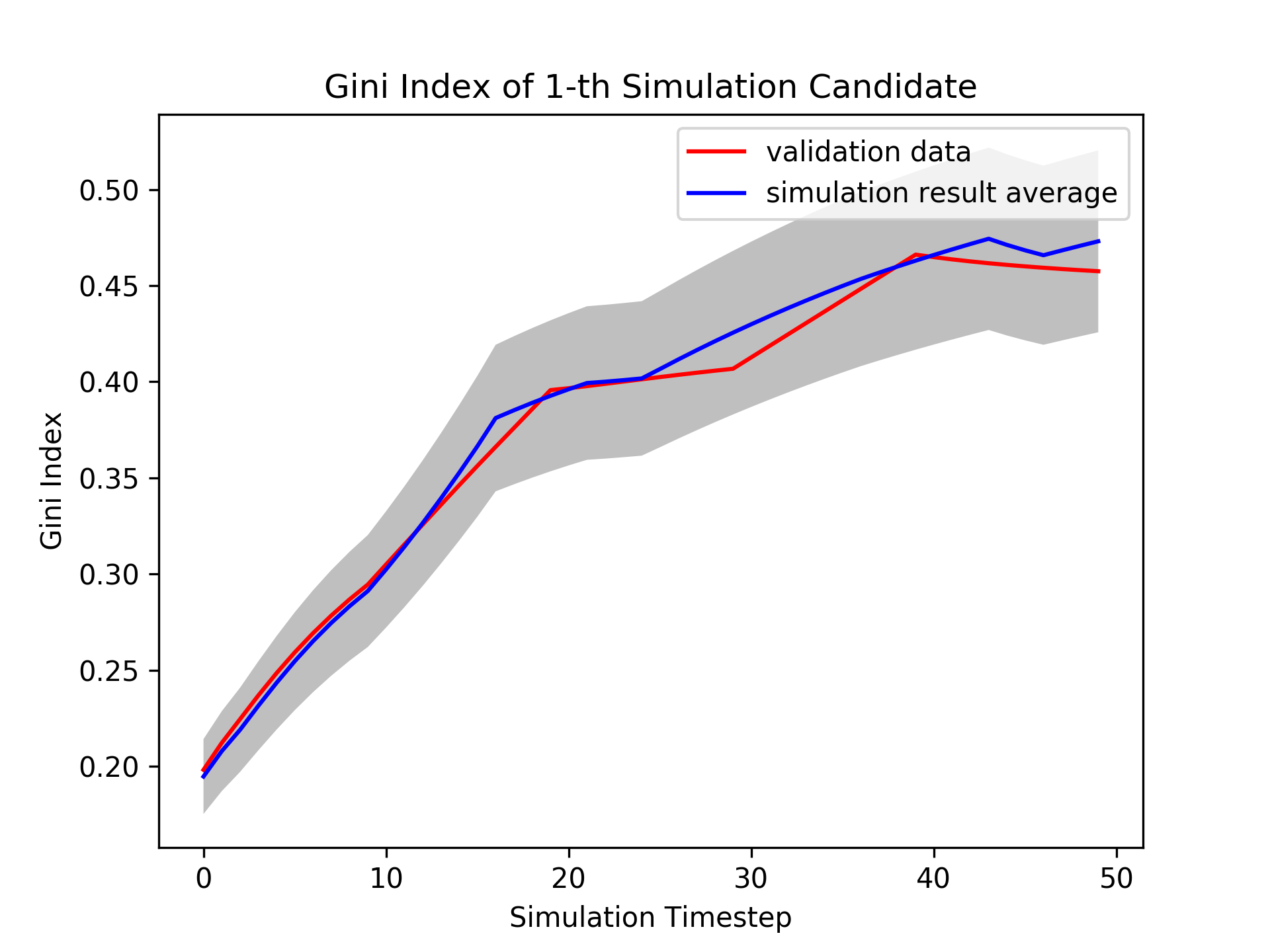}};
	
	\node[] at (-3.0,3.7) {(a)};
	\node[] at (-3.0,-0.3) {(b)};
	\node[] at (5.4,3.7) {(c)};
	\node[] at (5.4,-0.3) {(d)};
	\node[] at (5.4,-5.6) {(e)};
	\node[] at (-4.0,-5.6) {(f)};
	\node[] at (-4.0, -13.6) {(g)};
	\node[] at (5.4,-13.6) {(h)};
	
	\draw[myarrow] (node1.east) -- node [above] {\footnotesize{\makecell[l]{Normalized\\likelihoods $\hat{L}_{t}^{i}$}}} (node3.west);
	\draw[myarrow] (node2.east) -- node [above] {\footnotesize{\makecell[l]{Merged\\Regimes $MR_{u}$}}} (node4.west);
	\draw[myarrow] (node4.south) -- node [right] {\footnotesize{\makecell[l]{Beta Distribution\\for Regime $\text{Beta}(\alpha_{u},\beta_{u})$}}} (node6.north);
	\draw[myarrow] (node6.west) -- node [above] {\footnotesize{\makecell[l]{Generate Next\\Parameter $\mathcal{P}_{dyn,C+1}^{i}$}}} (node5.east);
	\draw[myarrow] (node5.south) -- node [right] {\footnotesize{\makecell[l]{Run Simulation\\ $\mathcal{M}(\mathcal{P}_{dyn,C+1}^{i}\cup\mathcal{P}_{het};\omega)$}}} (node7.north);
	\draw[ultra thick, dashed, dash phase=3pt, ->] (node7.east) -- node [above] {\footnotesize{\makecell[l]{After\\Iterations}}} (node8.west);
	
	\end{tikzpicture} 
	\medskip
	\caption{Dynamic parameter calibration mechanism is plotted. (a)-(g) presents a single dynamic calibration iteration, and (h) presents a final calibrated simulation result. (a) and (c) present calculating the simulation error, in Alg. \ref{alg:DynamicCalibration}-Line 12-15. (b) and (d) present the regime detection process, in Alg. \ref{alg:DynamicCalibration}-Line 17. (e) and (f) present the parameter generation process, in Alg. \ref{alg:DynamicCalibration}-Line 18}
	\label{fig:dynamic_algorithm} 
\end{figure*}

\begin{figure}
	\begin{tikzpicture}[node distance=-3cm, auto]  
	\tikzset{
		mynode/.style={rectangle,rounded corners,draw=black, top color=white, bottom color=yellow!50,very thick, inner sep=1em, minimum size=3em, text centered},
		myarrow/.style={->, >=latex', shorten >=1pt, thick},
		mylabel/.style={text width=7em, text centered}
	}
	\node[] (node1) at (0,0) {\includegraphics[scale=0.5]{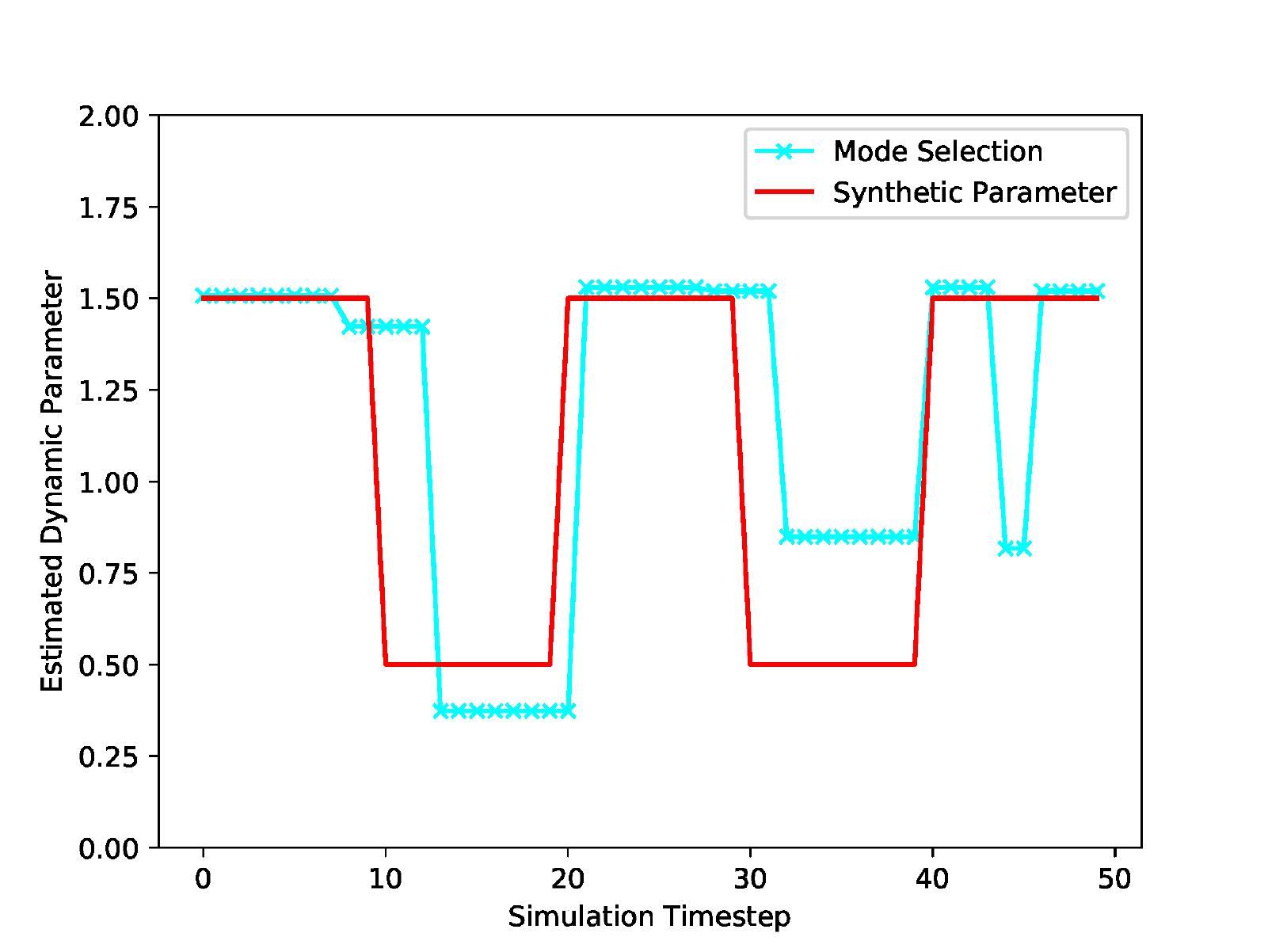}};
	\node[] (node2) at (8,0)
	{\includegraphics[scale=0.5]{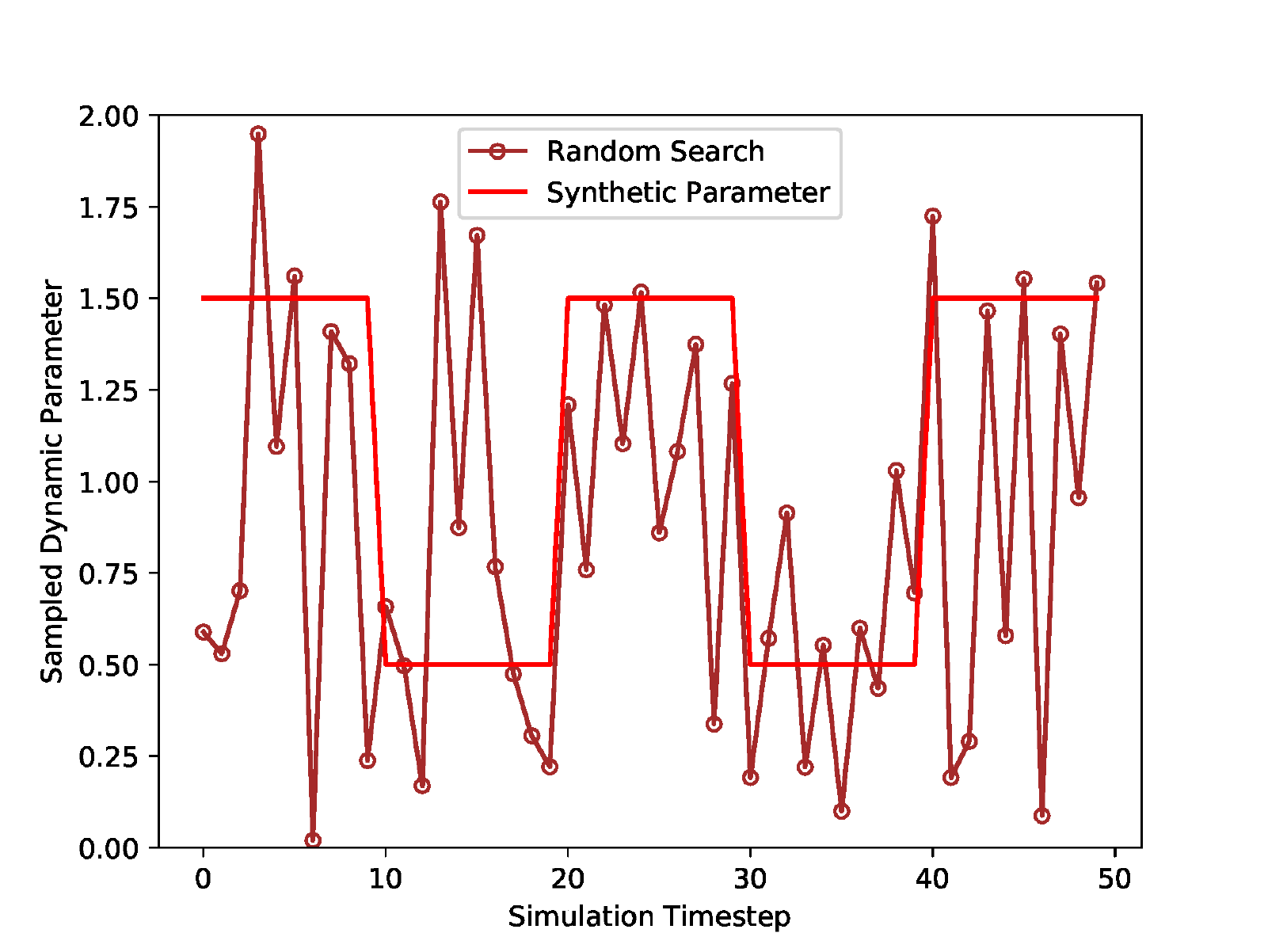}};
	
	\node[] at (-4.0,2.7) {(a)};
	\node[] at (4.0,2.7) {(b)};

	\end{tikzpicture} 
	\caption{(a) The estimated optimal dynamic parameter, x marked line, is plotted. Dynamic calibration generates a dynamic parameter regime-wisely to avoid overfitting in \textit{Mode Selection} update rule. (b) The sampled optimal dynamic parameter, o marked line, is plotted. \textit{Random Search} finds an overfitted dynamic parameter, which only fits the given validation data, without following the synthetic dynamic parameter trend}
	\label{fig:DynamicTrueComparison}
\end{figure}

\begin{figure}
	\begin{tikzpicture}[node distance=-3cm, auto]  
	\tikzset{
		mynode/.style={rectangle,rounded corners,draw=black, top color=white, bottom color=yellow!50,very thick, inner sep=1em, minimum size=3em, text centered},
		myarrow/.style={->, >=latex', shorten >=1pt, thick},
		mylabel/.style={text width=7em, text centered}
	}
	\node[] (node2) at (0,0)
	{\includegraphics[scale=0.5]{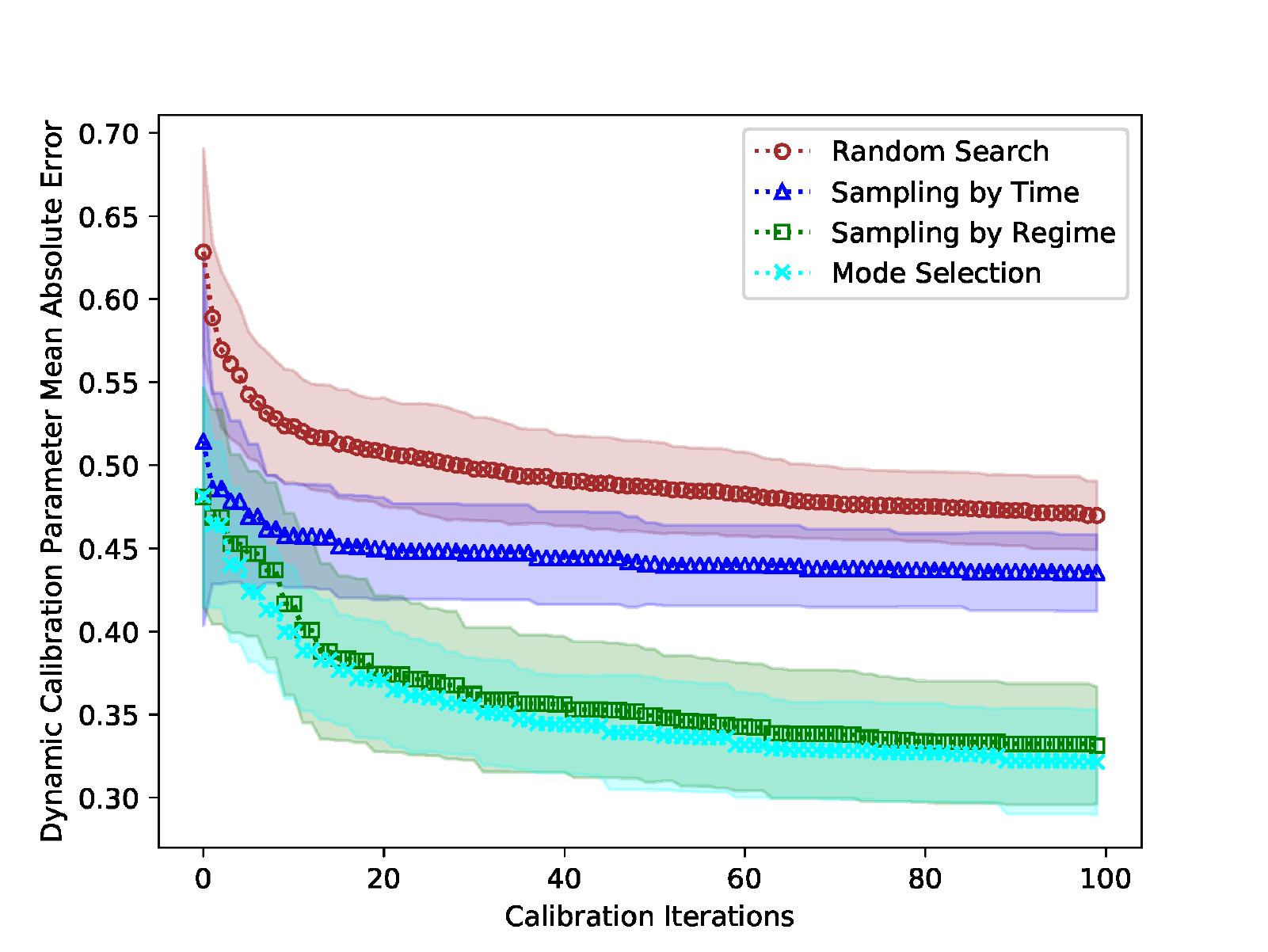}};
	\node[] (node1) at (8,0) {\includegraphics[scale=0.5]{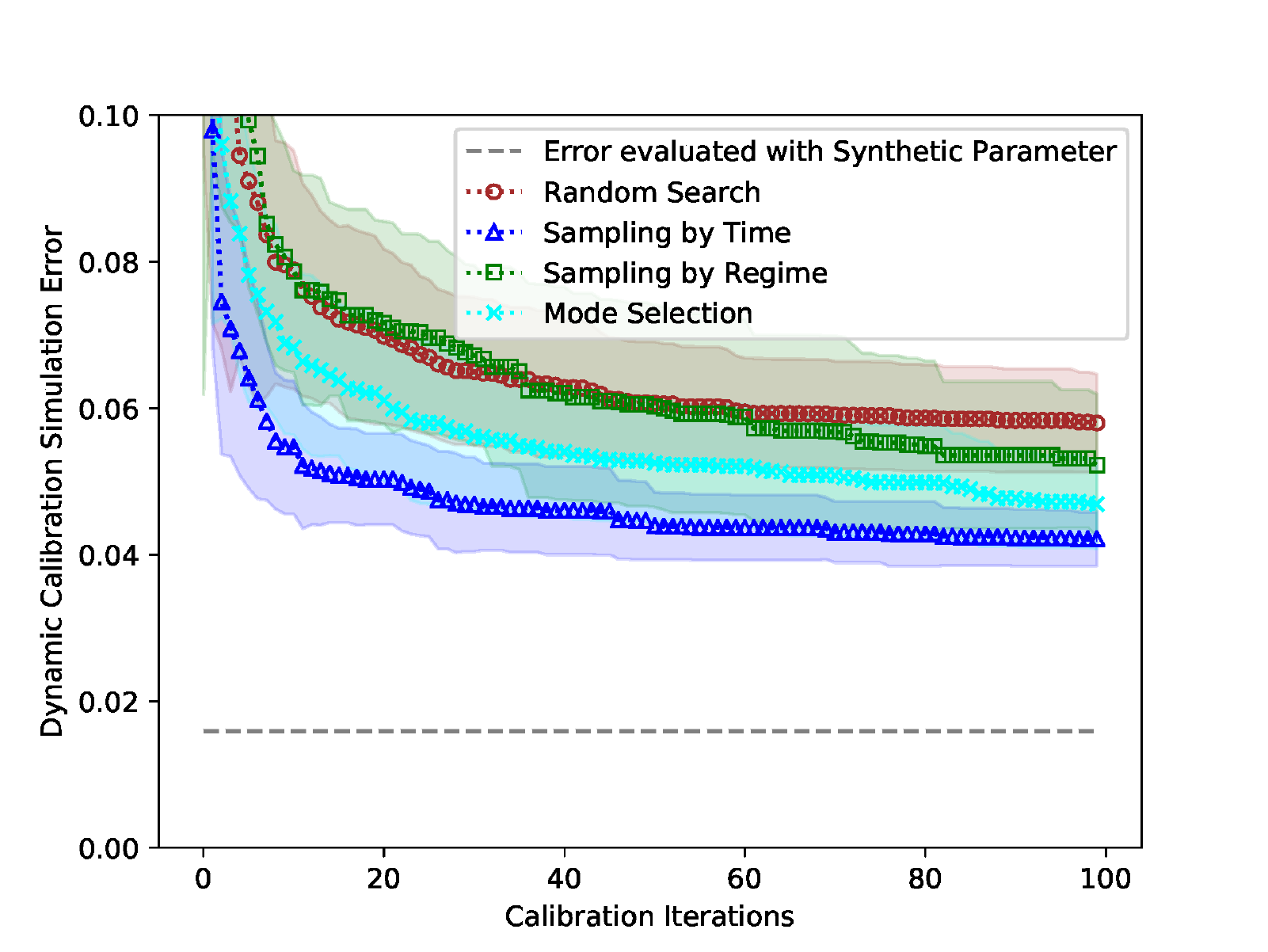}};
	
	\node[] at (-4.0,2.7) {(a)};
	\node[] at (4.0,2.7) {(b)};

	\end{tikzpicture} 
	\caption{(a) The calibrated dynamic parameter mean absolute errors are plotted. Parameter generation methods \textit{Sampling by Regime} and \textit{Mode Selection} find parameters closer to the synthetic parameter than the other methods. (b) Dynamic calibration simulation error is plotted by iterations. Parameter generation method \textit{Sampling by Time} performs the best in terms of simulation MAPE}
	\label{fig:DynamicError}
\end{figure}

\begin{figure}
	\centering
	\includegraphics[scale=0.5]{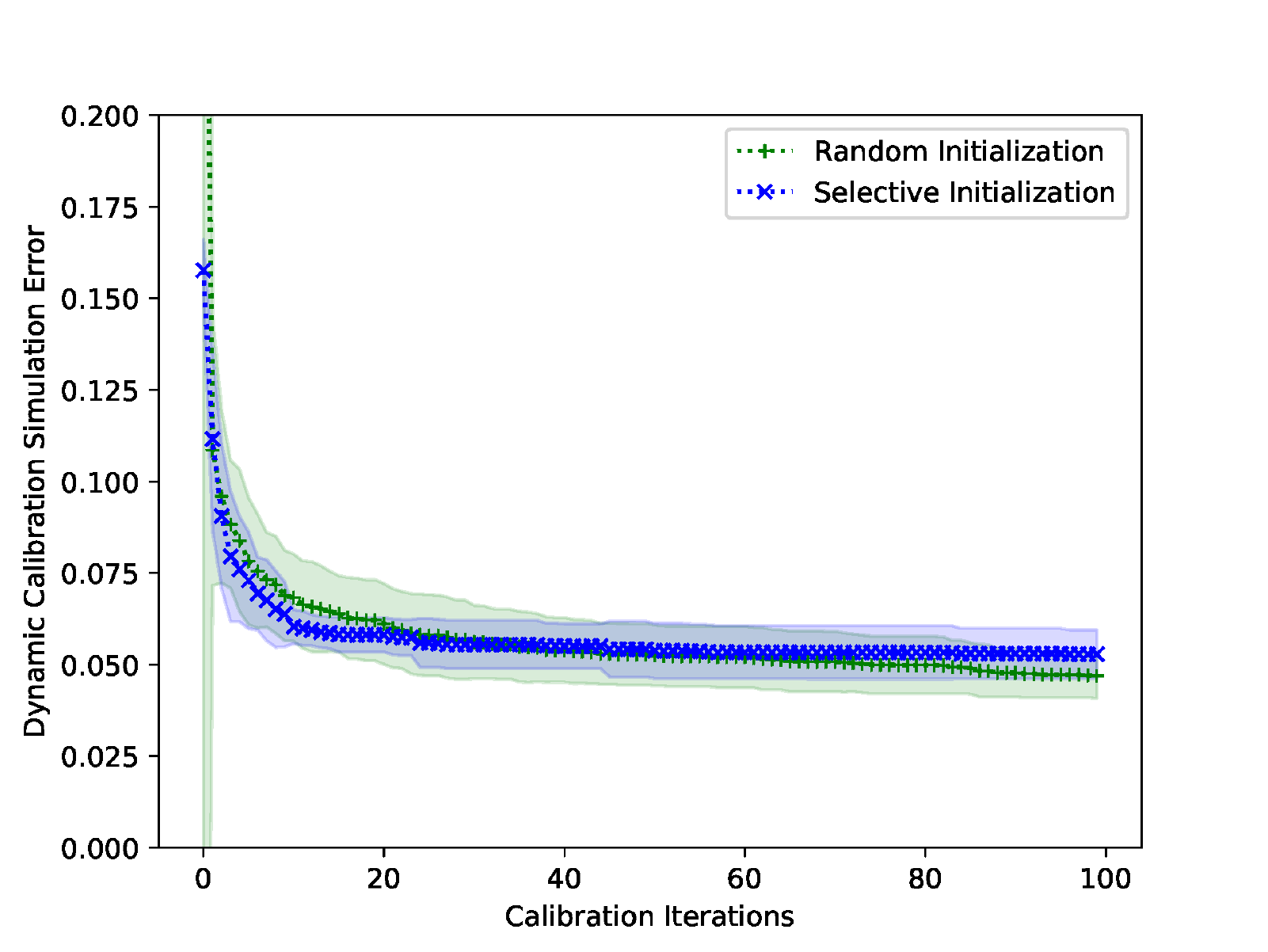}
	\caption{Dynamic calibration simulation errors with different initializations are plotted. The + marked line is the experimental result of the random initialization and the x marked line is that of the selective initialization, with equally distributed initial static parameters, having values 0.5, 1.0, and 1.5}
	\label{fig:DynamicCalibrationInitialization}
\end{figure}

\begin{figure*}
	\begin{tikzpicture}[node distance=0cm, auto]  
	\tikzset{
		mynode/.style={rectangle,rounded corners,draw=black, top color=white, bottom color=yellow!50,very thick, inner sep=1em, minimum size=3em, text centered},
		myarrow/.style={->, >=latex', shorten >=1pt, thick},
		mylabel/.style={text width=7em, text centered}
	}  
	\node[] (node1) at (0,0) {\includegraphics[scale=0.23]{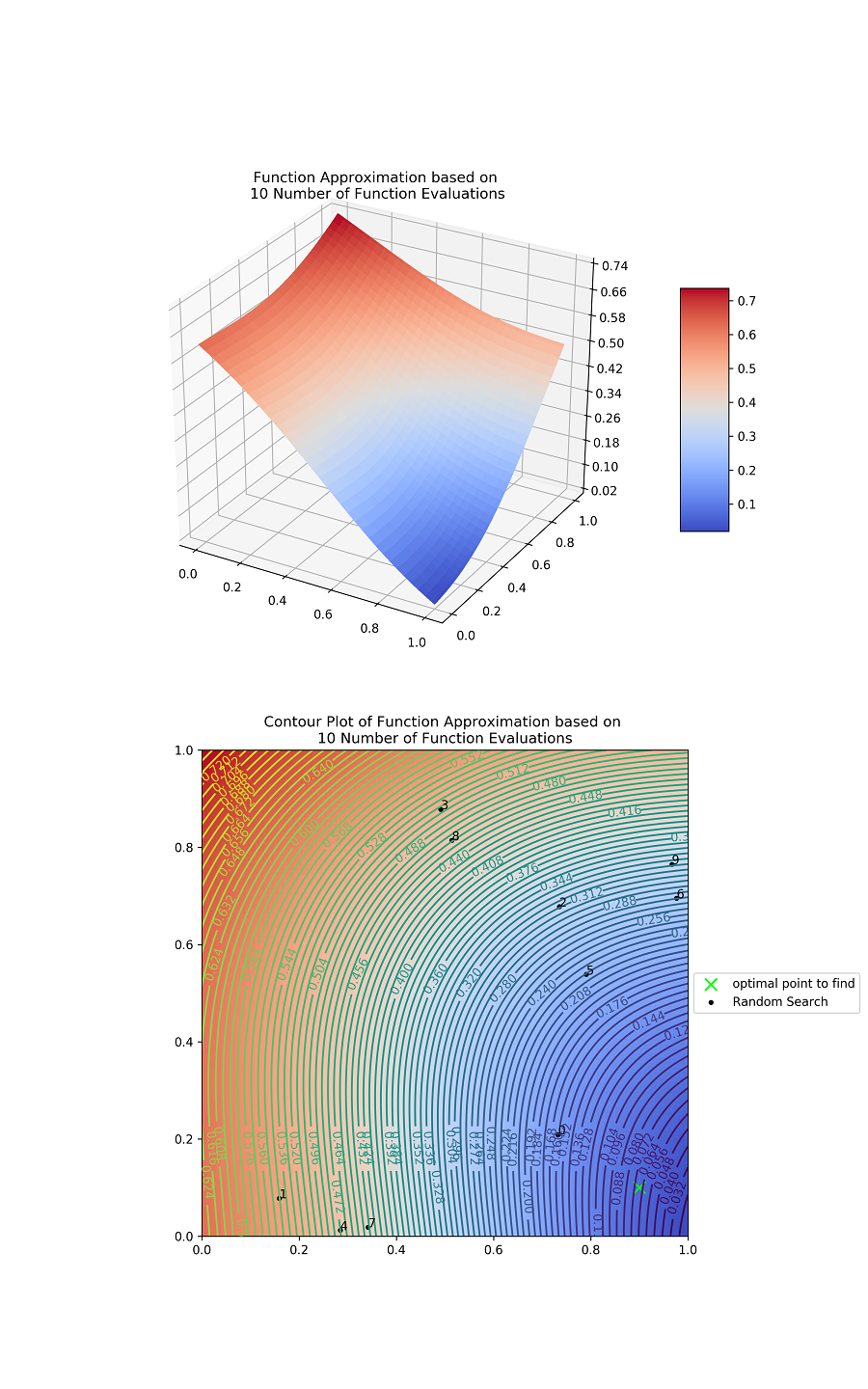}};
	\node[] (node2) at (5.8,0) {\includegraphics[scale=0.23]{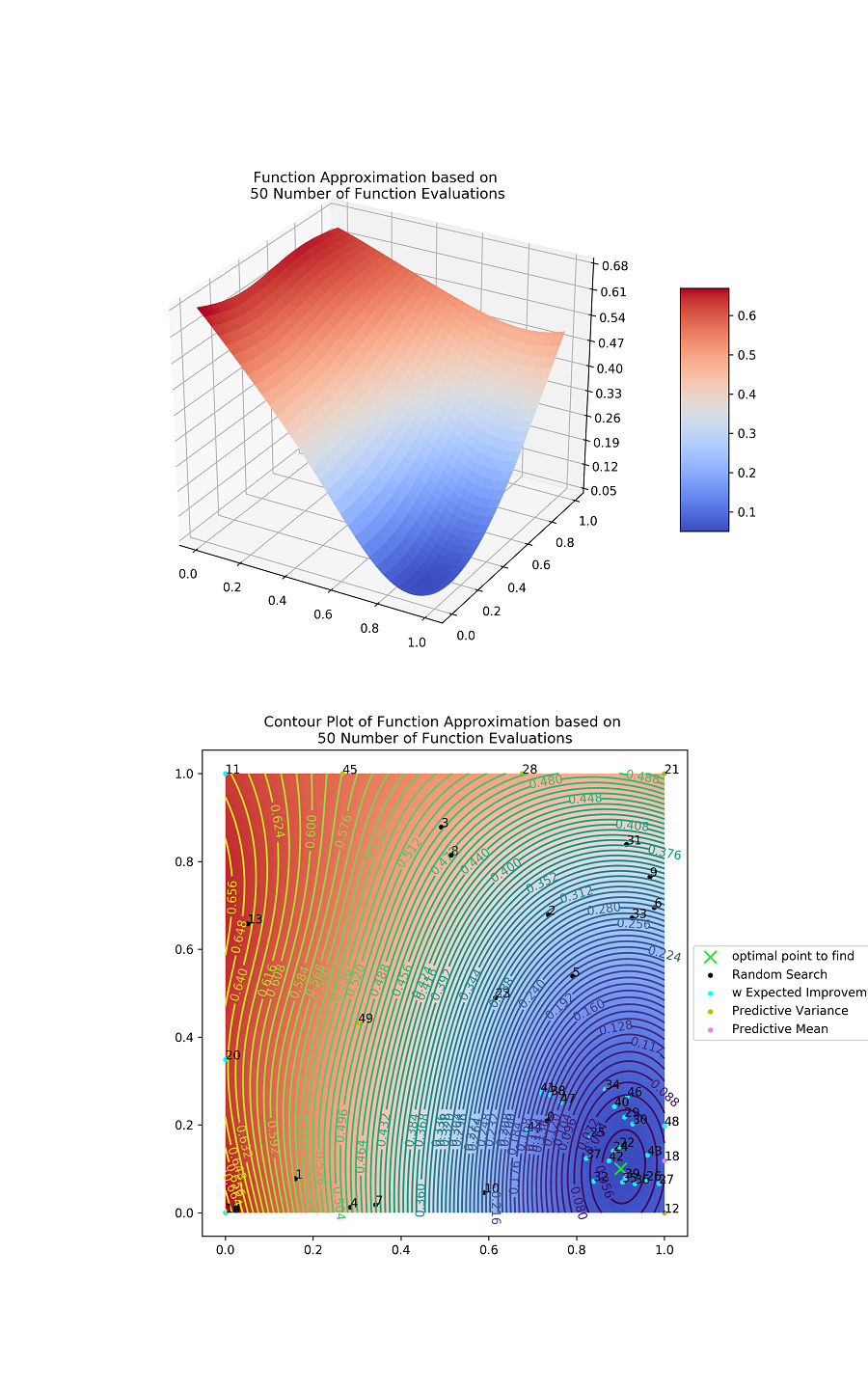}};  
	\node[] (node3) at (11.6,0) {\includegraphics[scale=0.23]{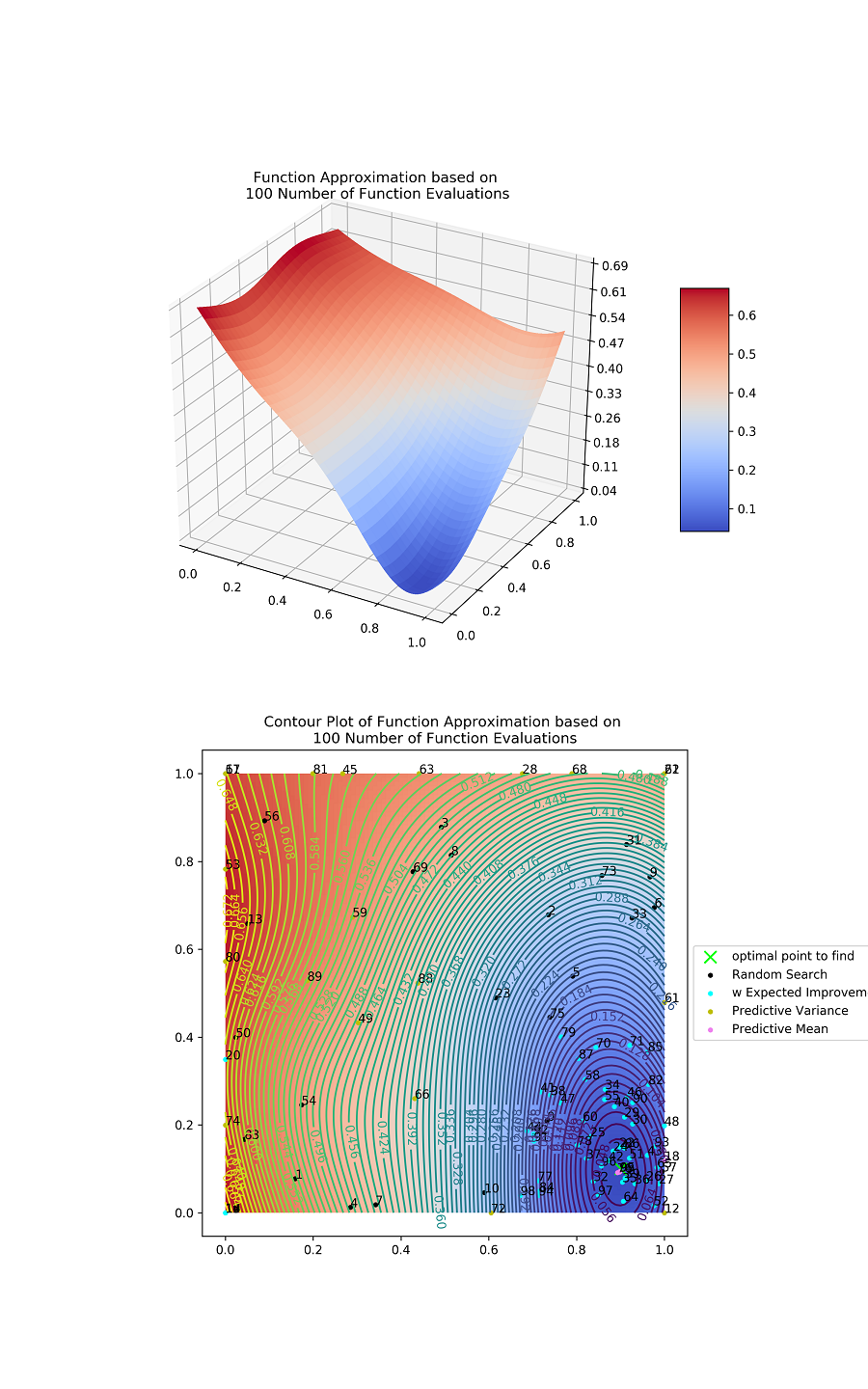}};  
	\node[] (node4) at (2,-8) {\includegraphics[scale=0.4]{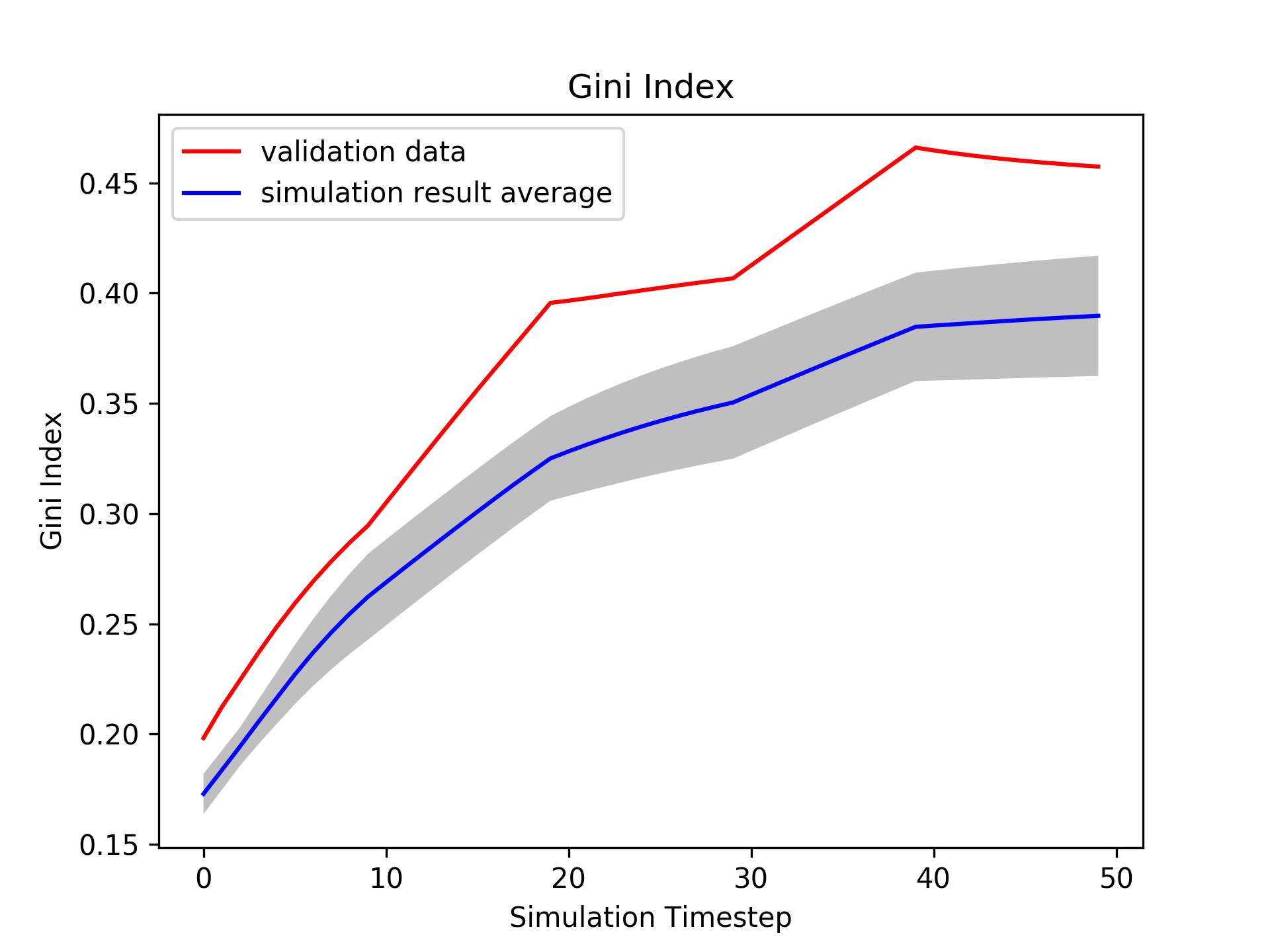}};  
	\node[] (node5) at (9,-8) {\includegraphics[scale=0.4]{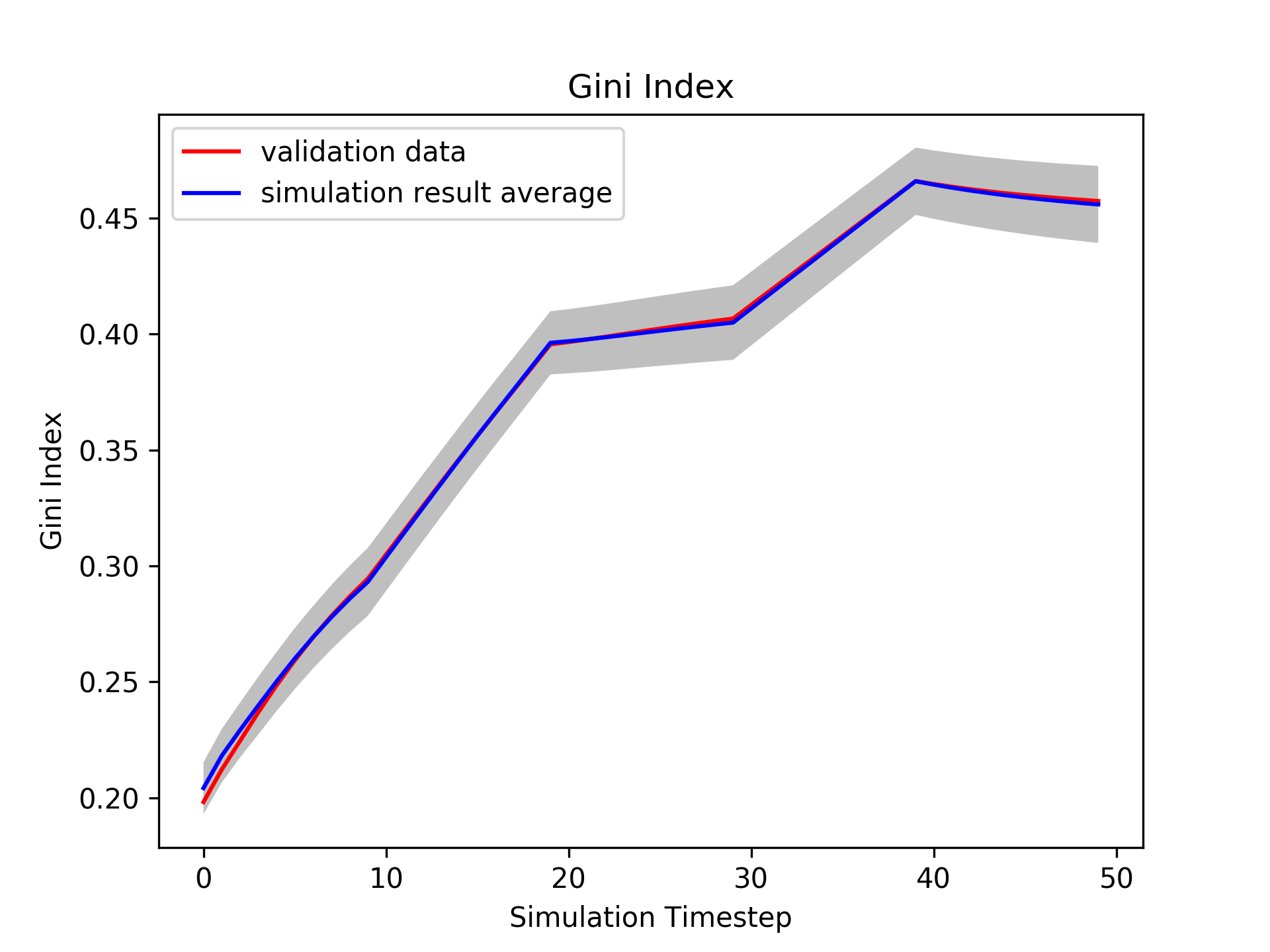}};
	
	\node[] at (-2.2,3.4) {(a)};
	\node[] at (3.55,3.4) {(b)};
	\node[] at (9.35,3.4) {(c)};
	\node[] at (-2.2,-0.5) {(d)};
	\node[] at (3.55,-0.5) {(e)};
	\node[] at (9.35,-0.5) {(f)};
	\node[] at (-1,-6) {(g)};
	\node[] at (6,-6) {(h)};
	
	\draw[myarrow] (0.5,-4.4) -- node [right] {\footnotesize{  Initial Result}} (2,-5.6);
	\draw[myarrow] (11.6,-4.4) -- node[right] {\footnotesize{   After Iterations}} (9,-5.6);
	%\draw[myarrow] (node2.east) -- node [above] {\small{Merged Regimes}} (node4.west);
	%\draw[myarrow] (node4.south) -- node [right] {\small{Beta Distribution for Regime}} (node6.north);
	%\draw[myarrow] (node6.west) -- node [above] {\small{Generate Next Parameter}} (node5.east);
	%\draw[myarrow] (node5.south) -- node [right] {\small{Run Simulation}} (node7.north);
	%\draw[ultra thick, dashed, dash phase=3pt, ->] (node7.east) -- node [above] {\small{After 7 Iterations}} (node8.west);
	
	\end{tikzpicture} 
	\medskip
	\caption{Heterogeneous parameter calibration mechanism is plotted. (a), (b), and (c) present the learning process of the response surface as iterations increase, in Alg. \ref{alg:HeterogeneousCalibration}-Line 28. (d), (e), and (f) are the contour plots of the response surface, with evaluation path, in Alg. \ref{alg:HeterogeneousCalibration}-Line 29. The search finds the global optimum as we expected. (g) and (h) present a single summary statistics, Gini index, to compare before and after in calibration}
	\label{fig:heterogeneous_results} 
\end{figure*}

Fig. \ref{fig:dynamic_algorithm} presents the overall mechanism of dynamic calibration. Fig. \ref{fig:dynamic_algorithm} (a) presents a single dimension of summary statistics, \textsc{Gini Index}, of a parameter candidate hypothesis. (b) presents the result of HMM, $R_{t}^{i}$ on Alg. \ref{alg:DynamicCalibration}-Line 3, that divides the well-fitted regime, in green points, with the poorly-fitted regime, in red and blue points. The joint likelihoods, $L_{t}^{i}$ on Alg. \ref{alg:DynamicCalibration}-Line 15, are presented in (c). The dotted lines are the thresholds in Eq. \ref{eq:PoorlyFitted}. Exploration is executed in the most of the simulation timesteps for the first few iterations. Fig. \ref{fig:dynamic_algorithm} (d) presents the regime detection results for all three candidate hypotheses, $\text{MR}_{u}$ on Alg. \ref{alg:DynamicCalibration}-Line 17. The estimated beta distribution of a single merged regime, $\text{Beta}(\alpha_{u},\beta_{u})$ on Alg. \ref{alg:DynamicCalibration}-Line 8, is presented in (e). Fig. \ref{fig:dynamic_algorithm} (f) presents one of the next candidate hypotheses, $\mathcal{P}_{dyn,C+1}^{i}$ on Alg. \ref{alg:CalibrationFramework}-Line 6, generated from the estimated beta distributions. The current dynamic parameter candidates, $\mathcal{P}_{dyn,C}^{i}$, are plotted as the grey lines, and the generated dynamic parameter evaluated in the next simulation is plotted as the black line, with different colors represent different merged regimes.

The well-fitted regime in Fig. \ref{fig:dynamic_algorithm} (b) makes the parameter estimation at the front period success, as the red colored dots in (f), because the second candidate hypothesis at the well-fitted regime has high likelihoods that the next parameter will be generated near the second hypothesis. The simulation has failed to recover the validation statistics afterwards, and this leads the algorithm to force explorations at the latter simulation timesteps. Fig. \ref{fig:dynamic_algorithm} (g) is a simulation result, $\mathcal{S}_{dyn}$ on Alg. \ref{alg:DynamicCalibration}-Line 14, evaluated with the estimated parameter in Fig. \ref{fig:dynamic_algorithm} (f). Iterative dynamic updates will change the simulation result to the synthetic generated validation data, as in Fig. \ref{fig:dynamic_algorithm} (h).

Fig. \ref{fig:DynamicTrueComparison} presents the best estimated parameters with two updating rules: \textit{Mode Selection} and \textit{Random Search}. \textit{Mode Selection} in (a) estimates the plausible parameter which shares the tendency with the synthetic parameter. However, \textit{Random Search} in (b) finds the highly fluctuating parameter, which is unable to interpret. Fig. \ref{fig:DynamicError} (a) plots the parameter Mean Absolute Error (MAE) for the four updating rules. \textit{Mode Selection} and \textit{Sampling by Regime} performs better than other updating rules in parameter error. However, Fig. \ref{fig:DynamicError} (b) shows that the updating rule \textit{Sampling by Time} performs the best in terms of simulation error.

Tab. \ref{tab:statisticsinTestCase1} shows the experimental statistics of dynamic calibration. \textit{Sampling by Time} performs the best in simulation error, but it has the highest parameter error, which implicates that \textit{Sampling by Time} estimates the overfitted parameter. \textit{Sampling by Regime} performance, on the other hand, is not significantly different from that of \textit{Random Search}. Lastly, \textit{Mode Selection} is the best updating rule in reducing the parameter error.

Fig. \ref{fig:DynamicCalibrationInitialization} shows the effect of the parameter initialization. The plus marked line is the calibration result with the randomly intialized dynamic parameters. Each parameter candidate has random static value at the initial calibration step, see gray lines in Fig. \ref{fig:dynamic_algorithm} (f). The x marked line is the simulation result of the selective initialization, where the parameter candidates are equally distributed in the parameter space, having values 0.5, 1.0, and 1.5. The selective dynamic parameters are believed to be initialized well, since the synthetic dynamic parameter is alternating between 0.5 and 1.5, see Fig. \ref{fig:DynamicTrueComparison}. Fig. \ref{fig:DynamicCalibrationInitialization} shows the selective initialization saturates faster than the random initialization, but at the same time, the random initialization performs as good as the selective initialization after iterations, due to the exploration in Eq. \ref{eq:PoorlyFitted}.

%Note that the performance improvement compare to the random search is not significant. This is because of the small lag of the estimated parameter. In Fig. \ref{fig:DynamicTrueComparison} (a), the estimated parameter capture the dynamic trend 2-3 simulation timesteps lately, and this discrepancy at the front simulation timesteps makes the entire simulation gap irreparable.

\subsubsection{Heterogeneous Calibration Result}
\label{sec:HeterogeneousCalibrationResultsTestCase1}

Fig. \ref{fig:heterogeneous_results} (a), (b), and (c) illustrate the surrogate error functions of Gaussian process regression with 10, 50, and 100 evaluation points, respectively. The surrogate function is gradually sophisticated as the iteration increases. Fig. \ref{fig:heterogeneous_results} (d), (e), and (f) present the contour plot to illustrate the evaluation paths of the suggested Bayesian optimization. The initial simulation result in Fig. \ref{fig:heterogeneous_results} (g) is transformed to be close to the validation data as the best simulation result in Fig. \ref{fig:heterogeneous_results} (h).

\begin{figure}
	\begin{tikzpicture}[node distance=-3cm, auto]  
	\tikzset{
		mynode/.style={rectangle,rounded corners,draw=black, top color=white, bottom color=yellow!50,very thick, inner sep=1em, minimum size=3em, text centered},
		myarrow/.style={->, >=latex', shorten >=1pt, thick},
		mylabel/.style={text width=7em, text centered}
	}
	\node[] (node2) at (0,0)
	{\includegraphics[scale=0.5]{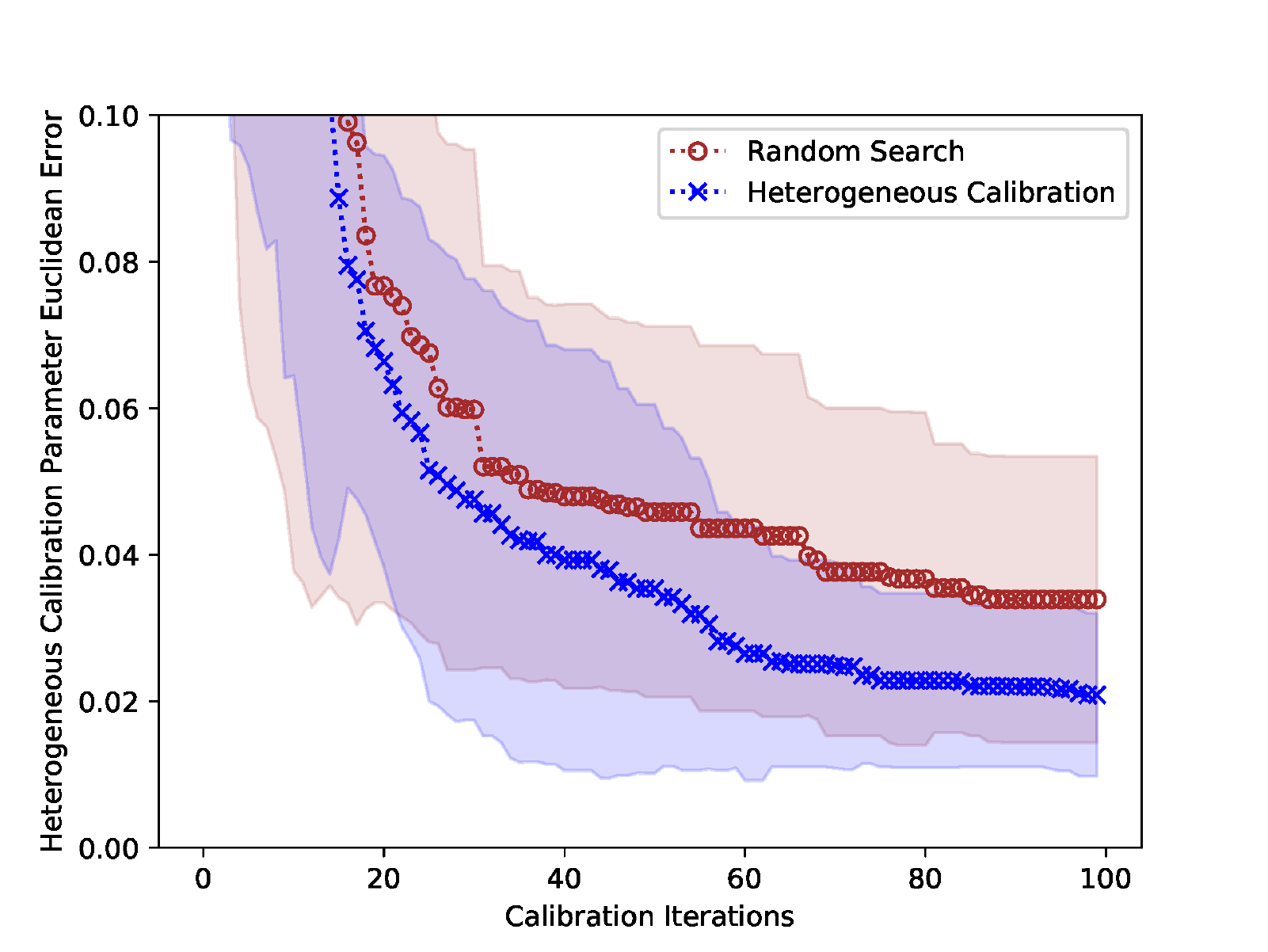}};
	\node[] (node1) at (8,0) {\includegraphics[scale=0.5]{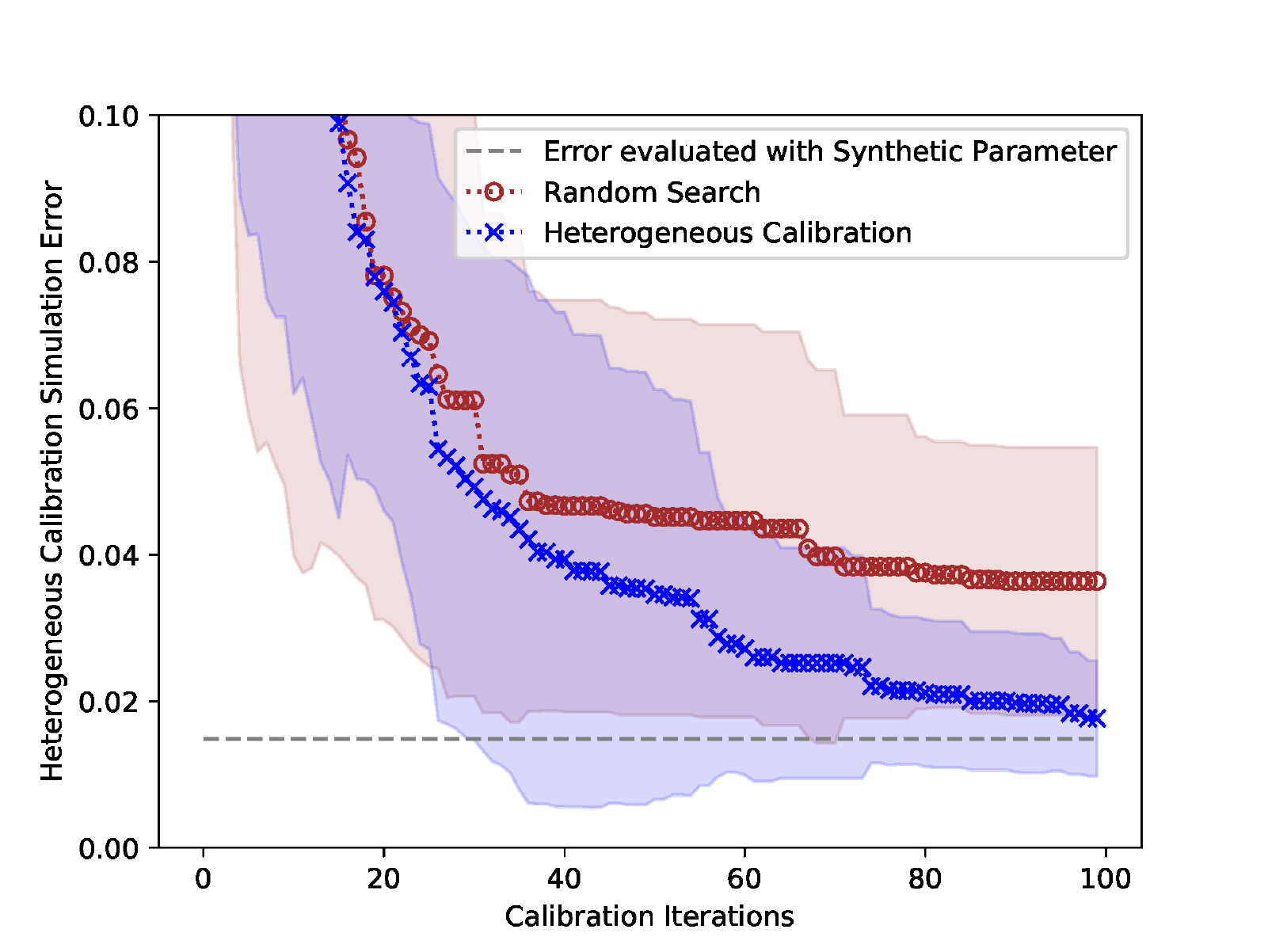}};
	
	\node[] at (-4.0,2.7) {(a)};
	\node[] at (4.0,2.7) {(b)};

	\end{tikzpicture} 
	\caption{(a) The calibrated heterogeneous parameter Euclidean errors are plotted. (b) Hetergeneous calibration simulation MAPE is plotted. Suggested calibration converges to the optimal lower bound, which is the simulation error with the synthetic parameters as input parameters. The optimal lower bound is not 0 since the simulation is stochastic}
	\label{fig:HeterogeneousResultTestCase1}
\end{figure}

Fig. \ref{fig:HeterogeneousResultTestCase1} (a) presents the parameter Euclidean error of heterogeneous calibration. The parameter Euclidean error of heterogeneous calibration saturates at 0.021, due to the observation error. In other words, the stochasticity of the simulation makes the surrogate function biased, and this biasness hinders the Bayesian optimization to estimate the exact synthetic parameter. However, in terms of the simulation error, heterogeneous calibration converges to the optimal error as in Fig. \ref{fig:HeterogeneousResultTestCase1} (b), where the optimal error is the simulation MAPE evaluated with the synthetic parameter.

\subsubsection{Calibration Framework Result}
\label{sec:CalibrationFrameworkResultsTestCase1}

\begin{figure}
	\centering
	\includegraphics[scale=0.8]{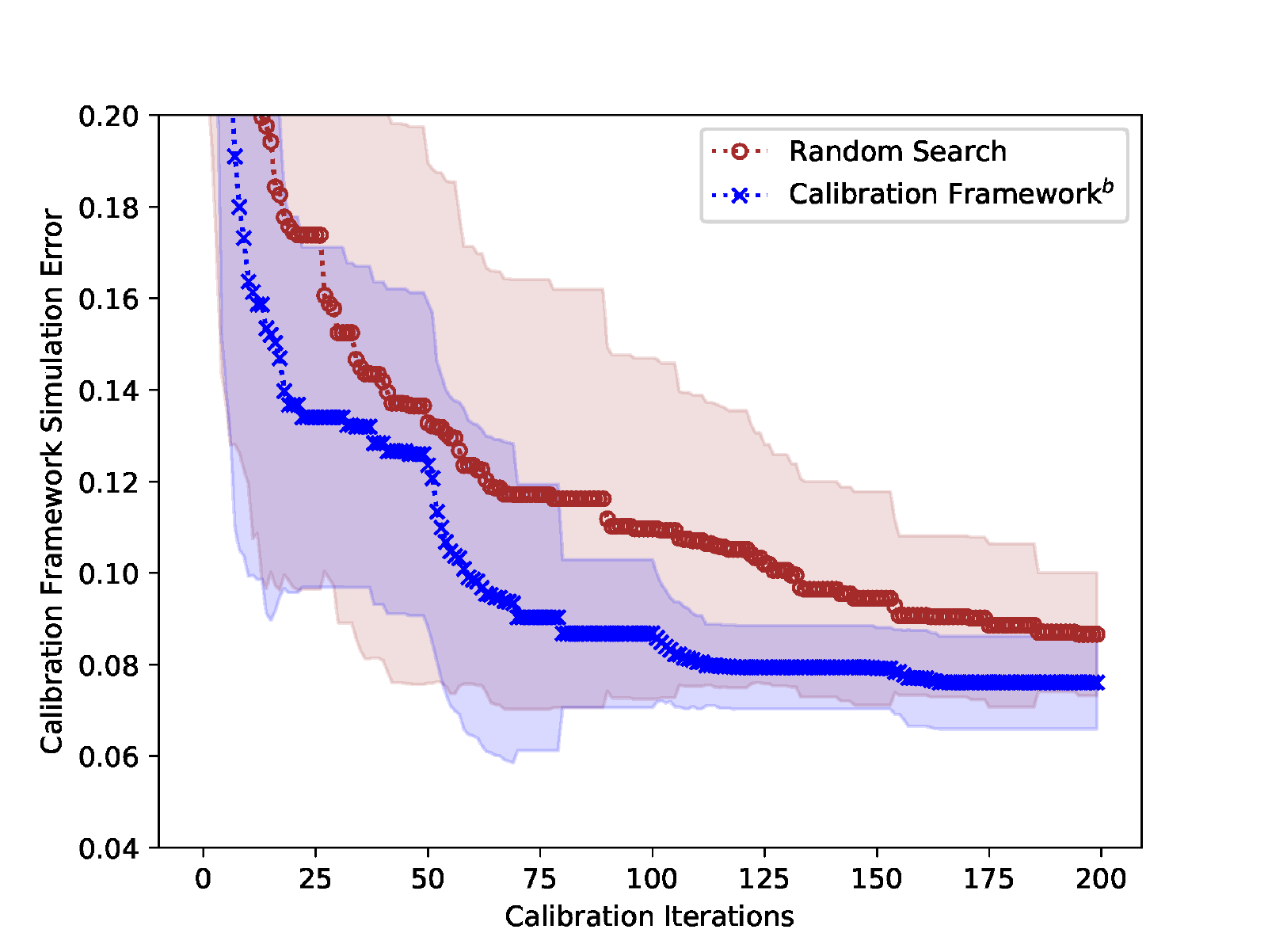}
	\caption{The calibration framework$^{\text{b}}$ simulation MAPE is plotted. The calibration framework$^{\text{b}}$ first updates the dynamic parameter for 20 iterations, with fixed heterogeneous parameter, and next updates the heterogeneous parameter for 30 iterations, with previously estimated optimal dynamic parameter}
	\label{fig:CalibrationFramework}
\end{figure}

Tab. \ref{tab:statisticsinTestCase1} demonstrates the calibration framework$^{\text{b}}$, with $C_{dyn}=20, C_{het}=30$, outperforms the calibration framework$^{\text{a}}$, in terms of the simulation error. Although the calibration framework$^{\text{a}}$ has the least dynamic parameter error among all the experiments in test case 1, the simulation error is large, due to the inaccuracy in the estimation of the heterogeneous parameter. The calibration framework$^{\text{b}}$, on the other hand, performs better than the calibration framework$^{\text{a}}$ in terms of the simulation error, since the calibration framework$^{\text{b}}$ estimates closer heterogeneous parameter than the calibration framework$^{\text{a}}$. Fig. \ref{fig:CalibrationFramework} presents the experimental result of the calibration framework$^{\text{b}}$.

\subsection{Test Case 2: Real Estate Market ABM}
\label{sec:TestCase2}

\subsubsection{Model Description}
\label{sec:ModelDescriptionTestCase2}

\begin{figure}
	\centering
	\includegraphics[width=\textwidth]{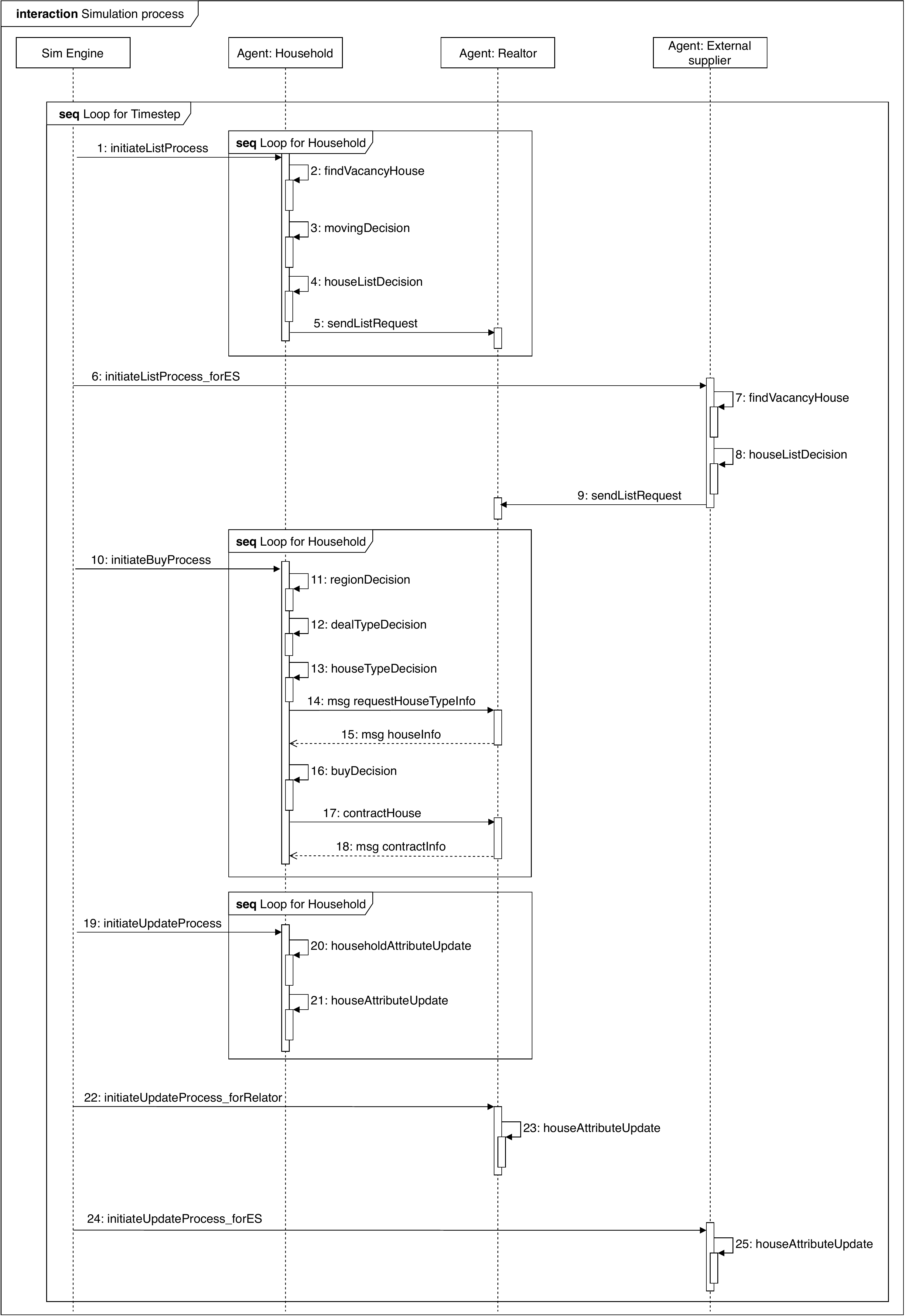}
	\caption{Real Estate Market Model flowchart of a single timestep is plotted. A single simulation timestep consists of the three processes: \textit{List Process}, \textit{Buy Process}, and \textit{Update Process}. Each process has sequential interactions between agents, as presented in the flowchart}
	\label{fig:ABM3}
\end{figure}

Real Estate Market Model is an agent-based model to predict the housing price and the number of house transactions. % analyze the impact of mortgage loan regulation and house supply on the housing market. From a macroscopic point of view, the model purpose is predicting the price and the transaction dynamics. % This model is a realistic model that it is particularly designed to align with the Korean residential system. For example, Jeonse is the Korean specific residential form of living 2 years without monthly payment, by paying a large amount of deposits at the initial contract, which is returned at the end of contract period. The input scenario and the economic variables follow from the real data of South Korea that each agent is initiated according to the real data at the starting time.
The model consists of three types of agents: households, an external supplier and a realtor. Fig. \ref{fig:ABM3} visualizes the interactions between agents. A household buy or sell houses to meet the residential needs. A household registers the owned houses in \textit{List Process} and decides to buy a listed house in \textit{Buy Processes}.

In \textit{List Process}, a household register its vacant houses to the market through a realtor agent. A house is empty if the lease contract\footnote{Lease contract includes Jeonse and rental contract, where Jeonse is the lump-sum housing lease that is Korean specific residential form of living 2 years by paying a large amount of deposits at the initial contract, which is returned at the end of contract period, instead of paying rental fee monthly.} is over, and no other bargain is made. When no household lives in a vacant house, and if a house owner does not move to the house, then a household register its house in a market looking for a leaseholder or a buyer. The external supplier lists its houses automatically in the housing market, since it acts as a house supplier in the model.

In \textit{Buy Process}, a household purchase houses in three steps: participation step, selection step, and contract step. First, a household make a decision to participate in the market. A household without any house to remain unconditionally engage in the housing market, looking for a new house to reside. A household with a house for the residence participates with probability \textit{Market Participation Rate}, to search for a new house to invest. Second, a household in the market selects a house out of the registered houses, according to its preferences. The preference of a household consists of area, contract type, and house type. A household first decides between capital area and noncapital area for a new house. A household, then, decides the contract type between lease and purchase, and the household decides the house type, among a house, an apartment and a condo. Leaseholders, who have remaining contract period, choose always to purchase a house if they make a deal with other house owners. Others, including 1) the house owners living in their house and 2) leaseholders with contract termination, purchase or lease a house with probability \textit{Purchase Rate} or $(1-\textit{Purchase Rate})$. A realtor receives the participated agents' preferences, and the realtor replies the candidate houses to each household with matching preferences. Last, in contract step, a household makes a deal on a house within its budget from a replied house list. If other household contracts the house before, a household chooses other house to contract. The heterogeneous parameter, \textit{Willing to Pay}, controls the budget, which is the maximum ratio out of the total asset an agent would raise, using savings and loan services.

\textit{Update Process} updates all of required simulation state variables. A household saves the rest of its monthly salary after excluding the expenses, such as consumption, taxes, rental fee, and loan repayment. Housing price is decreased to
\begin{equation}
100(1-\textit{Price Decrease Rate}+\text{Inflation Rate})\%    
\end{equation}
up to ten times, if it is listed in the market, but no agents make a contract for the house at the timestep. Housing price is increased to
\begin{equation}
100(1+\textit{Price Increase Rate}+\text{Inflation Rate})\%    
\end{equation}
if other houses with the same conditions are favored at the timestep. After updating state variables, an external supplier supplies new houses in the market. The newly produced houses have the state variables sampled from the real dataset.

\subsubsection{Virtual Experimental Design}
\label{sec:VirtualExperimentTestCase2}

\paragraph{Calibration Parameters}
\label{sec:CalibrationParameters}

Real Estate Market ABM has three dynamic parameters and two heterogeneous parameters: \textit{Market Participation Rate}, \textit{Market Price Increase Rate} and \textit{Market Price Decrease Rate} for the dynamic parameters; and \textit{Willing to Pay} and \textit{Purchase Rate} for the heterogeneous parameters. The economic trends directly influence the supply and the demand in the reality, and the dynamic parameters are selected to reflect this seasonal bull and bear periods. A model uses an external supplier to reflect the supply, using a real dataset, and the model adopts \textit{Market Participation Rate} to represent the demand. The seasonal outcome from the law of supply and demand is modeled by other dynamic parameters, \textit{Market Price Increase Rate} and \textit{Market Price Decrease Rate}. The agent heterogeneity is embedded in the heterogeneous parameters, \textit{Willing to Pay} and \textit{Purchase Rate}, of the investment portfolio. For instance, agents favor in real estate will be represented to have high \textit{Purchase Ratio}.

\begin{table*}
	\begin{center}
		\centering
		\caption{Calibration parameters in the second test case are listed. The dynamic parameters and the heterogeneous parameters are the key control parameters for the model dynamics. The dynamic parameters represents the seasonal effect, or business cycle, and the heterogeneous parameters stand for the agent investment portfolios}
		\label{tab:ListofParametersinTestCase2}
		\begin{tabu}{||l|l|l||}
			\hline
			%			\multicolumn{6}{||c||}{Test Case 1: Wealth Distribution ABM}\\
			%			\hline\hline
			\multicolumn{1}{||p{4cm}|}{Parameter} & \multicolumn{1}{p{3cm}|}{Parameter Type} & \multicolumn{1}{p{3cm}||}{Parameter Range}\\
			\hline\hline
			
			\multicolumn{1}{||p{4cm}|}{\textit{Market Participation Rate}} & \multicolumn{1}{p{2cm}|}{Dynamic} & \multicolumn{1}{p{1.5cm}||}{0-0.05}\\
			\multicolumn{1}{||p{4cm}|}{\textit{Market Price Increase Rate}} & \multicolumn{1}{p{2cm}|}{Dynamic} & \multicolumn{1}{p{1.5cm}||}{0-0.1}\\
			\multicolumn{1}{||p{4cm}|}{\textit{Market Price Decrease Rate}} & \multicolumn{1}{p{2cm}|}{Dynamic} & \multicolumn{1}{p{1.5cm}||}{0-0.1}\\\hline
			\multicolumn{1}{||p{4cm}|}{\textit{Willing to Pay}} & \multicolumn{1}{p{3cm}|}{Heterogeneous} & \multicolumn{1}{p{1.5cm}||}{0.3-0.9}\\
			\multicolumn{1}{||p{4cm}|}{\textit{Purchase Rate}} & \multicolumn{1}{p{3cm}|}{Heterogeneous} & \multicolumn{1}{p{1.5cm}||}{0.3-0.9}\\
			\hline
		\end{tabu}
	\end{center}
\end{table*}

\paragraph{Summary Statistics}
\label{sec:SummaryStatisticsTestCase2}

The main interest in the model is predicting the housing price and the number of housing transactions closer to the validation dataset. Government official data on the price index and the transaction number is released in Korea Appraisal Board (KAB). We count the Jevons index \cite{silver2007elementary} as the housing price index, following the KAB index rule. We scale up the simulation transaction numbers by multipling the ratio of the real population, $2\times 10^{7}$, over the simulation population, $10^{4}$, to adjust the simulation results compatible with the validation transaction data.

KAB releases 24 types summary statistics, with each types are having the following different housing characteristics: two house regions (Capital/Noncapital), three house types (Detached/Apartment/Multiplex), two transaction types (Sales/Lease) and two summary statistics types (Index/Transaction number). We select eight Apartment summary statistics, as listed in Tab. \ref{tab:ListofSummaryStatisticsinTestCase2}, because Apartment forms the $70\%$ of the housing transactions in Korea.

In agent clustering, total seven simulation agent-level state variables $\mathcal{S}_{agent}$ are considered. Agents living in the capital area and the noncapital area are assigned separately in \textsc{Living Region}. \textsc{Savings, Income} and \textsc{Loan} variables are normalized to have values from zero to one. The one-hot encoding is executed for the variables \textsc{House Type} and \textsc{Living Type}, where the one-hot encoding is transforming the integer $i$ into a multi-dimensional zero vector but having one at the $i^{\text{th}}$ element.

\begin{table*}
	\begin{center}
		\centering
		\caption{The validation summary statistics and the agent-level summary statistics in the second test case are listed. The macro-level validation dataset is collected from the Korea Appraisal Board. Apartment typed validation data, apartment price and transaction numbers, are used, since the apartment forms $70\%$ of the housing transactions in Korea. The agent-level summary statistics are the average of the simulation replications}
		\label{tab:ListofSummaryStatisticsinTestCase2}
		\begin{tabu}{||l|l|l|l||}
			\hline
			%			\multicolumn{6}{||c||}{Test Case 1: Wealth Distribution ABM}\\
			%			\hline\hline
			\multicolumn{1}{||p{2cm}|}{Types of Summary Statistics} & \multicolumn{1}{p{7cm}|}{Name of Summary Statistics} & \multicolumn{1}{p{3cm}|}{Variable Description} & \multicolumn{1}{p{3cm}||}{Variable Value}\\
			\hline\hline
			
			\multirow{8}{*}[-1em]{\parbox{2cm}{Validation-level Summary Statistics}} & \multicolumn{1}{p{7cm}|}{\textsc{Apartment Sales Price Index in Capital}} & \multirow{2}{*}{\parbox{3cm}{Jevons price index of Apartment sales price}} & \multirow{4}{*}{\parbox{3cm}{Housing Price is converted into a percentage, with base value as 100 at the initial timestep.}} \\\cline{2-2}
			\multicolumn{1}{||p{1cm}|}{} & \multicolumn{1}{p{7cm}|}{\textsc{Apartment Sales Price Index in Noncapital}} & \multicolumn{1}{p{3cm}|}{} &\\\cline{3-3}\cline{2-2}
			\multicolumn{1}{||p{1cm}|}{} & \multicolumn{1}{p{7cm}|}{\textsc{Apartment Lease Price Index in Capital}} & \multirow{2}{*}{\parbox{3cm}{Jevons price index of Apartment lease price.}} &\\\cline{2-2}
			\multicolumn{1}{||p{1cm}|}{} & \multicolumn{1}{p{7cm}|}{\textsc{Apartment Lease Price Index in Noncapital}} & \multicolumn{1}{p{3cm}|}{} &\\\cline{3-4}\cline{2-2}
			\multicolumn{1}{||p{1cm}|}{} & \multicolumn{1}{p{7cm}|}{\textsc{Apartment Sales Transaction Number in Capital}} & \multirow{2}{*}{\parbox{3cm}{Transaction numbers of Apartment sales.}} & \multirow{4}{*}{\parbox{3cm}{Simulation transaction number is scaled up to be compatible with the validation transaction number.}}\\\cline{2-2}
			\multicolumn{1}{||p{1cm}|}{} & \multicolumn{1}{p{7cm}|}{\textsc{Apartment Sales Transaction Number in Noncapital}} & \multicolumn{1}{p{3cm}|}{} &\\\cline{3-3}\cline{2-2}
			\multicolumn{1}{||p{1cm}|}{} & \multicolumn{1}{p{7cm}|}{\textsc{Apartment Lease Transaction Number in Capital}} & \multirow{2}{*}{\parbox{3cm}{Transaction numbers of Apartment lease.}} &\\\cline{2-2}
			\multicolumn{1}{||p{1cm}|}{} & \multicolumn{1}{p{7cm}|}{\textsc{Apartment Lease Transaction Number in Noncapital}} & \multicolumn{1}{p{3cm}|}{} &\\\cline{2-2}
			\hline
			\multirow{7}{*}[2.3em]{\parbox{2cm}{Agent-level Summary Statistics}} & \multicolumn{1}{p{7cm}|}{\textsc{Living Region}} & \multicolumn{1}{p{3cm}|}{Agent living region between capital/noncapital area.} & \multicolumn{1}{p{3cm}||}{1: Capital, 0: Noncapital}\\\cline{2-4}
			\multicolumn{1}{||p{2cm}|}{} & \multicolumn{1}{p{7cm}|}{\textsc{Savings}} & \multicolumn{1}{p{3cm}|}{Total savings.} & \multicolumn{1}{p{3cm}||}{1 unit/1000 KRW}\\\cline{2-4}
			\multicolumn{1}{||p{2cm}|}{} & \multicolumn{1}{p{7cm}|}{\textsc{Income}} & \multicolumn{1}{p{3cm}|}{Sum of the labor income and transfer income.} & \multicolumn{1}{p{3cm}||}{1 unit/1000 KRW}\\\cline{2-4}
			\multicolumn{1}{||p{2cm}|}{} & \multicolumn{1}{p{7cm}|}{\textsc{Loan}} & \multicolumn{1}{p{3cm}|}{Total amount of money agent have borrowed from bank.} & \multicolumn{1}{p{3cm}||}{1 unit/1000 KRW}\\\cline{2-4}
			\multicolumn{1}{||p{2cm}|}{} & \multicolumn{1}{p{7cm}|}{\textsc{House Type}} & \multicolumn{1}{p{3cm}|}{Type of house where an agent lives.} & \multicolumn{1}{p{3cm}||}{1: Detached House, 2: Apartment, 3: Multiplex House, 4: No House}\\\cline{2-4}
			\multicolumn{1}{||p{2cm}|}{} & \multicolumn{1}{p{7cm}|}{\textsc{Living Type}} & \multicolumn{1}{p{3cm}|}{Type of living where an agent lives} & \multicolumn{1}{p{3cm}||}{1: Owner, 2: Lease, 3: No House}\\\cline{2-4}
			\multicolumn{1}{||p{2cm}|}{} & \multicolumn{1}{p{7cm}|}{\textsc{Number of Own Houses}} & \multicolumn{1}{p{3cm}|}{Number of houses agent owns} & \multicolumn{1}{p{3cm}||}{1 unit/1 House} \\\cline{2-4}
			\hline
		\end{tabu}
	\end{center}
\end{table*}

\paragraph{Experimental Cases}
\label{sec:ExperimentalCasesTestCase2}

There are three experimental cases: dynamic calibration, heterogeneous calibration, and the combined calibration framework. Dynamic calibration experiments three subcases: the first subcase calibrates the first parameter, the next subcase calibrates the first and the second parameter, and the last subcase calibrates all the dynamic parameters. Dynamic calibration accepts the manually calibrated heterogeneous parameter, in the calibration process. Heterogeneous calibration evaluates 15 subcases: first five subcases calibrate the first parameter, next five subcases calibrate the second parameter, and the last five subcases calibrate all heterogeneous parameters. Each of five subcases tests the different number of clusters: the first subcase calibrates the undivided parameters, the next four subcases calibrate the divided parameters, with first three subcases for the parametric clustering, and the last subcase for the nonparametric clustering. Heterogeneous parameter uses the manually human calibrated dynamic parameter, in the calibration process. The calibration framwork, combining dynamic and heterogeneous calibrations, investigates the effects of the combination of each calibration. The framework iteratively calibrates both the dynamic and the heterogeneous parameters, as in Tab. \ref{tab:ListofExperimentalVariablesinTestCase2}.

\begin{table*}
	\begin{center}
		\centering
		\caption{Experimental variables of each experiment in the second test case are listed. Dynamic calibration calibrates the dynamic parameters, with fixed human calibrated heterogeneous parameters. Heterogeneous calibration calibrates the heterogeneous parameters, with fixed human calibrated dynamic parameters. The calibration framework calibrates all the dynamic and the heterogeneous parameters, with two subcases as in the Table}
		\label{tab:ListofExperimentalVariablesinTestCase2}
		\begin{tabu}{||l|c|c|c|c|c|c|c|c|c||}
			\hline
			\diagbox{Experiments}{Variable} & $C_{cal}$ & $C_{dyn}$ & $C_{het}$ & $K_{dyn}$  & $K_{het}$ & $A$ & $T$ & $R$ & $I$\\\hline\hline
			Dynamic Calibration & 100 & 1 & 0 & 3 & 1 & 10000 & 24 & 10 & 3\\\hline
			Heterogeneous Calibration & 100 & 0 & 1 & 3 & \multicolumn{1}{c|}{\makecell[c]{1,2,4,8,\\ nonparametric}} & 10000 & 24 & 10 & 1\\\hline
			\multirow{2}{*}{Calibration Framework} & 200 & 2 & 3 & 3 & 2, nonparametric & 10000 & 24 & 10 & 3\\\cline{2-10}
			& 200 & 20 & 30 & 3 & 2, nonparametric & 10000 & 24 & 10 & 3\\
			
			\hline
		\end{tabu}
	\end{center}
\end{table*}

\begin{sidewaystable}
\resizebox{\textwidth}{!}{
	\begin{threeparttable}
		\caption{Performance evaluations of the proposed calibration models in test case 2 are listed. Numbers are the mean and the standard deviation of experiments by replicating 30 times. The boldface indicates the smallest error in each experimental case. One tailed Welch's t-tests on the suggested calibration methodologies are implemented with the baseline human calibration results.}
		\label{tab:statisticsinTestCase2}

			\begin{tabular}{||l|l|cc|ccccccccc||}
				%\toprule
				\hline
				
				\multicolumn{13}{||c||}{Test Case 2: Real Estate Market ABM} \\
				
				\hline
				
				\multirow{1}{*}{\parbox{2cm}{Experiments}} &
				\multirow{1}{*}{\parbox{2cm}{Calibration Parameters}} &
				\multicolumn{2}{c|}{\multirow{1}{*}{Number of Clusters}} &
				\multicolumn{1}{p{2cm}|}{\textsc{Apartment Sales Price Index in Capital} MAPE} &
				\multicolumn{1}{p{2cm}|}{\textsc{Apartment Lease Price Index in Capital} MAPE} &
				\multicolumn{1}{p{2cm}|}{\textsc{Apartment Sales Price Index in Noncapital} MAPE} &
				\multicolumn{1}{p{2cm}|}{\textsc{Apartment Lease Price Index in Noncapital} MAPE} &
				\multicolumn{1}{p{2cm}|}{\textsc{Apartment Sales Transaction Number in Capital} MAPE} &
				\multicolumn{1}{p{2cm}|}{\textsc{Apartment Lease Transaction Number in Capital} MAPE} &
				\multicolumn{1}{p{2cm}|}{\textsc{Apartment Sales Transaction Number in Noncapital} MAPE} &
				\multicolumn{1}{p{2cm}|}{\textsc{Apartment Lease Transaction Number in Noncapital} MAPE} &
				\multirow{1}{*}{\parbox{2cm}{Total MAPE}}\\
				
				\hline\hline
				
				\multirow{2}{*}{\parbox{2cm}{Human Calibration}} &
				\multirow{2}{*}{\parbox{2cm}{All Parameters}} &
				\multicolumn{2}{c|}{\multirow{2}{*}{---}} &
				0.012 & 0.020 & 0.007 & 0.003 & 0.765 & 0.159 & 0.284 & 0.463 & 0.214\\
				
				&&&& ($\pm$0.001) & ($\pm$0.001) & ($\pm$0.000) & ($\pm$0.000) & ($\pm$0.027) & ($\pm$0.012) & ($\pm$0.016) & ($\pm$0.017) & ($\pm$0.005)\\\hline
				
				\multirow{2}{*}{\parbox{2cm}{Random Search}} &
				\multirow{2}{*}{\parbox{2cm}{All Parameters}} &
				\multicolumn{2}{c|}{\multirow{2}{*}{---}}&
				0.025 &
				0.011 &
				0.009 &
				0.008 &
				0.554 &
				0.291 &
				0.338 &
				0.397 &
				0.204\\
				
				&&&&
				($\pm$0.010) &
				($\pm$0.006) &
				($\pm$0.005) &
				($\pm$0.004) &
				($\pm$0.181) &
				($\pm$0.085) &
				($\pm$0.105) &
				($\pm$0.127) &
				($\pm$0.008)\\

				\hline\hline

				\multirow{6}{*}{\parbox{2cm}{Dynamic Calibration}} &
				\multirow{2}{*}{\parbox{2cm}{First Parameter}} &
				\multicolumn{2}{c|}{\multirow{2}{*}{---}} &
				\textbf{0.005} &
				0.025 &
				\textbf{0.003} &
				\textbf{0.005} &
				\textbf{0.266} &
				0.277 &
				0.343 &
				\textbf{0.409} &
				\textbf{0.167}\tnote{*}\\
				
				&
				&
				&
				&
				($\pm$0.001) &
				($\pm$0.002) &
				($\pm$0.000) &
				($\pm$0.001) &
				($\pm$0.015) &
				($\pm$0.012) &
				($\pm$0.016) &
				($\pm$0.018) &
				($\pm$0.003)
				\\
				
				&
				\multirow{2}{*}{\parbox{2cm}{First and Second Parameters}} &
				\multicolumn{2}{c|}{\multirow{2}{*}{---}} &
				0.009 &
				0.024 &
				0.005 &
				0.006 &
				0.273 &
				0.280 &
				0.337 &
				0.410 &
				0.168\tnote{*}\\
				
				&
				&
				&
				&
				($\pm$0.005) &
				($\pm$0.008) &
				($\pm$0.003) &
				($\pm$0.002) &
				($\pm$0.034) &
				($\pm$0.018) &
				($\pm$0.023) &
				($\pm$0.018) &
				($\pm$0.003)
				\\
				
				&
				\multirow{2}{*}{\parbox{2cm}{All Dynamic Parameters}} &
				\multicolumn{2}{c|}{\multirow{2}{*}{---}} &
				0.015 &
				\textbf{0.020} &
				0.008 &
				0.012 &
				0.281 &
				\textbf{0.272} &
				\textbf{0.335} &
				0.452 &
				0.174\tnote{*}\\
				
				&
				&
				&
				&
				($\pm$0.009) &
				($\pm$0.010) &
				($\pm$0.003) &
				($\pm$0.004) &
				($\pm$0.027) &
				($\pm$0.018) &
				($\pm$0.017) &
				($\pm$0.029) &
				($\pm$0.004)
				\\
				
				\hline\hline
				
				\multirow{30}{*}{\parbox{2cm}{Heterogeneous Calibration}} &
				\multirow{10}{*}{\parbox{2cm}{First Parameter}} &
				\multirow{2}{*}{\parbox{2cm}{Undivided}} &
				\multirow{2}{*}{1} &
				0.012 &
				\textbf{0.020} &
				0.007 &
				\textbf{0.003} &
				0.667 &
				\textbf{0.142} &
				0.279 &
				0.469 &
				0.200\tnote{*}\\
				
				&
				&
				&
				&
				($\pm$0.001) &
				($\pm$0.001) &
				($\pm$0.001) &
				($\pm$0.000) &
				($\pm$0.080) &
				($\pm$0.018) &
				($\pm$0.056) &
				($\pm$0.066) &
				($\pm$0.004)
				\\\cline{3-3}
				
				&
				&
				\multirow{6}{*}{\parbox{2cm}{Parametric}} &
				\multirow{2}{*}{2} &
				0.006 &
				0.027 &
				\textbf{0.004} &
				0.006 &
				\textbf{0.259} &
				0.255 &
				\textbf{0.166} &
				\textbf{0.150} &
				0.109\tnote{*}\\
				
				&
				&
				&
				&
				($\pm$0.002) &
				($\pm$0.004) &
				($\pm$0.001) &
				($\pm$0.002) &
				($\pm$0.012) &
				($\pm$0.010) &
				($\pm$0.011) &
				($\pm$0.020) &
				($\pm$0.003)
				\\
				
				&
				&
				&
				\multirow{2}{*}{4} &
				0.006 &
				0.026 &
				0.004 &
				0.005 &
				0.266 &
				0.244 &
				0.202 &
				0.161 &
				0.114\tnote{*}\\
				
				&
				&
				&
				&
				($\pm$0.002) &
				($\pm$0.002) &
				($\pm$0.001) &
				($\pm$0.001) &
				($\pm$0.030) &
				($\pm$0.005) &
				($\pm$0.020) &
				($\pm$0.015) &
				($\pm$0.002)
				\\
				
				&
				&
				&
				\multirow{2}{*}{8} &
				\textbf{0.004} &
				0.029 &
				0.004 &
				0.006 &
				0.263 &
				0.164 &
				0.178 &
				0.208 &
				\textbf{0.107}\tnote{*}\\
				
				&
				&
				&
				&
				($\pm$0.003) &
				($\pm$0.007) &
				($\pm$0.003) &
				($\pm$0.003) &
				($\pm$0.099) &
				($\pm$0.044) &
				($\pm$0.057) &
				($\pm$0.051) &
				($\pm$0.013)
				\\\cline{3-3}
				
				&
				&
				\multirow{2}{*}{\parbox{2cm}{Nonparametric}} &
				15.3 &
				0.005 &
				0.026 &
				0.004 &
				0.005 &
				0.265 &
				0.164 &
				0.357 &
				0.177 &
				0.125\tnote{*}\\
				
				&
				&
				&
				($\pm$1.4) &
				($\pm$0.000) &
				($\pm$0.000) &
				($\pm$0.000) &
				($\pm$0.000) &
				($\pm$0.181) &
				($\pm$0.129) &
				($\pm$0.064) &
				($\pm$0.146) &
				($\pm$0.020)
				\\
				
				\cline{2-13}
				
				&
				\multirow{10}{*}{\parbox{2cm}{Second Parameter}} &
				\multirow{2}{*}{\parbox{2cm}{Undivided}} &
				\multirow{2}{*}{1} &
				0.012 &
				0.020 &
				0.007 &
				\textbf{0.003} &
				0.764 &
				0.218 &
				0.249 &
				0.360 &
				0.204\tnote{*}\\
				
				&
				&
				&
				&
				($\pm$0.001) &
				($\pm$0.000) &
				($\pm$0.000) &
				($\pm$0.000) &
				($\pm$0.039) &
				($\pm$0.033) &
				($\pm$0.021) &
				($\pm$0.057) &
				($\pm$0.004)
				\\\cline{3-3}
				
				&
				&
				\multirow{6}{*}{\parbox{2cm}{Parametric}}&
				\multirow{2}{*}{2} &
				0.013 &
				\textbf{0.019} &
				0.007 &
				0.003 &
				0.835 &
				\textbf{0.104} &
				\textbf{0.184} &
				0.256 &
				0.178\tnote{*}\\
				
				&
				&
				&
				&
				($\pm$0.001) &
				($\pm$0.000) &
				($\pm$0.000) &
				($\pm$0.000) &
				($\pm$0.038) &
				($\pm$0.029) &
				($\pm$0.020) &
				($\pm$0.044) &
				($\pm$0.009)
				\\
				
				&
				&
				&
				\multirow{2}{*}{4} &
				0.013 &
				0.020 &
				0.007 &
				0.003 &
				0.827 &
				0.137 &
				0.201 &
				0.255 &
				0.183\tnote{*}\\
				
				&
				&
				&
				&
				($\pm$0.003) &
				($\pm$0.002) &
				($\pm$0.001) &
				($\pm$0.001) &
				($\pm$0.099) &
				($\pm$0.164) &
				($\pm$0.038) &
				($\pm$0.067) &
				($\pm$0.023)
				\\
				
				&
				&
				&
				\multirow{2}{*}{8} &
				\textbf{0.011} &
				0.021 &
				0.006 &
				0.003 &
				0.567 &
				0.164 &
				0.185 &
				\textbf{0.205} &
				0.145\tnote{*}\\
				
				&
				&
				&
				&
				($\pm$0.006) &
				($\pm$0.015) &
				($\pm$0.008) &
				($\pm$0.008) &
				($\pm$0.097) &
				($\pm$0.113) &
				($\pm$0.041) &
				($\pm$0.080) &
				($\pm$0.013)
				\\\cline{3-3}
				
				&
				&
				\multirow{2}{*}{\parbox{2cm}{Nonparametric}} &
				15.9 &
				0.012 &
				0.020 &
				\textbf{0.006} &
				0.003 &
				\textbf{0.414} &
				0.144 &
				0.192 &
				0.211 &
				\textbf{0.125}\tnote{*}\\
				
				&
				&
				&
				($\pm$1.2) &
				($\pm$0.001) &
				($\pm$0.000) &
				($\pm$0.000) &
				($\pm$0.000) &
				($\pm$0.102) &
				($\pm$0.085) &
				($\pm$0.038) &
				($\pm$0.058) &
				($\pm$0.021)
				\\
				
				\cline{2-13}
				
				&
				\multirow{10}{*}{\parbox{2cm}{All Heterogeneous Parameters}} &
				\multirow{2}{*}{\parbox{2cm}{Undivided}} &
				\multirow{2}{*}{1} &
				0.008 &
				\textbf{0.012} &
				\textbf{0.004} &
				0.004 &
				0.398 &
				0.255 &
				0.424 &
				0.461 &
				0.196\tnote{*}\\
				
				&
				&
				&
				&
				($\pm$0.002) &
				($\pm$0.003) &
				($\pm$0.001) &
				($\pm$0.002) &
				($\pm$0.126) &
				($\pm$0.078) &
				($\pm$0.072) &
				($\pm$0.098) &
				($\pm$0.006)
				\\\cline{3-3}
				
				&
				&
				\multirow{6}{*}{\parbox{2cm}{Parametric}}&
				\multirow{2}{*}{2} &
				0.011 &
				0.021 &
				0.006 &
				\textbf{0.003} &
				0.303 &
				\textbf{0.127} &
				0.200 &
				\textbf{0.179} &
				\textbf{0.106}\tnote{*}\\
				
				&
				&
				&
				&
				($\pm$0.001) &
				($\pm$0.001) &
				($\pm$0.000) &
				($\pm$0.000) &
				($\pm$0.092) &
				($\pm$0.028) &
				($\pm$0.028) &
				($\pm$0.054) &
				($\pm$0.015)
				\\
				
				&
				&
				&
				\multirow{2}{*}{4} &
				0.009 &
				0.023 &
				0.005 &
				0.004 &
				0.289 &
				0.241 &
				\textbf{0.168} &
				0.181 &
				0.115\tnote{*}\\
				
				&
				&
				&
				&
				($\pm$0.003) &
				($\pm$0.002) &
				($\pm$0.001) &
				($\pm$0.001) &
				($\pm$0.054) &
				($\pm$0.106) &
				($\pm$0.039) &
				($\pm$0.077) &
				($\pm$0.017)
				\\
				
				&
				&
				&
				\multirow{2}{*}{8} &
				\textbf{0.007} &
				0.025 &
				0.004 &
				0.005 &
				0.243 &
				0.180 &
				0.218 &
				0.230 &
				0.114\tnote{*}\\
				
				&
				&
				&
				&
				($\pm$0.006) &
				($\pm$0.015) &
				($\pm$0.008) &
				($\pm$0.008) &
				($\pm$0.097) &
				($\pm$0.113) &
				($\pm$0.041) &
				($\pm$0.080) &
				($\pm$0.013)
				\\\cline{3-3}
				
				&
				&
				\multirow{2}{*}{\parbox{2cm}{Nonparametric}} &
				15.5 &
				0.011 &
				0.021 &
				0.007 &
				0.003 &
				\textbf{0.232} &
				0.199 &
				0.198 &
				0.186 &
				0.107\tnote{*}\\
				
				&
				&
				&
				($\pm$1.2) &
				($\pm$0.002) &
				($\pm$0.001) &
				($\pm$0.000) &
				($\pm$0.000) &
				($\pm$0.198) &
				($\pm$0.151) &
				($\pm$0.065) &
				($\pm$0.176) &
				($\pm$0.030)
				\\
				
				\hline\hline
								
				\multirow{4}{*}{\parbox{2cm}{Calibration Framework\tnote{a}}}&
				\multirow{4}{*}{\parbox{2cm}{All Model Parameters}} &
				\multirow{2}{*}{\parbox{2cm}{Parametric}} &
				\multirow{2}{*}{2} &
				0.028 &
				\textbf{0.016} &
				0.020 &
				0.021 &
				0.286 &
				\textbf{0.178} &
				0.232 &
				0.202 &
				0.123\tnote{*}\\
				
				&
				&
				&
				&
				($\pm$0.012) &
				($\pm$0.015) &
				($\pm$0.009) &
				($\pm$0.013) &
				($\pm$0.073) &
				($\pm$0.067) &
				($\pm$0.055) &
				($\pm$0.048) &
				($\pm$0.016)
				\\\cline{3-3}
				
				&
				&
				\multirow{2}{*}{\parbox{2cm}{Nonparametric}} &
				15.6&
				0.031 &
				0.022 &
				0.019 &
				0.018 &
				0.324 &
				0.183 &
				0.215 &
				0.264 &
				0.135\tnote{*}\\
				
				&
				&
				&
				($\pm$1.3)&
				($\pm$0.018) &
				($\pm$0.016) &
				($\pm$0.015) &
				($\pm$0.014) &
				($\pm$0.108) &
				($\pm$0.033) &
				($\pm$0.063) &
				($\pm$0.116) &
				($\pm$0.019)\\\hline
				
				\multirow{4}{*}{\parbox{2cm}{Calibration Framework\tnote{b}}}&
				\multirow{4}{*}{\parbox{2cm}{All Model Parameters}}&
				\multirow{2}{*}{\parbox{2cm}{Parametric}} &
				\multirow{2}{*}{2} &
				\textbf{0.016} &
				0.016 &
				\textbf{0.005} &
				\textbf{0.004} &
				\textbf{0.219} &
				0.195 &
				0.252 &
				\textbf{0.136} &
				\textbf{0.105}\tnote{*}\\
				
				&
				&
				&
				&
				($\pm$0.001) &
				($\pm$0.001) &
				($\pm$0.003) &
				($\pm$0.001) &
				($\pm$0.007) &
				($\pm$0.018) &
				($\pm$0.049) &
				($\pm$0.000) &
				($\pm$0.002)
				\\\cline{3-3}
				
				&
				&
				\multirow{2}{*}{\parbox{2cm}{Nonparametric}} &
				15.4&
				0.024 &
				0.021 &
				0.009 &
				0.011 &
				0.320 &
				0.197 &
				\textbf{0.195} &
				0.265 &
				0.130\tnote{*}\\
				
				&
				&
				&
				($\pm$1.1)&
				($\pm$0.023) &
				($\pm$0.012) &
				($\pm$0.006) &
				($\pm$0.005) &
				($\pm$0.084) &
				($\pm$0.031) &
				($\pm$0.004) &
				($\pm$0.083) &
				($\pm$0.001)\\
				
				\hline
				%\bottomrule
			\end{tabular}
%		}
		\begin{tablenotes}
			\item[*] $P<0.05$
			\item[a] Experiment with $C_{dyn}=2$, and $C_{het}=3$
			\item[b] Experiment with $C_{dyn}=20$, and $C_{het}=30$
		\end{tablenotes}
	\end{threeparttable}
}
	%\end{table*}
\end{sidewaystable}

\subsubsection{Dynamic Calibration Result}
\label{sec:DynamicCalibrationResultTestCase2}

The simulation price indices are more stable and better aligned with the validation dataset than the alignment of the simulation transaction numbers. Therefore, most of the error improvement is achieved by aligning the simulation transaction numbers to the validation, which heavily depends on the first dynamic parameter, \textit{Market Participation Rate}. Comparing with the human calibration, Tab. \ref{tab:statisticsinTestCase2} shows a significant error reduction in \textsc{Total MAPE}, which is gained from dynamic calibration experiments. 

The second experimental subcase calibrates the first and the second dynamic parameters, where the second dynamic parameter influences to the housing prices. Comparing with the human calibration, Tab. \ref{tab:statisticsinTestCase2} demonstrates that the analogous performance, with respect to the housing price indices, is achieved by counting the second parameter into the calibration parameter. %Given that the initial random setting, This is remarkable, since the human calibration already recovers the validation housing price indices accurately.

The last experimental subcase, calibrating all three dynamic parameters, provides an insight when parameters with similar effects, the second and the third parameters, are calibrated. The price indices are less correct than that of the second subcase. Also, \textsc{total MAPE} is slightly worse than the second experiment, which is mainly due to \textsc{Apartment Lease Transaction Number in Noncaptial MAPE}. The transaction numbers, relatively irrelavant with the second and the third dynamic parameters, are disturbed by adding the third parameter to calibrate, since the problem complexity is increased, without attaining any supplementary power to control the model output.

\subsubsection{Heterogeneous Calibration Result}
\label{sec:HeterogeneousCalibrationResultTestCase2}

\begin{figure*}
	\begin{tikzpicture}[node distance=-3cm, auto]  
	\tikzset{
		mynode/.style={rectangle,rounded corners,draw=black, top color=white, bottom color=yellow!50,very thick, inner sep=1em, minimum size=3em, text centered},
		myarrow/.style={->, >=latex', shorten >=1pt, thick},
		mylabel/.style={text width=7em, text centered}
	}  
	\node[] (node2) at (0,0) {\includegraphics[scale=0.5]{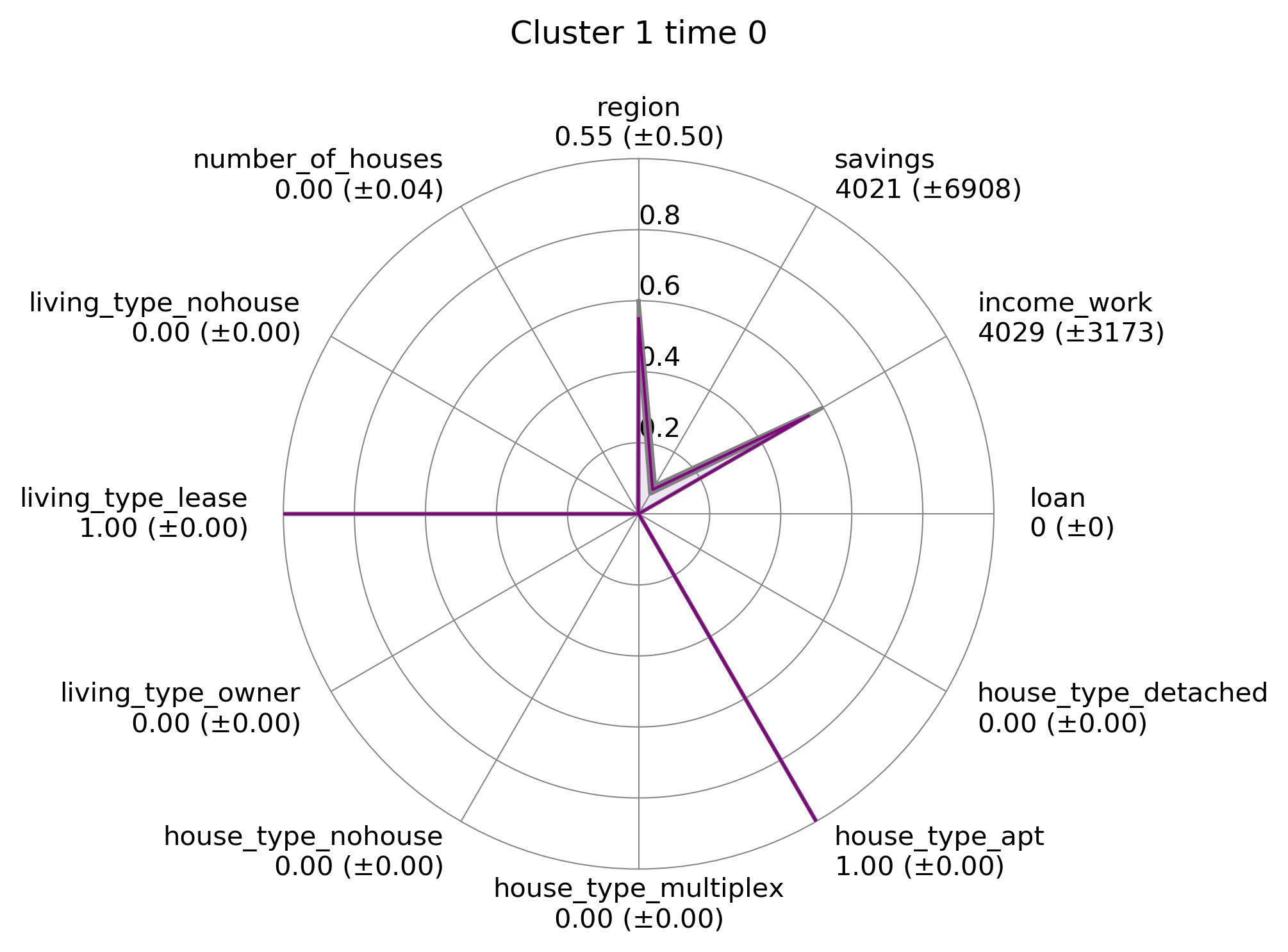}};  
	\node[] (node4) at (8,0) {\includegraphics[scale=0.5]{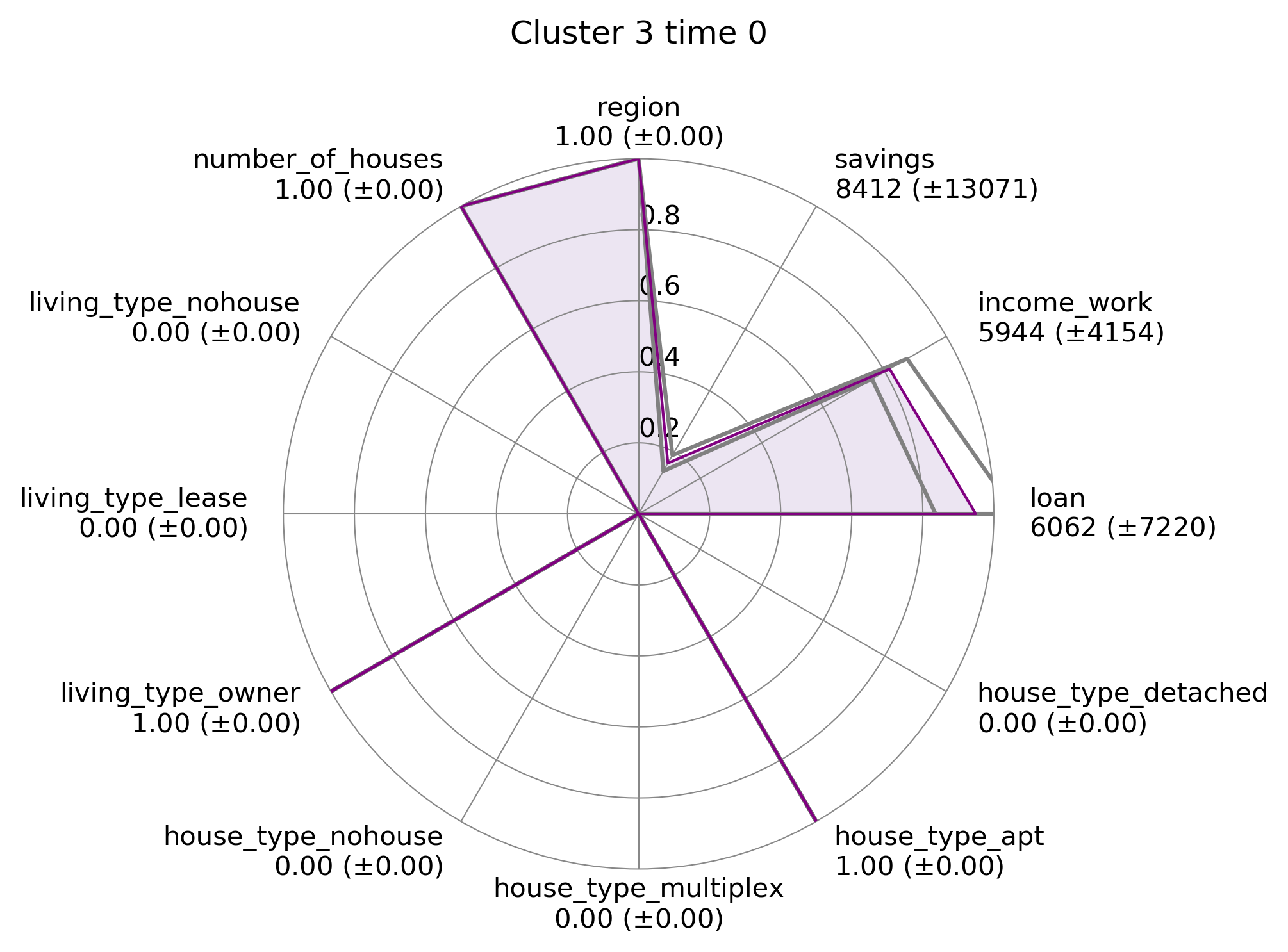}};  
	
%	\node[] at (-4,-3) {\makecell{Calibrated\\Heterogeneous\\Parameter}};
%	\node[] at (0,-3) {\makecell{Cluster 1\\(0.9,0.9)}};
%	\node[] at (8,-3) {\makecell{Cluster 3\\(0.3,0.9)}};
	\node[] at (-4.0,2.7) {(a)};
	\node[] at (4.0,2.7) {(b)};
	\end{tikzpicture} 
	\medskip
	\caption{The clustered agent-level simulation results are plotted in the radar charts. The first cluster is the group of leaseholders with middle-class salary, no debt, and low savings, who live in apartment. The third cluster is the group of apartment-owners in capital area with high-class salary, high debt, and low savings.}
	\label{fig:Cluster_Statistics} 
	%\end{sidewaysfigure*}
\end{figure*}

\begin{figure}
	\centering
	\includegraphics[scale=0.8]{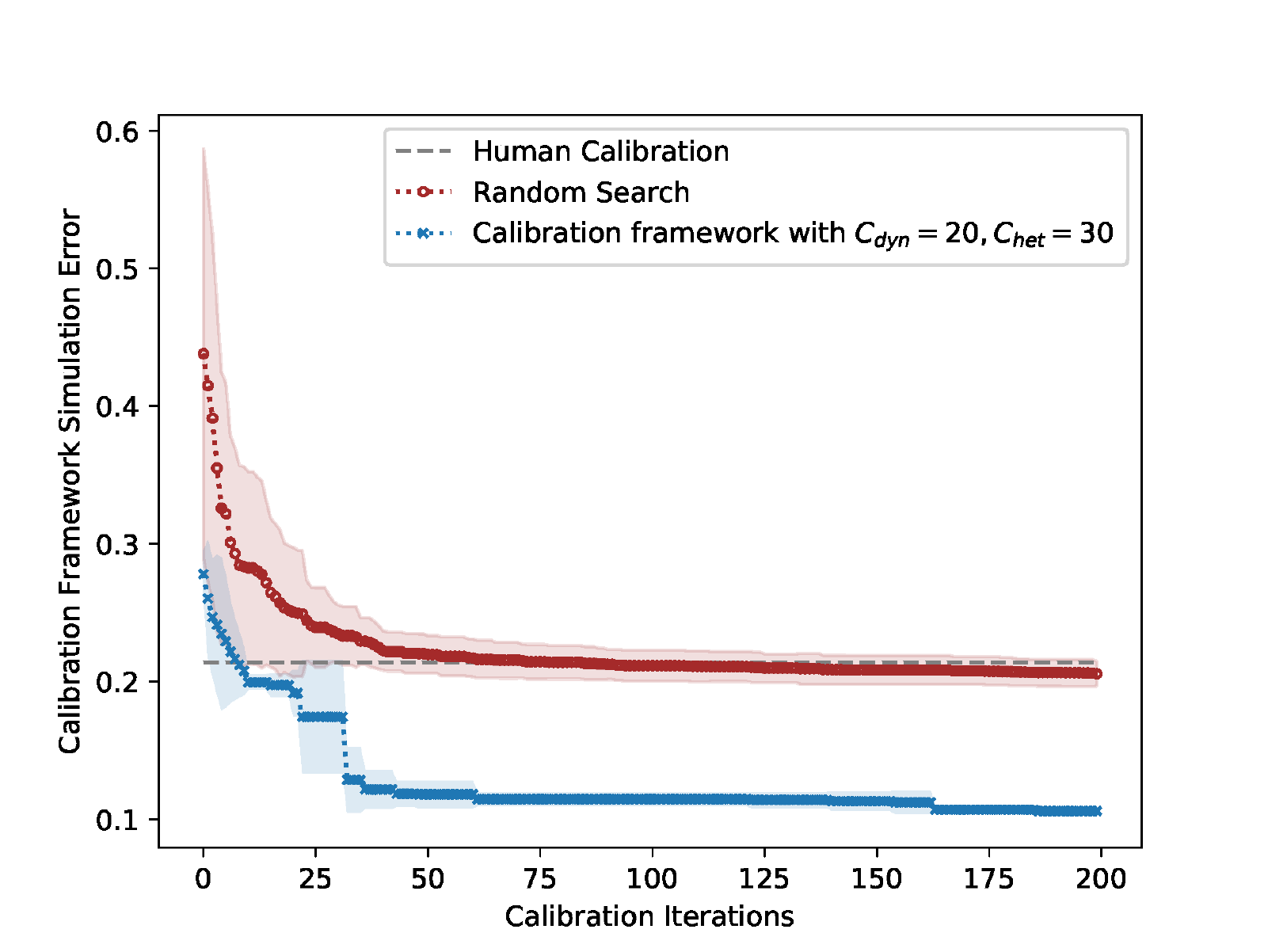}
	\caption{The calibration framework$^{\text{b}}$ simulation MAPE is plotted in the x marked line. The calibration framework$^{\text{b}}$ first calibrates all three dynamic parameters for 20 iterations, with fixed heterogeneous parameters, and next calibrates all two heterogeneous parameters for 30 iterations, with perviously calibrated optimal dynamic parameters. The random search reaches to the human calibration error after 80 iterations, and saturates on the afterward iterations. The calibration framework achieves the human calibration error in short iterations, and saturates to nearly half of the human calibration error}
	\label{fig:Combined_Calibration}
\end{figure}

\begin{figure*}
	\begin{tikzpicture}[node distance=-3cm, auto]  
	\tikzset{
		mynode/.style={rectangle,rounded corners,draw=black, top color=white, bottom color=yellow!50,very thick, inner sep=1em, minimum size=3em, text centered},
		myarrow/.style={->, >=latex', shorten >=1pt, thick},
		mylabel/.style={text width=7em, text centered}
	}  
	\node[] (node1) at (0,0) {\includegraphics[scale=0.5]{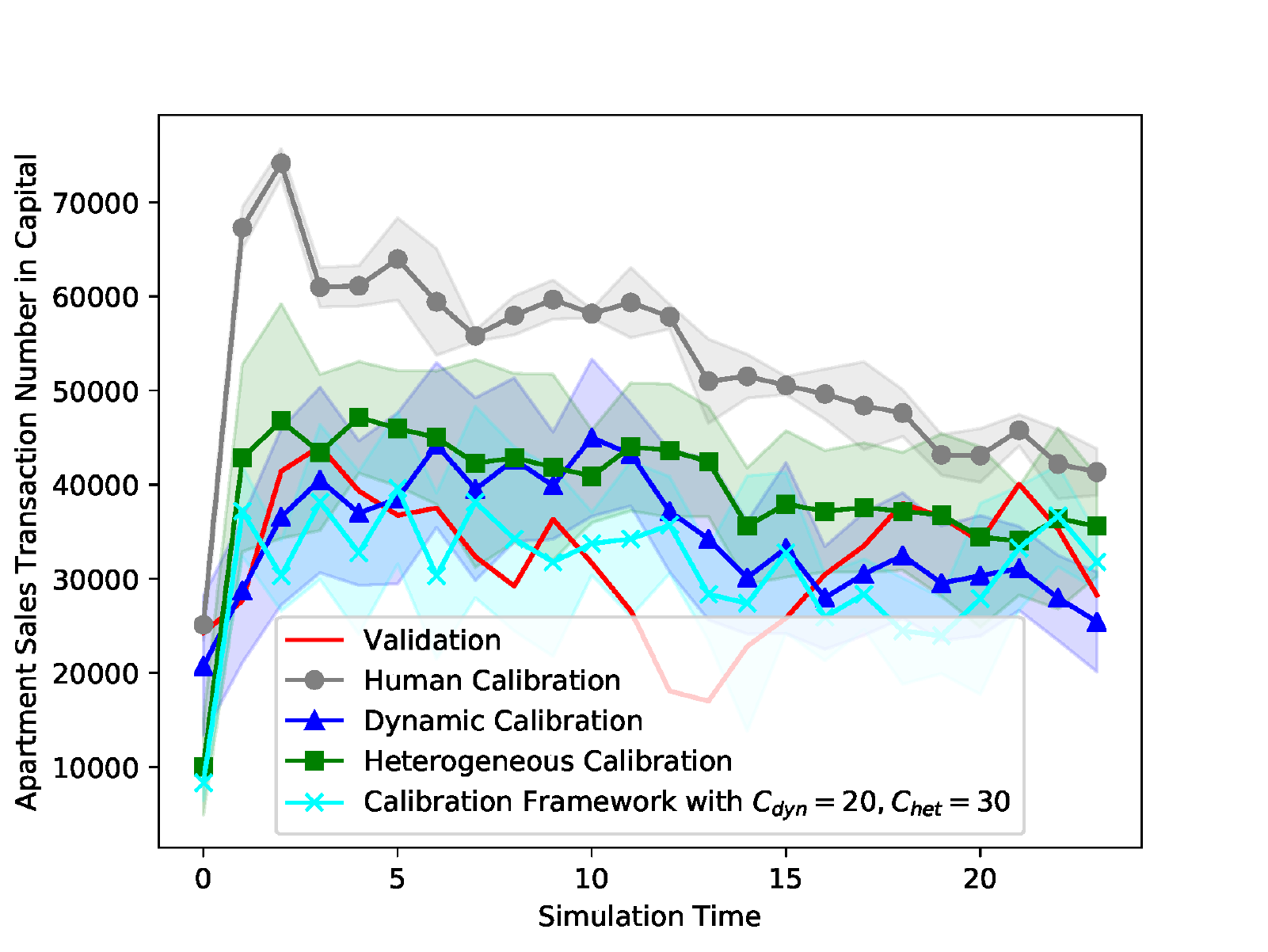}};
	\node[] (node2) at (0,-6) {\includegraphics[scale=0.5]{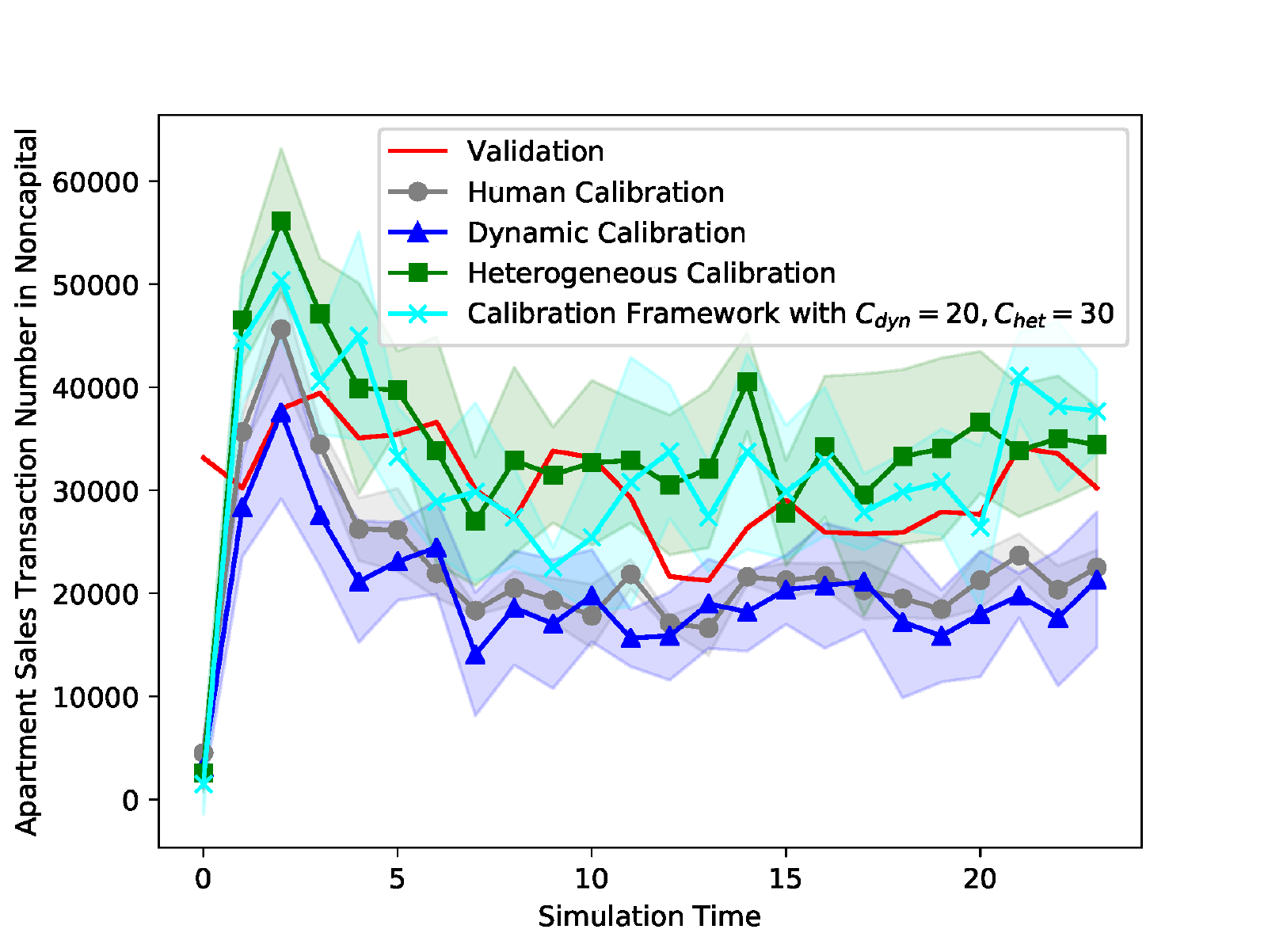}};  
	\node[] (node3) at (8,0) {\includegraphics[scale=0.5]{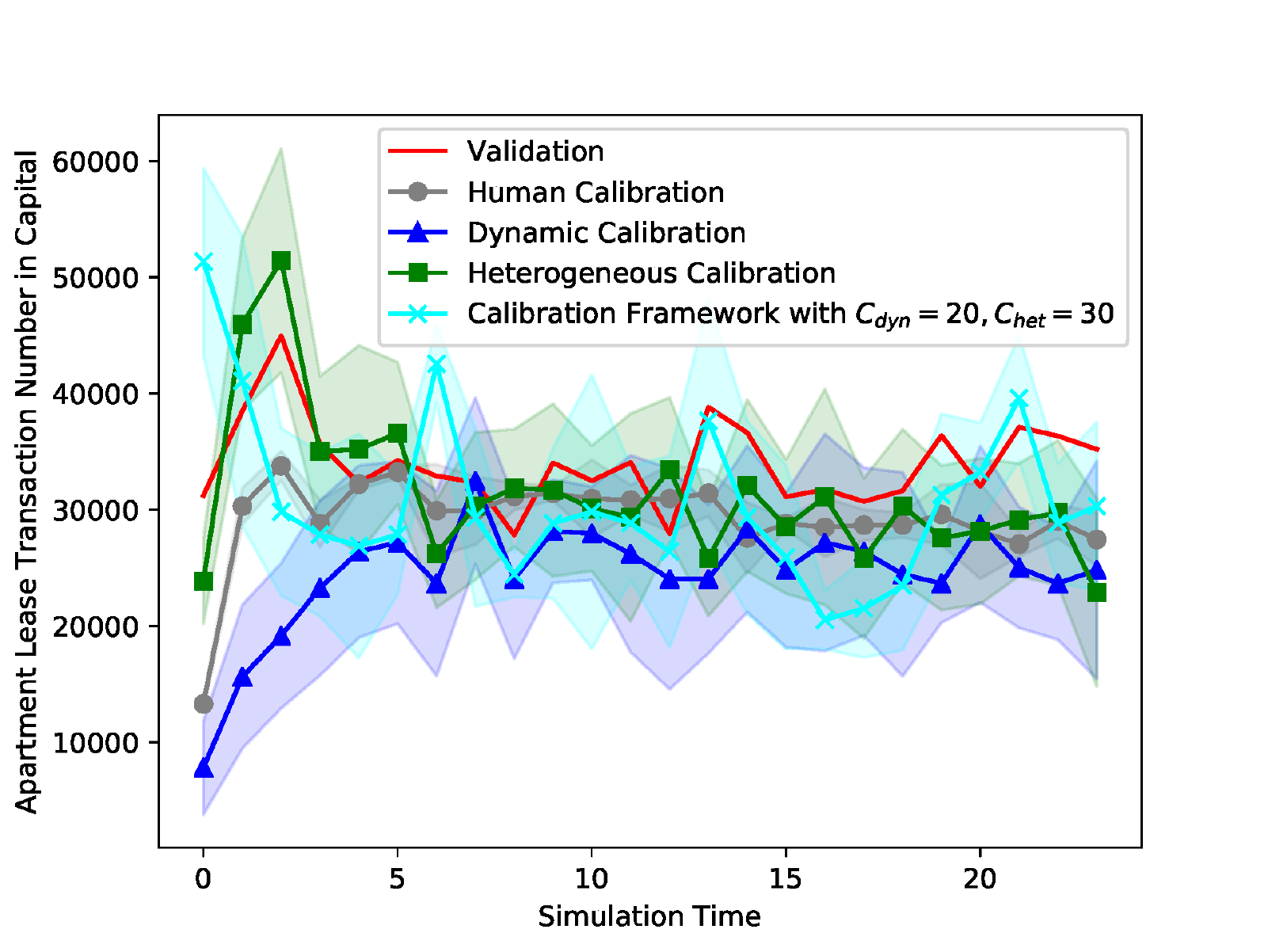}};
	\node[] (node4) at (8,-6) {\includegraphics[scale=0.5]{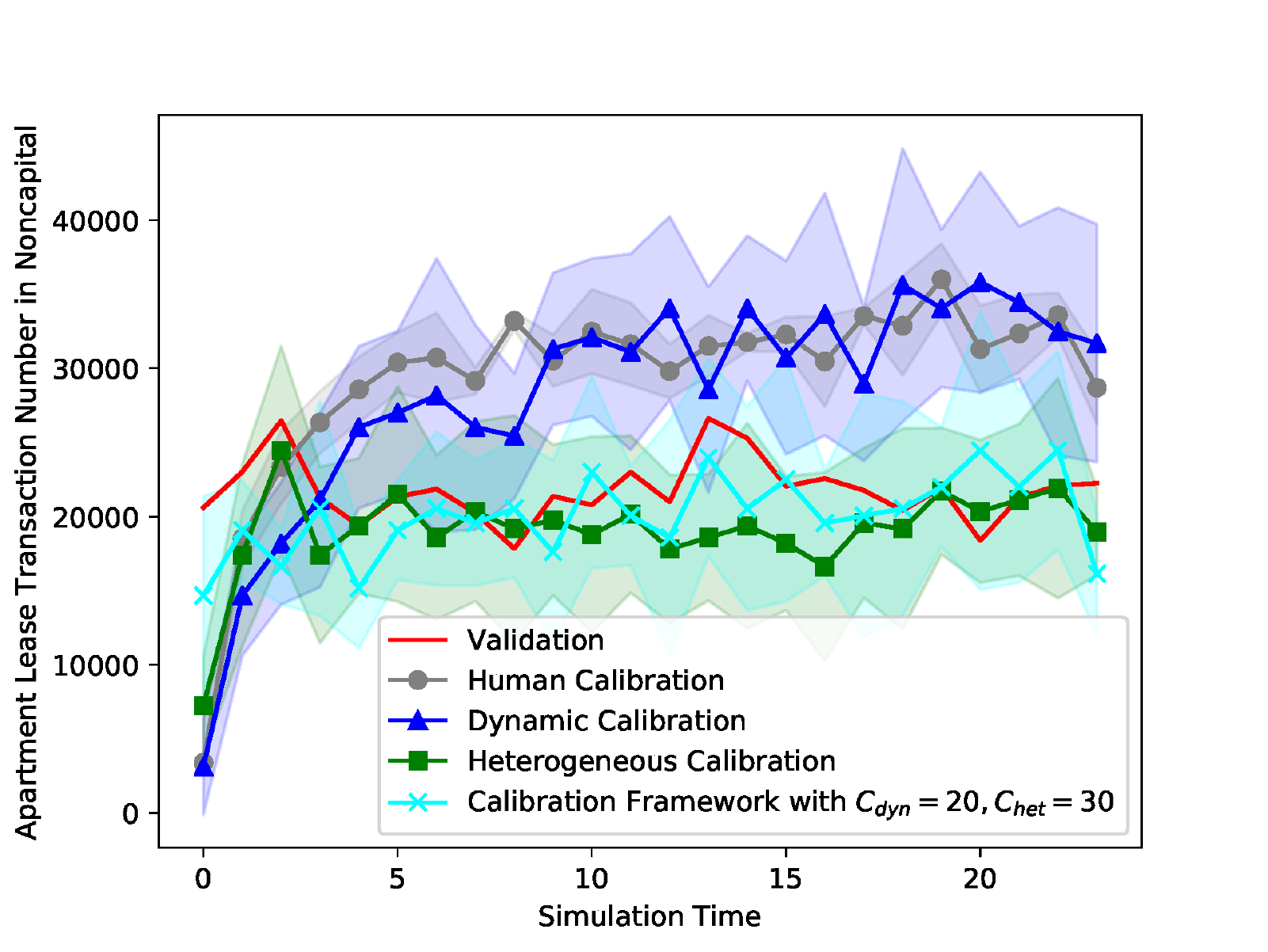}};   
	
	\node[] at (-4.0,2.7) {(a)};
	\node[] at (4.0,2.7) {(b)};
	\node[] at (-4.0,-3.3) {(c)};
	\node[] at (4.0,-3.3) {(d)};
	
	\end{tikzpicture} 
	\medskip
	\caption{Dynamic calibration (triangle marked), heterogeneous calibration (square marked), the calibration framework$^{\text{b}}$ (x marked) experiments are illustrated, with the human calibration result (circle marked) for comparison. (a) and (b) are the apartment transaction numbers of the capital area for sales and lease, respectively. (c) and (d) are the apartment transaction numbers of the noncapital area for sales and lease, respectively. All calibration methodologies outperforms the human calibration result in any of (a)-(d). Similarly, heterogeneous calibration outperforms dynamic calibration in all (a)-(d). The calibration framework$^{\text{b}}$ outperforms heterogeneous calibration in (a) and (d)}
	\label{fig:Heterogeneous_Results} 
	%\end{sidewaysfigure*}
\end{figure*}
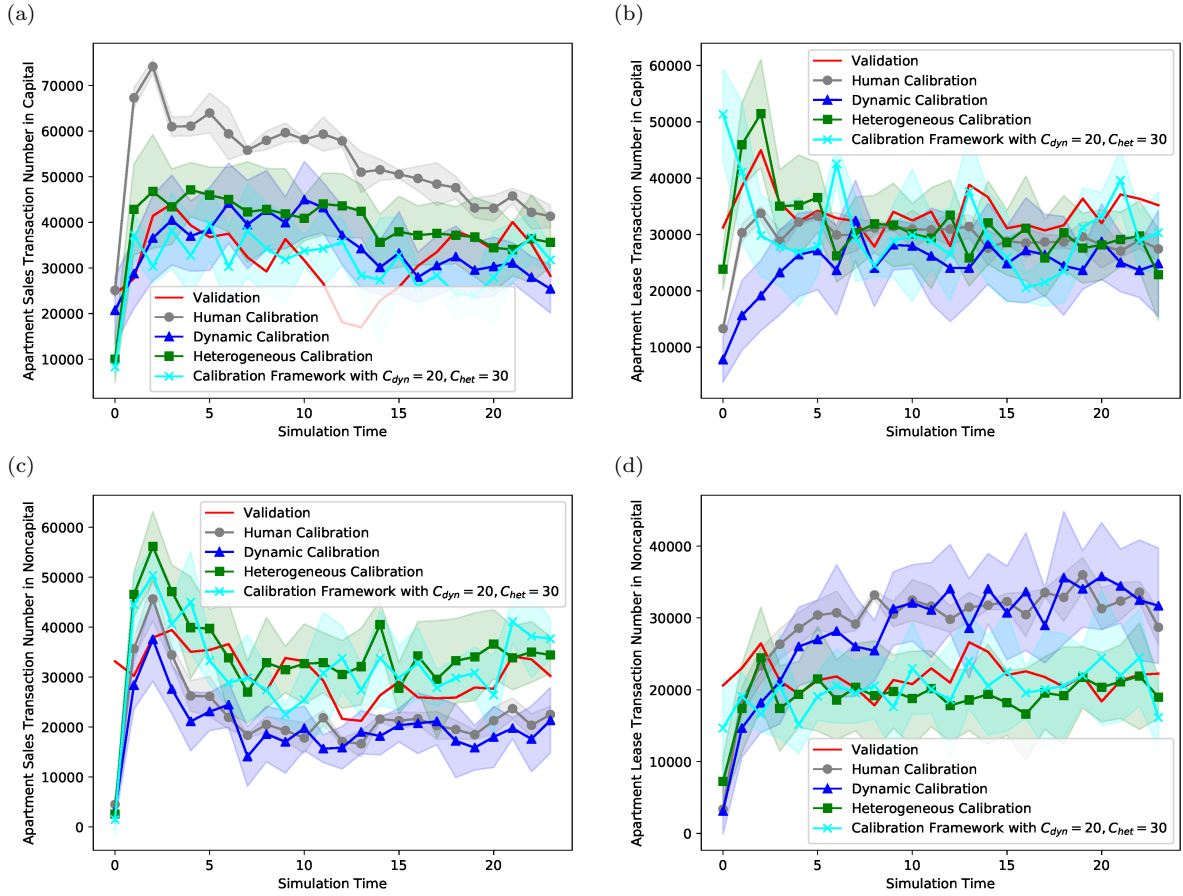

%\begin{sidewaysfigure*}

%In nonparametric clustering, the effect of the scaling parameter $\alpha$ in Eq. \ref{eq:DPMM}, which controls the number of clusters, is not evident. This is because the simulation state variables are not normally distributed. However, we apply $k$ to be 5$\%$ over the whole population in k-NN algorithm to handle small clusters occurred in outliers, which results in the moderate number of interpretable clusters.

Comparing with the human calibration, no significant improvement is attained by calibrating the heterogeneous parameters without agent clustering. However, when the heterogeneous parameters are differentiated with the agent sub-populations, errors are considerably improved. This is because one of the error source, agent heterogeneity, is mediated through differentiating parameter values by clusters, which generates a micro controller for the elaboration of the agent behaviors. %Compared to the dynamic calibration, the error improvement is even bigger.

Fig. \ref{fig:Cluster_Statistics} illustrates the detailed characteristics of agent clusters with radar charts. For example, the first cluster is the group of the leaseholder agents with middle-class salary, no debt, and low savings, who live in the apartments. As another example cluster, the third cluster is consisted of the house owners with high salary, high debt, and low savings, who live in the apartments located at the capital area.

When we calibrate all the heterogeneous parameters with agent clusters in Fig. \ref{fig:Cluster_Statistics}, the first and the second calibrated heterogeneous parameters have the same values in the first cluster, 0.9, which indicates the agents are trying to buy their houses sooner or later. In the third cluster, \textit{Willing to Pay} and \textit{Purchase Rate} have their values 0.3 and 0.9, respectively. Since the third gouped agents have their own house with high debts, they are unlikely to have high-risk investment propensity. The low \textit{Willing to Pay} indeed confirms their low-risk tendency, and high \textit{Purchase Rate} reflects the fact that they are already living in their own house.

Calibrating the first parameter enhances the simulation performance, since the first parameter, \textit{Willing to Pay}, directly controls the overall up and down of the transaction numbers. 
%The effect for the number of the clusters seems negligible in terms of the simulation error, but the interpretable parameters are only attainable from experimentations with enough number of clusters. 
The error improvement from the second parameter is not as significant as from the first parameter. The second parameter, \textit{Purchase Rate}, is only applicable for the minor agents whose lease contracts are finished. The simulation error gradually decreases as the number of the agent clusters increases, leading the nonparamtric clustering result performs the best. Although the two heterogeneous parameters plays different roles, there are considerable overlaps in their ramification on the simulation result. Thus, the calibration result with all the heterogeneous parameters reports the negligible error improvement, compare to the first parameter calibration.

\subsubsection{Calibration Framework Result}
\label{sec:CalibrationFrameworkResultTestCase2}

Tab. \ref{tab:statisticsinTestCase2} shows that the calibration framework outperforms both dynamic calibration alone and heterogeneous calibration alone. Also, the calibration framework$^{\text{b}}$ with two clusters is remarkably stable in its error improvement. However, despite of the lowest error, calibrating all the parameters are not as effective as we expected, since the first heterogeneous parameter calibration with eight clusters reports the analogous error, $0.107$, with the calibration framework$^{\text{b}}$ error, $0.105$. The reasons are two-folds. First, the price indices are relatively insensitive to the change of the parameter values, which leads that the error reduction relies mostly on the improvement in the transaction numbers. Second, the combination of the controller parameters on the transaction numbers hinders the response variables to be close to the validation, due to the increased dimensionality.

Fig. \ref{fig:Combined_Calibration} presents the simulation errors of the calibration frameowork$^{\text{b}}$ by calibration iterations. First, nearly 80 iterations for the random search are required to reach the human calibration result, and no significant improvement is made on the afterward iterations. Second, the calibration framework$^{\text{b}}$ defeats the human calibration within 20 iterations, in average. After superseding the human calibration, the calibration framework$^{\text{b}}$ saturates fast.

Fig. \ref{fig:Heterogeneous_Results} illustrates the calibration results of dynamic calibration, heterogeneous calibration, and the calibration framework$^{\text{b}}$. All the calibration results fit to the validation data better than the human calibration result. Heterogeneous calibration generates better fitted simulation results than dynamic calibration, in all transaction numbers. The calibration framework$^{\text{b}}$ fits better than heterogeneous calibration in (a) and (d).

\section{Conclusion}
\label{sec:Conclusion}

This paper proposes an automatic data-driven calibration framework for the dynamic and the heterogeneous parameters, by extracting the hidden structures from the simulation outputs. Dynamic calibration controls the dynamically switching parameters, by extracting the hidden dynamic regimes, and by treating each regimes separately. Heterogeneous calibration calibrates the agent cluster-wise heterogeneous parameters, by extracting the hidden agent-subpopulations. The experimental results on dynamic calibration and heterogeneous calibration demonstrate that the proposed separate calibrations reduce simulation error, with plausible estimated parameters. The calibration framework, the combination of dynamic calibration and heterogeneous calibration, exhibits the expected results in both test cases with both Keep-It-Simple-and-Stupid (KISS) model as well as elaborated and complex model of housing markets.

%\begin{acknowledgements}
%If you'd like to thank anyone, place your comments here
%and remove the percent signs.
%\end{acknowledgements}

% Authors must disclose all relationships or interests that 
% could have direct or potential influence or impart bias on 
% the work: 
%
% \section*{Conflict of interest}
%
% The authors declare that they have no conflict of interest.

% BibTeX users please use one of
\bibliographystyle{spbasic}      % basic style, author-year citations
\bibliography{reference}   % name your BibTeX data base

% Non-BibTeX users please use
%\begin{thebibliography}{}
%%
%% and use \bibitem to create references. Consult the Instructions
%% for authors for reference list style.
%%
%\bibitem{RefJ}
%% Format for Journal Reference
%Author, Article title, Journal, Volume, page numbers (year)
%% Format for books
%\bibitem{RefB}
%Author, Book title, page numbers. Publisher, place (year)
%% etc
%\end{thebibliography}

\end{document}